\documentclass[12pt]{article}

\usepackage{amsmath,amssymb,amsfonts,amsthm,hyperref,bbm,latexsym,epsfig}
\usepackage{graphicx,multirow}
\usepackage{cite,caption}

\newcommand{\bs}{\begin{subequations}}
\newcommand{\es}{\end{subequations}}
\newcommand{\be}{\begin{equation}}
\newcommand{\ee}{\end{equation}}
\newcommand{\ba}{\begin{eqnarray}}
\newcommand{\ea}{\end{eqnarray}}
\newcommand{\no}{\nonumber \\}

\allowdisplaybreaks
\newcommand{\Ymin}{Y_{\mathrm{min}}}
\newcommand{\Ymax}{Y_{\mathrm{max}}}

\begin{document}

\title{
\normalsize \hfill CFTP/21-011
\\[6mm]
\LARGE Two-body lepton-flavour-violating decays \\
in a 2HDM with soft family-lepton-number breaking}

\author{
\addtocounter{footnote}{2}
Darius~Jur\v{c}iukonis$^{(1)}$\thanks{\tt darius.jurciukonis@tfai.vu.lt}
\ {\normalsize and}
Lu\'is~Lavoura$^{(2)}$\thanks{\tt balio@cftp.tecnico.ulisboa.pt}
\\*[3mm]
$^{(1)} \! $
\small University of Vilnius,
\small Institute of Theoretical Physics and Astronomy, \\
\small Saul\.{e}tekio av.~3, LT-10222 Vilnius, Lithuania 
\\[2mm]
$^{(2)} \! $
\small Universidade de Lisboa, Instituto Superior T\'ecnico, CFTP, \\
\small Av.~Rovisco Pais~1, 1049-001 Lisboa, Portugal
\\*[2mm]
}

\date{\today}

\maketitle

\begin{abstract}
  We evaluate the decays $\ell_1^\pm \to \ell_2^\pm \gamma$,
  $Z \to \ell_1^+ \ell_2^-$,
  and $h \to \ell_1^+ \ell_2^-$,
  where $\ell_1$ and $\ell_2$ are charged leptons with different flavours
  and $h$ is the scalar particle with mass 125.25\,GeV,
  in a two-Higgs-doublet model
  where all the Yukawa-coupling matrices conserve the lepton flavours
  but the Majorana mass terms of the right-handed neutrinos
  break the flavour lepton numbers.
  We find that
  (1) the decays $\ell_1^\pm \to \ell_2^\pm \gamma$
  require large Yukawa couplings and very light right-handed neutrinos
  in order to be visible,
  (2) the decays $Z \to \ell_1^+ \ell_2^-$ will be invisible
  in all the planned experiments,
  except in a very restricted range of circumstances,
  but (3) the decays $h \to \ell_1^+ \ell_2^-$ 
  may be detected in future experiments
  for rather relaxed sets of input parameters.
\end{abstract}

\newpage
\tableofcontents

\newpage

\section{Introduction}
\label{Sect_Introd}

The well-established phenomenon
of neutrino oscillations~\cite{Fukuda:1998mi,Fukuda:2000np} implies
that the family lepton numbers are not unbroken symmetries of Nature.
Therefore,
other processes that violate those symmetries,
like the two-body decays $\ell_1^\pm \to \ell_2^\pm \gamma$,
$Z \to \ell_1^+ \ell_2^-$,
and $h \to \ell_1^+ \ell_2^-$ may occur.
(Here,
$h$ is the recently discovered scalar particle with mass $m_h = 125.25$\,GeV,
and $\ell_1$ and $\ell_2$ are charged leptons with different flavours.)
In the Standard Model (SM) those decays only appear at the one-loop level
and they are suppressed by a GIM-like mechanism~\cite{Glashow:1970gm},
due to the light-neutrino masses being very small
and almost identical when compared to the Fermi scale.
As a consequence,
in the SM those lepton-flavour-violating (LFV) decays
have very small rates and are,
in practice,
invisible.
This fact renders them all the more inviting to explore both experimentally,
as windows to New Physics,
and theoretically,
in extensions of the SM.

In table~\ref{LFV_limits} we display the nine LFV two-body decays,
the present upper bounds on their branching ratios,
and the expected sensitivity of some future experiments.
\begin{table}[h!]
  \centering
  \begin{tabular}{|c|cc|cc|}
    \hline
    Decay & Present & Experiment & Future & Experiment \\ \hline \hline
    $\tau^\pm \to \mu^\pm \gamma$ &
    $4.4 \times 10^{-8}$ & BABAR (2010) \cite{Aubert:2009ag} &
    $1\times 10^{-9}$ & BELLE-II \cite{Kou:2018nap}
    \\
    &  &  & $2\times 10^{-9}$ & FCC-ee \cite{Dam:2018rfz,Abada:2019lih}
    \\ \hline
    $\tau^\pm \to e^\pm \gamma$  &
    $3.3 \times 10^{-8}$ & BABAR (2010) \cite{Aubert:2009ag} &
    $3\times 10^{-9}$ & BELLE-II \cite{Kou:2018nap,Aushev:2010bq}
    \\ \hline
    $\mu^\pm \to e^\pm \gamma$ &
    $4.2 \times 10^{-13}$ & MEG (2016) \cite{TheMEG:2016wtm} &
    $6 \times 10^{-14}$ & MEG-II \cite{MEGII:2018kmf}
    \\ \hline \hline
    $Z \to \tau^\pm \mu^\mp$  &
    $1.2 \times 10^{-5}$ & DELPHI (1997) \cite{Abreu:1996mj}  &
    $\sim 10^{-6}$ & HL-LHC \cite{Dam:2018rfz}
    \\
    &  &  & $\sim 10^{-9}$ & FCC-ee \cite{Dam:2018rfz,Abada:2019lih}
    \\ \hline
    $Z \to \tau^\pm e^\mp$ &
    $9.8 \times 10^{-6}$ & OPAL (1995) \cite{Akers:1995gz} &
    $\sim 10^{-6}$ & HL-LHC \cite{Dam:2018rfz} \\
    &  &  & $\sim 10^{-9}$ & FCC-ee \cite{Dam:2018rfz,Abada:2019lih}
    \\ \hline
    $Z \to \mu^\pm e^\mp$ &
    $7.5 \times 10^{-7}$ & ATLAS (2014) \cite{Aad:2014bca}  &
    $\sim 7 \times 10^{-8}$ & HL-LHC \cite{Dam:2018rfz} \\
    &  &  & $\sim 10^{-10}$ & FCC-ee \cite{Dam:2018rfz,Abada:2019lih}
    \\ \hline \hline
    $h \to \tau^\pm \mu^\mp$ &
    $2.5 \times 10^{-3}$ & CMS (2018) \cite{Sirunyan:2017xzt}  &
    $1.4 \times 10^{-4}$ & FCC-ee \cite{Qin:2017aju} \\ \hline
    $h \to \tau^\pm e^\mp$ &
    $4.7 \times 10^{-3}$ & ATLAS (2019) \cite{Aad:2019ugc} &
    $1.6 \times 10^{-4}$ & FCC-ee \cite{Qin:2017aju} \\ \hline
    $h \to \mu^\pm e^\mp$ &
    $6.1 \times 10^{-5}$ & ATLAS (2019) \cite{ATLAS:2019xlq}    &
    $1.2 \times 10^{-5}$ & FCC-ee \cite{Qin:2017aju}
    \\ \hline
  \end{tabular}
  \caption{Present upper bounds and future sensitivities
    for the branching ratios of LFV decays.}
  \label{LFV_limits}   
\end{table}
Note that,
according to Ref.~\cite{Dam:2018rfz},
the HL-LHC experiment for $Z$ decays
will lead to improvements of about one order of magnitude
on the branching ratios from the full LHC samples;
we have indicated those general indications through signs $\sim$
in table~\ref{LFV_limits}.

In this paper we numerically compute
the above-mentioned decays
in a simple extension of the SM.
That extension is a particular case
of the scheme proposed in Ref.~\cite{Grimus:2002ux},
which is characterized by the following features:
\begin{itemize}
\item It is a multi-Higgs-doublet model.
\item It has three right-handed neutrinos (RH$\nu$),
  with Majorana masses that enable a type-I seesaw mechanism.
\item All the Yukawa-coupling matrices are diagonal in lepton-flavour space,
  because of the invariance of the dimension-four terms in the Lagrangian
  under the lepton-flavour symmetries.
\item The violation of the family lepton numbers arises only \emph{softly},
  through the dimension-three Majorana mass terms
  of the RH$\nu$.
\end{itemize}
In Ref.~\cite{Grimus:2002ux} the above-mentioned decays
have been computed analytically within that general scheme.
In this paper
we check that analytical computation,
but express the amplitudes through Passarino--Veltman (PV) functions.
That allows us to use the resulting formulas
in high-precision numerical computations
and to establish the impact of the separate amplitudes
on the branching ratios (BRs) of the LFV decays. 
Although our analytical results allow one to study the LFV decays
in a model with an arbitrary number of scalar doublets,
in this paper we only perform the numerical computation
in the context of a simple version of the two-Higgs-doublet model (2HDM).

The branching ratios of the LFV decays predicted by seesaw models like ours
are usually small due to the strong suppression from the very large
RH$\nu$ Majorana
masses~\cite{Ilakovac:1994kj,Arganda:2004bz}.
A recent paper~\cite{Marcano:2019rmk} about the LFV Higgs decays
in the framework of a general type-I seesaw model
with mass-insertion approximation
concludes that the maximal decay rates
are far from the current experimental bounds.
The inverse seesaw model,
a specific realization of low-scale seesaw models,
might yield larger decay
rates~\cite{Arganda:2014dta,Arganda:2017vdb,DeRomeri:2016gum,Herrero:2018luu}.

There is a large number of theoretical papers on the LFV decays,
therefore we refer only to some of them,
grouping them according to the decays under consideration,
since most of the research has been conducted on individual types of decays:
\begin{itemize}
\item LFV decays of charged leptons were analyzed
  in the context of the inverse seesaw model~\cite{Arganda:2014dta},
  of effective field
  theory~\cite{Davidson:2012cwa,Pruna:2014asa,Davidson:2016utf,Dekens:2018pbu},
  of 2HDMs~\cite{Paradisi:2005tk,Davidson:2010xv,Davidson:2016utf},
  and of the flipped 3--3--1 model~\cite{Hong:2020qxc}.
  The current experimental and theoretical situation for these decays
  is reviewed in Ref.~\cite{Calibbi:2017uvl}.
\item The LFV $Z$ decay rates have been computed in frameworks
  with massive Majorana neutrinos~\cite{Korner:1992an,Illana:2000ic,Hernandez-Tome:2019lkb},
  in the inverse seesaw model~\cite{DeRomeri:2016gum,Herrero:2018luu},
  effective field theory~\cite{FloresTlalpa:2001sp,Davidson:2012wn,Coy:2018bxr,Calibbi:2021pyh},
  a general 2HDM~\cite{Iltan:2001au},
  and the minimal 3--3--1 model~\cite{CortesMaldonado:2011uh}.
\item LFV Higgs decays were analyzed in the framework of the inverse seesaw
  model~\cite{Pilaftsis:1992st,Arganda:2014dta,Arganda:2017vdb,Thao:2017qtn,Arganda:2015naa},
  2HDM~\cite{Sierra:2014nqa,Bizot:2015qqo,Crivellin:2015hha,Liu:2015oaa,Botella:2015hoa,Omura:2015nja,Tobe:2016qhz,Hou:2020tgl},
  effective field theory~\cite{deLima:2015pqa},
  3--3--1 models~\cite{Hong:2020qxc,Nguyen:2018rlb},
  in models with TeV sterile neutrinos~\cite{Hernandez-Tome:2020lmh}
  and models with $L_\mu - L_\tau$
  symmetry~\cite{Altmannshofer:2016oaq,Chen:2017cic}.
  We also refer to the recent
  review
  by Vicente~\cite{Vicente:2019ykr}.
\end{itemize}
As mentioned above,
the three types of LFV decays have been analyzed mostly separately,
but there are also studies
that endeavour to combine
all three types
together~\cite{Crivellin:2013wna,Benbrik:2015evd}.
Correlations among separate decay rates may exist,
and some LFV decays may be constrained by other LFV decays.
Some constraints could appear in several models,
while other constraints operate only in specific models.
For example,
Ref.~\cite{DeRomeri:2016gum} shows that
$Z \to \tau^\pm \mu^\mp$ is constrained by
$\tau^\pm \to \mu^\pm \gamma$ in the inverse seesaw model;
constraints on the $Z$ decays from the LFV decays of charged leptons
also emerge in the 2HDM~\cite{Iltan:2001au}
and in the minimal 3--3--1 model~\cite{CortesMaldonado:2011uh}.
The authors of Ref.~\cite{Crivellin:2015hha} claim that
$h \to \tau^\pm \mu^\mp$
is constrained by $\tau^\pm \to \mu^\pm \gamma$ in their specific 2HDM,
but in Ref.~\cite{Sierra:2014nqa} no such constraints have been found
in the type-III 2HDM.

In this paper we perform the numerical study of all nine LFV two-body decays
($\tau^\pm \to \mu^\pm \gamma$,
$\tau^\pm \to e^\pm \gamma$,
$\mu^\pm \to e^\pm \gamma$,
$Z / h \to \tau^\pm \mu^\mp$,
$Z / h \to \tau^\pm e^\mp$,
and $Z / h\to \mu^\pm e^\mp$)
in the context of the 2HDM with seesaw mechanism
and flavour-conserving Yukawa couplings.
Our purpose is to see under which circumstances
the decay rates might be close
to their present experimental upper bounds---namely,
whether one has to resort to
either very large or very small
Yukawa couplings,
to a very low mass of the charged scalar
of the 2HDM,
or to very low RH$\nu$ masses.
We want to elucidate which are the very relevant
and the less relevant parameters of that model for the LFV decays.

In section~\ref{Sect_ourModelGeneral}
we review both the scalar and the leptonic sectors of our model.
Section~\ref{Sect_NumResults} contains our main numerical results.
The findings of this paper
are summarized in section~\ref{Sect_Conclusions}.
The Passarino--Veltman functions relevant for the analytic computation
are expounded in appendix~\ref{Sect_PVfunctions}.
The full one-loop analytical formulas for the LFV decays
in terms of PV functions are collected in appendices~\ref{Sect_taumugamma},
\ref{Sect_taumuZ},
and~\ref{Sect_taumuS}.
Appendix~\ref{Sect_Z}
makes a digression
on the invisible $Z$ decay width
and appendix~\ref{Sect_mH+} reviews some literature
on lower bounds on the charged-Higgs mass.

\section{The model}
\label{Sect_ourModelGeneral}

\subsection{Scalar sector}
\label{Sect_scalarSect}

\subsubsection{The matrices $\mathcal{U}$ and $\mathcal{V}$}

In general,\footnote{Soon we shall restrict the model to $n_d = 2$.}
we assume the existence of $2 n_d$ scalar $SU(2)$ doublets
\be
\Phi_k = \left( \begin{array}{c} \varphi_k^+ \\*[1mm]
  \varphi_k^0 \end{array} \right),
\quad \quad
\tilde \Phi_k = \left( \begin{array}{c} {\varphi_k^0}^\ast \\
  - {\varphi_k^+}^\ast \end{array} \right)
\quad \quad
(k = 1, \ldots, n_d).
\ee
We assume that no other scalar fields exist,
except possibly $SU(2)$ singlets of charge either $0$ or $\pm1$. 
The neutral fields $\varphi_k^0$ have vacuum expectation values (VEVs)
$v_k \left/ \sqrt{2} \right.$ that may be complex.
We use the formalism of Ref.~\cite{Grimus:1989pu},
that was further developed in Refs.~\cite{Grimus:2002ux}
and~\cite{Grimus:2007if}.
The scalar eigenstates of mass are
$n$ charged scalars $H_a^+$ ($a = 1, \ldots, n$)
and $m$ real neutral scalars $S^0_b$ ($b = 1, \ldots, m$),
with $n \ge n_d$ and $m \ge 2 n_d$.
The fields $\varphi_k^+$ and $\varphi_k^0$
are superpositions of the eigenstates of mass according to
\be
\label{mvjdkfo}
\varphi_k^+ = \sum_{a=1}^n \mathcal{U}_{ka} H_a^+,
\quad \quad
\varphi_k^0 = \frac{1}{\sqrt{2}} \left( v_k
+ \sum_{b=1}^m \mathcal{V}_{kb} S_b^0 \right).
\ee
The matrix $\mathcal{U}$ is $n_d \times n$
and the matrix $\mathcal{V}$ is $n_d \times m$.
In general,
they are not unitary;
however,
there are matrices
\be
\tilde{\mathcal{U}} = \left( \begin{array}{c}
  \mathcal{U} \\ \mathcal{T}
\end{array} \right),
\quad \quad
\tilde{\mathcal{V}} = \left( \begin{array}{c}
  \mathrm{Re}\, \mathcal{V} \\
  \mathrm{Im}\, \mathcal{V} \\
  \mathcal{R}
\end{array} \right)
\label{RT}
\ee
that are $n \times n$ unitary and $m \times m$ real orthogonal,
respectively.
The matrices $\mathcal{T}$ and $\mathcal{R}$
account for the possible presence in the model
of charged-scalar $SU(2)$ singlets and of scalar gauge invariants,
respectively.
The unitarity of $\tilde{\mathcal{U}}$ and the orthogonality
of $\tilde{\mathcal{V}}$ imply
\be
\mathcal{U} \mathcal{U}^\dagger =
\mathrm{Re}\, \mathcal{V}\ \mathrm{Re}\, \mathcal{V}^T =
\mathrm{Im}\, \mathcal{V}\ \mathrm{Im}\, \mathcal{V}^T =
1_{n_d \times n_d},
\quad \quad
\mathrm{Re}\, \mathcal{V}\ \mathrm{Im}\, \mathcal{V}^T = 0_{n_d \times n_d}.
\ee

By definition,
$H_1^+ := G^+$ and $S_1^0 := G^0$ are the `would-be Goldstone bosons'.
Hence~\cite{Grimus:2007if},
\be
\label{k1}
\mathcal{U}_{k1} = \frac{v_k}{v},
\quad \quad
\mathcal{V}_{k1} = i\, \frac{v_k}{v},
\ee
where
\be
\label{v}
v := \sqrt{\sum_{k=1}^{n_d} \left| v_k \right|^2}
= \frac{2 s_w m_W}{e} = \frac{2 c_w s_w m_Z}{e}.
\ee
In Eq.~\eqref{v},
$s_w$ and $c_w$ are the sine and the cosine,
respectively,
of the weak mixing angle,
and $e$ is the electric charge of the proton.
Clearly,
because of Eqs.~\eqref{k1} and~\eqref{v},
\be
\label{jvifdosp}
\sum_{k=1}^{n_d} \left| \mathcal{U}_{k1} \right|^2
= \sum_{k=1}^{n_d} \left| \mathcal{V}_{k1} \right|^2 = 1.
\ee
Thus,
$\left( \mathcal{V}^\dagger \mathcal{V} \right)_{11} = 1$.

In Eq.~\eqref{RT},
$\tilde{\mathcal{U}}$ is unitary and $\tilde{\mathcal{V}}$ is orthogonal.
Hence,
because of Eq.~\eqref{jvifdosp},
the first columns of $\mathcal{T}$ and $\mathcal{R}$ are identically zero.
Therefore the orthogonality of $\tilde{\mathcal{V}}$ implies that,
for $b \neq 1$,
\ba
0 &=& \sum_{k=1}^{n_d} \left(
\mathrm{Re}\, \mathcal{V}_{k1}\, \mathrm{Re}\, \mathcal{V}_{kb}
+
\mathrm{Im}\, \mathcal{V}_{k1}\, \mathrm{Im}\, \mathcal{V}_{kb}
\right)
\no &=&
\sum_{k=1}^{n_d} \mathrm{Re} \left( \mathcal{V}_{k1}^\ast \mathcal{V}_{kb} \right)
\no &=&
\sum_{k=1}^{n_d} \mathrm{Re} \left( - i\, \frac{v_k^\ast}{v}\,
\mathcal{V}_{kb} \right)
\no &=&
\sum_{k=1}^{n_d} \mathrm{Im} \left( \frac{v_k^\ast}{v}\,
\mathcal{V}_{kb} \right).
\ea
Thus,
\be
\label{xb}
x_b := \frac{1}{v}\, \sum_{k=1}^{n_d} v_k^\ast \mathcal{V}_{kb}
\quad \quad (b \neq 1)
\ee
is real.
So,
$\left( \mathcal{V}^\dagger \mathcal{V} \right)_{1b} = - i x_b$
is imaginary for all $b \neq 1$.

\subsubsection{Some interactions of the scalars}

The parameters $x_b$ in Eq.~\eqref{xb}
are important because they appear in the interaction
of the neutral scalars $S^0_b$ with two $W$ gauge bosons~\cite{Grimus:2007if}:
\be
\label{vuffsid9}
\mathcal{L} = \cdots + \frac{e m_W}{s_w}\,
W_\mu^- W^{\mu +} \sum_{b=2}^{m} x_b S_b^0.
\ee

Another important interaction
is the one of a $W$ gauge boson with one neutral scalar
and one charged scalar.
It is given by~\cite{Grimus:2002ux}
\bs
\label{vfi493586}
\ba
\mathcal{L} &=& \cdots + i\, \frac{e}{2 s_w}\, \sum_{a=1}^n \sum_{b=1}^m \left[
  \left( \mathcal{U}^\dagger \mathcal{V} \right)_{ab} W_\mu^+
  \left( H_a^-\, \partial^\mu S_b^0 - S_b^0\, \partial^\mu H_a^- \right)
  \right. \\ & & \left.
  + \left( \mathcal{V}^\dagger \mathcal{U} \right)_{ba} W_\mu^-
  \left( S_b^0\, \partial^\mu H_a^+ - H_a^+\, \partial^\mu S_b^0 \right)
  \right].
\ea
\es

Also relevant in this paper is the interaction of a neutral scalar
with two charged scalars.
We parameterize it as
\be
\label{3vifogp}
\mathcal{L} = \cdots + \sum_{a, a^\prime = 1}^n \sum_{b=1}^m
\lambda_{a a^\prime b} H_a^- H_{a^\prime}^+ S_b^0,
\ee
where the coefficients obey $\lambda_{a a^\prime b} = \lambda_{a^\prime a b}^\ast$
because of the Hermiticity of $\mathcal{L}$.
Equation~\eqref{3vifogp} corresponds,
when either $a=1$ or $a^\prime = 1$,
to an interaction of the charged would-be Goldstone bosons.
The coefficients for those interactions
may be shown---either by gauge invariance or indeed through an analysis
of the scalar potential~\cite{Grimus:2002ux}---to be
\bs
\label{vjfifdop0}
\ba
\lambda_{1ab} &=& \frac{e \left( m_a^2 - m_b^2 \right)}{2 s_w m_W}\,
\left( \mathcal{V}^\dagger \mathcal{U} \right)_{ba},
\\
\lambda_{a1b} &=& \frac{e \left( m_a^2 - m_b^2 \right)}{2 s_w m_W}\,
\left( \mathcal{U}^\dagger \mathcal{V} \right)_{ab},
\\
\lambda_{11b} &=& \frac{- e m_b^2}{2 s_w m_W}\, x_b,
\ea
\es
where $m_a$ is the mass of the charged scalar $H_a^+$
and $m_b$ is the mass of the neutral scalar $S_b^0$.

\subsection{Leptonic sector}
\label{Sect_leptonSect}

\subsubsection{The matrices $U$ and $X$}

We assume the existence of three right-handed neutrinos $\nu_{\ell R}$,
where $\ell = e, \mu, \tau$.
We assume that \emph{the flavour lepton numbers are conserved
in the Yukawa Lagrangian of the leptons}:
\be
\mathcal{L}_\mathrm{Y} = - \sum_{k=1}^{n_d}\ \sum_{\ell = e, \mu, \tau}
\left[ \Phi_k^\dagger\, \bar{\ell}_R \left( \Gamma_k \right)_{\ell \ell}
  + \tilde{\Phi}_k^\dagger\, \bar{\nu}_{\ell R}
  \left( \Delta_k \right)_{\ell \ell} \right]
\left( \begin{array}{c} \nu_{\ell L} \\*[1mm] \ell_L
\end{array} \right)
+ \mathrm{H.c.}
\ee
All the $2 n_d$ matrices $\Gamma_k$ and $\Delta_k$ are diagonal by assumption.
The charged-lepton mass matrix $M_\ell$
and the neutrino Dirac mass matrix $M_D$ are
\be
\label{MlMD}
M_\ell = \sum_{k=1}^{n_d}\, \frac{v_k^\ast}{\sqrt{2}}\, \Gamma_k,
\quad \quad
M_D = \sum_{k=1}^{n_d}\, \frac{v_k}{\sqrt{2}}\, \Delta_k,
\ee
respectively.
The matrices $M_\ell$ and $M_D$ are diagonal just as the matrices $\Gamma_k$
and $\Delta_k$,
respectively.
Without loss of generality,
we choose the phases of the fields $\ell_R$
in such a way that the diagonal matrix elements of $M_\ell$ are real and positive,
\textit{viz.}\ they are the charged-lepton masses;
thus,
\be
\sum_{k=1}^{n_d} v_k^\ast \left( \Gamma_k \right)_{\ell \ell} = \sqrt{2} m_\ell
\quad \quad (\ell = e, \mu, \tau).
\ee
The neutrino mass terms are
\ba
\mathcal{L}_{\nu\, \mathrm{mass}} &=&
- \left( \begin{array}{ccc} \bar \nu_{eR}, & \bar \nu_{\mu R}, &
  \bar \nu_{\tau R} \end{array} \right)
  M_D \left( \begin{array}{c} \nu_{eL} \\ \nu_{\mu L} \\
    \nu_{\tau L} \end{array} \right)
  \nonumber \\*[1mm] & &
  - \frac{1}{2} \left( \begin{array}{ccc} \bar \nu_{eR}, & \bar \nu_{\mu R}, &
    \bar \nu_{\tau R} \end{array} \right) M_R
  \left( \begin{array}{c} C \bar \nu_{eR}^T \\*[1mm] C \bar \nu_{\mu R}^T \\*[1mm]
  C \bar \nu_{\tau R}^T \end{array} \right)
+ \mathrm{H.c.},
\ea
where $C$ is the charge-conjugation matrix in Dirac space.
The flavour-space matrix $M_R$ is \emph{non-diagonal}\/ and symmetric;
it is the sole origin of lepton mixing in this model.

There are six physical Majorana neutrino fields $\nu_i = C \bar{\nu}_i^T$
($i = 1, \ldots, 6$).
The three $\nu_{\ell L}$ and the three $\nu_{\ell R}$
are superpositions thereof~\cite{Grimus:2002ux}:
\be
\nu_{\ell L} = \sum_{i=1}^6 U_{\ell i} P_L \nu_i,
\quad \quad
\nu_{\ell R} = \sum_{i=1}^6 X_{\ell i} P_R \nu_i,
\ee
where $P_L := \left. \left( 1 - \gamma_5 \right) \right/ 2$
and $P_R := \left. \left( 1 + \gamma_5 \right) \right/ 2$
are the projectors of chirality.
The matrices $U$ and $X$ are $3 \times 6$.
The matrix
\be
\label{U6}
U_6 := \left( \begin{array}{c} U \\ X^\ast \end{array} \right)
\ee
is $6 \times 6$ and unitary,
hence
\be
\label{vklspac1}
U U^\dagger = 1_{3 \times 3},
\quad \quad
U^\dagger U + X^T X^\ast = 1_{6 \times 6}.
\ee
The matrix $U_6$ diagonalizes the full $6 \times 6$ neutrino mass matrix as
\be
\label{dnksps}
U_6^T \left( \begin{array}{cc} 0_{3 \times 3} & M_D^T
  \\ M_D & M_R \end{array} \right) U_6
= \hat{m} := \mathrm{diag} \left( m_1, \ldots, m_6 \right).
\ee
In Eq.~\eqref{dnksps},
the $m_i$ ($i = 1, \ldots, 6$) are non-negative real;
$m_i$ is the mass of the neutrino $\nu_i$.
From Eq.~\eqref{dnksps},
\be
\label{vklspac2}
U \hat{m} U^T = 0_{3 \times 3},
\quad \quad
X \hat{m} U^\dagger = M_D.
\ee
The matrix $M_D$ is diagonal.
Therefore,
$M_D^\dagger M_D
= U \hat m \left( 1_{6 \times 6} - U^T U^\ast \right) \hat m U^\dagger
= U \hat{m}^2 U^\dagger$ is diagonal.
It follows from Eqs.~\eqref{vklspac1} and~\eqref{vklspac2} that
\be
X^\dagger M_D U = \hat{m} U^\dagger U.
\label{idpqow}
\ee
Equation~\eqref{dnksps} implies $M_D^T X^\ast = U^\ast \hat{m}$.
Therefore,
\be
\label{nvkdoso}
X_{\ell i} = U_{\ell i}\, \frac{m_i}{\left( M_D^\ast \right)_{\ell \ell}}.
\ee

\subsubsection{The interactions of the leptons}

The charged-current Lagrangian is
\be
\label{ncksod}
\mathcal{L}_\mathrm{cc} = \frac{e}{\sqrt{2} s_w}
\sum_{\ell = e, \mu, \tau} \sum_{i=1}^6 \left(
  W_\sigma^- U_{\ell i}\, \bar{\ell} \gamma^\sigma P_L \nu_i
  + W_\sigma^+ U_{\ell i}^\ast\, \bar{\nu}_i \gamma^\sigma P_L \ell \right).
\ee
The neutral-current Lagrangian is
\bs
\ba
\mathcal{L}_\mathrm{nc} &=& \frac{e Z_\sigma}{2 c_w s_w}
\sum_{\ell = e, \mu, \tau} \bar{\ell}\, \gamma^\sigma
\left[ \left( s_w^2 - c_w^2 \right) P_L + 2 s_w^2 P_R \right] \ell
\label{n2} \\ & &
+ \frac{e Z_\sigma}{4 c_w s_w}
\sum_{i, j = 1}^6 \bar{\nu}_i\, \gamma^\sigma
\left( q_{ij} P_L - q_{ji} P_R \right) \nu_j,
\label{Z}
\ea
\es
where
\be
\label{qij}
q_{ij} := \left( U^\dagger U \right)_{ij}.
\ee
When extracting the Feynman rule for the vertex from line~\eqref{Z},
one must multiply by a factor 2 because the $\nu_i$ are Majorana fields.

The charged scalars interact with the charged leptons
and the neutrinos through
\be
\mathcal{L}_\mathrm{Y} \supset
\sum_{a=1}^n \sum_{\ell = e, \mu, \tau} \sum_{i=1}^6 \left[
H_a^-\, \bar \ell \left( R_{a \ell i} P_R - L_{a \ell i} P_L \right)
\nu_i
+ H_a^+\, \bar \nu_i \left( R^\ast_{a \ell i} P_L - L^\ast_{a \ell i} P_R \right)
\ell \right].
\label{mvjckp}
\ee
The coefficients in Eq.~\eqref{mvjckp} are given by
\be
R_{a \ell i} = \sum_{k=1}^{n_d} \mathcal{U}_{ka}^\ast
\left( \Delta_k^\dagger X \right) _{\ell i},
\quad \quad
L_{a \ell i} = \sum_{k=1}^{n_d} \mathcal{U}_{ka}^\ast
\left( \Gamma_k U \right)_{\ell i}.
\ee
The neutral scalars interact with the charged leptons and with the neutrinos
through
\bs
\label{mvkxpso}
\ba
\mathcal{L}_\mathrm{Y} &\supset&
- \sum_{b=1}^m \sum_{\ell = e, \mu, \tau} \frac{S_b^0}{\sqrt{2}}\
\bar{\ell}
\left( g_{b \ell} P_L + g_{b \ell}^\ast P_R \right)
\ell
\label{nvjfui} \\ & &
- \sum_{b=1}^m \sum_{i, j = 1}^6 \frac{S_b^0}{2 \sqrt{2}}\
\bar{\nu_i} \left( f_{bij} P_L + f_{bij}^\ast P_R \right) \nu_j.
\label{jdsiao}
\ea
\es
When extracting the Feynman rule for the vertex from line~\eqref{jdsiao},
one must multiply by a factor 2 because the $\nu_i$ are Majorana fields.
The coefficients in Eq.~\eqref{mvkxpso} are given by
\be
g_{b \ell} = \sum_{k=1}^{n_d} \mathcal{V}_{kb}^\ast
\left( \Gamma_k \right)_{\ell \ell},
\quad \quad
f_{bij} = \sum_{k=1}^{n_d} \mathcal{V}_{kb}
\left( X^\dagger \Delta_k U + U^T \Delta_k X^\ast \right)_{ij}.
\label{fjdofd0}
\ee
Notice that $f_{bij} = f_{bji}$.

The reader may now appreciate the practical computation of the amplitudes for
$\ell_1^\pm \to \ell_2^\pm \gamma$ (appendix~\ref{Sect_taumugamma}),
$Z \to \ell_1^+ \ell_2^-$ (appendix~\ref{Sect_taumuZ}),
and $S^0_b \to \ell_1^+ \ell_2^-$ (appendix~\ref{Sect_taumuS}).
Those amplitudes are expressed in terms of the Passarino--Veltman functions
defined
in appendix~\ref{Sect_PVfunctions}.

\subsection{Restriction to a two-Higgs-doublet model}
\label{Sect_2HDM}

In the numerical computations in this paper,
we work in the context of a two-Higgs-doublet model
without any scalar $SU(2)$ singlets.
We use,
without loss of generality,
the `Higgs basis',
wherein only the first scalar doublet has a VEV,
and moreover that VEV is real and positive:
\be
\Phi_1 = \left( \begin{array}{c} G^+ \\
  \left( v + \rho_1 + i G^0 \right) \left/ \sqrt{2} \right.
\end{array} \right),
\quad \quad
\Phi_2 = \left( \begin{array}{c} H^+ \\
  \left( \rho_2 + i \eta \right) \left/ \sqrt{2} \right.
\end{array} \right).
\ee
In this basis,
$G^+ = H_1^+$ is the charged would-be Goldstone boson
and $H^+ = H_2^+$ is a physical charged scalar.
Thus,
the matrix $\mathcal{U}$ defined through Eq.~\eqref{mvjdkfo}
is the $2 \times 2$ unit matrix.
Moreover,
$G^0 = S^0_1$ is the neutral would-be Goldstone boson,
and~\cite{Lavoura:1994fv}
\be
\left( \begin{array}{c} S_2^0 \\ S_3^0 \\ S_4^0 \end{array} \right)
= T
\left( \begin{array}{c} \rho_1 \\ \rho_2 \\ \eta \end{array} \right),
\ee
where $T$ is a real orthogonal $3 \times 3$ matrix.
Without loss of generality,
we restrict $T_{11}$,
$T_{21}$,
and $T_{31}$ to be non-negative---this corresponds to
a choice for the signs of $S_2^0$,
$S_3^0$,
and $S_4^0$,
respectively.
Without loss of generality,
we choose the phase of the doublet $\Phi_2$ in such a way that
$T_{12} + i T_{13}$ is real and non-negative;
thus,
$T_{12} + i T_{13} = \sqrt{1 - T_{11}^2}$.
The matrix $\mathcal{V}$ defined through Eq.~\eqref{mvjdkfo} is given by
\be
\mathcal{V} = \left( \begin{array}{cccc}
  i & T_{11} & T_{21} & T_{31} \\
  0 & \sqrt{1 - T_{11}^2} & T_{22} + i T_{23} & T_{32} + i T_{33}
\end{array} \right).
\ee
Then,
\be
\mathcal{V}^\dagger \mathcal{V} = \left( \begin{array}{cccc}
  1 & - i T_{11} & - i T_{21} & - i T_{31} \\
  i T_{11} & 1 & \pm i T_{31} & \mp i T_{21} \\
  i T_{21} & \mp i T_{31} & 1 & \pm i T_{11} \\
  i T_{31} & \pm i T_{21} & \mp i T_{11} & 1
\end{array} \right)
\quad \quad \mbox{for}\ \det{T} = \pm 1.
\ee

We are interested only in $S_2^0$,
\textit{viz.}\ in the index $b=2$.
Through the definition~\eqref{xb},
\be
T_{11} = x_2.
\ee
%
From now on we will only use $x_2$
and we will not mention $T$ and its matrix elements again.
We use the notation $m_h$ for the mass of $S_2^0$;
since $S_2^0$ is supposed to be the scalar discovered at the LHC,
$m_h = 125.25$\,GeV.
We use the notation $m_{H^+}$ for the mass of the charged scalar $H^+$.
The neutral scalars $S_3^0$ and $S_4^0$
are unimportant in this paper.

We use the following notation~\cite{Aeikens:2020acn}
for the matrix elements of $\Gamma_{1,2}$ and $\Delta_{1,2}$:

\be
\left( \Gamma_2 \right)_{\ell \ell} = \gamma_\ell,
\quad \quad
\left( \Delta_1 \right)_{\ell \ell} = d_\ell,
\quad \quad
\left( \Delta_2 \right)_{\ell \ell} = \delta_\ell,
\ee
while $\left( \Gamma_1 \right)_{\ell \ell} = \sqrt{2} m_\ell / v$.
Clearly,
according to Eq.~\eqref{MlMD} with $v_k = v \delta_{k1}$,
\be
\label{ivufo}
M_D = \frac{v}{\sqrt{2}}\ \mathrm{diag} \left( d_e,\ d_\mu,\ d_\tau \right).
\ee
From Eq.~\eqref{nvkdoso},
\be
\label{mbkfpso}
X_{\ell i} = \frac{\sqrt{2}}{d_\ell^\ast}\, U_{\ell i}\, \frac{m_i}{v}.
\ee

We use both Eqs.~\eqref{fjdofd0} and the definition~\eqref{qij} to derive
\bs
\label{2ert9}
\ba
g_{2 \ell} &=& x_2\, \frac{\sqrt{2} m_\ell}{v}
+ \sqrt{1 - x_2^2}\ \gamma_\ell,
\\
f_{2ij} &=& x_2\, \frac{\sqrt{2} \left( m_i q_{ij} + m_j q_{ji} \right)}{v}
+ \sqrt{1 - x_2^2}\,
\left( X^\dagger \Delta_2 U + U^T \Delta_2 X^\ast \right)_{ij}.
\hspace*{7mm}
\ea
\es

\begin{figure}[ht]
\begin{center}
\includegraphics[width=1.0\textwidth]{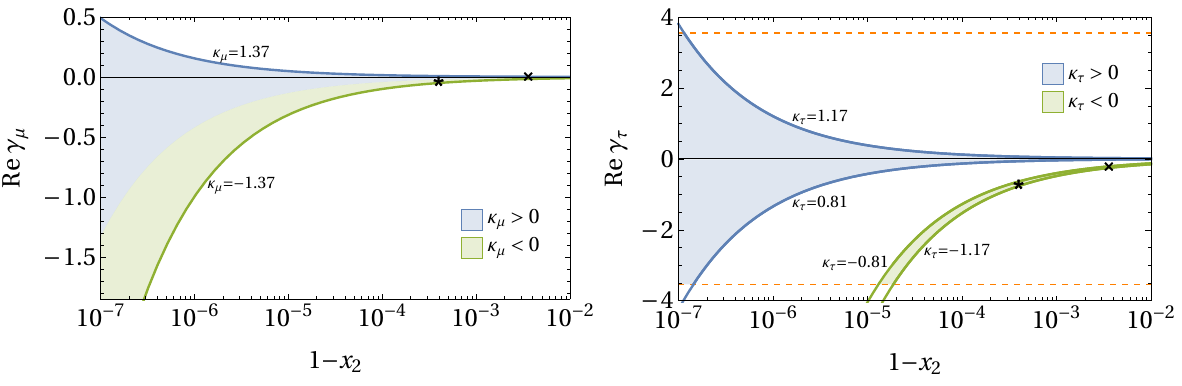}
\end{center}
\caption{The possible values of the real parts
  of the Yukawa couplings $\gamma_\mu$ (left panel)
  and $\gamma_\tau$ (right panel)
  \textit{versus} $1 - x_2$,
  according to inequalities~\eqref{SM10conditions-d}
  and~\eqref{SM10conditions-b},
  respectively.
  In the right panel,
  the dashed orange lines indicate the upper bound
  $\left| \mathrm{Re}\, \gamma_\tau \right| \le \sqrt{4 \pi} \approx 3.5$
  that we impose in our fits.
  The locations of the benchmark points of table~\ref{table-bench}
  are marked by crosses (for~\protect\hyperlink{BP1}{BP-1})
  and stars (for~\protect\hyperlink{BP2}{BP-2}).}
\label{fig-kappas}
\end{figure}

The scalar $S_2^0$ couples to pairs of gauge bosons
according to the Lagrangian~\cite{Grimus:2007if}
\be
\label{jg03932f}
\mathcal{L} = \cdots + \frac{e}{s_w}\, S_2^0 \left( m_W W_\mu^+ W^{\mu -}
+ \frac{m_Z}{2 c_w}\, Z_\mu Z^\mu \right) x_2.
\ee
It couples to the $\tau$ and $\mu$ leptons
through the Lagrangian---\textit{cf.}\ Eq.~\eqref{nvjfui}---
\bs
\label{it94304329k}
\ba
\mathcal{L} &=& \cdots - S_2^0 \left[
  \bar \tau \left( \frac{x_2 e m_\tau}{2 s_w m_W}
  + \frac{\sqrt{1 - x_2^2}\ \mathrm{Re}\, \gamma_\tau}{\sqrt{2}}
  - \frac{\sqrt{1 - x_2^2}\ \mathrm{Im}\, \gamma_\tau}
  {\sqrt{2}}\, \gamma_5 \right) \tau
  \right. \\ & & \left.
  + \bar \mu \left( \frac{x_2 e m_\mu}{2 s_w m_W}
  + \frac{\sqrt{1 - x_2^2}\ \mathrm{Re}\, \gamma_\mu}{\sqrt{2}}
  - \frac{\sqrt{1 - x_2^2}\ \mathrm{Im}\, \gamma_\mu}
  {\sqrt{2}}\, \gamma_5 \right) \mu
  \right].
\ea
\es
Experimentalists
usually
write
\be
\mathcal{L} = \cdots + \frac{e}{s_w}\, S_2^0
\left( \kappa_W\, m_W W_\mu^+ W^{\mu -}
+ \kappa_Z\ \frac{m_Z}{2 c_w}\, Z_\mu Z^\mu \right)
- \frac{e}{2 s_w m_W}\, S_2^0 \left( \kappa_\tau\, \bar \tau m_\tau \tau
+ \kappa_\mu\, \bar \mu m_\mu \mu \right),
\ee
\textit{viz.}\ with factors $\kappa_{W,Z,\tau,\mu}$
that parameterize the deviations from the SM.
Detailed limits on those factors have been derived from experiment,
see for instance Refs.~\cite{PDG2020,ATLAS:2019nkf,CMS:2018uag,CMS:2018vqh}.
In our fits we enforce
the conditions~\cite{CMS:2018uag}\footnote{The LHC results also suggest that
the couplings of the Higgs particle to the top and bottom quarks
should be quite close to the SM ones.
However,
since in our model we do not specify the Yukawa couplings of the quarks,
we refrain from imposing any constraint arising from the quark sector.}
\bs
\label{SM10conditions}
\ba
0.93 \le \left| \kappa_W \right| = \left| \kappa_Z \right|
= x_2 &\le& 1, \label{SM10conditions-a}
\\
0.81 \le \left| \kappa_\tau \right| =
\left| x_2 + \frac{\sqrt{2} s_w m_W}{e m_\tau}\, \sqrt{1-x_2^2}\
\mathrm{Re}\,\gamma_\tau \right| &\le& 1.17, \label{SM10conditions-b}
\\
\left| \frac{m_W}{m_\tau}\, \sqrt{1-x_2^2}\
\mathrm{Im}\,\gamma_\tau \right| &\le& 0.3, \label{SM10conditions-c}
\\
\left| \kappa_\mu \right| =
\left| x_2 + \frac{\sqrt{2} s_w m_W}{e m_\mu}\, \sqrt{1-x_2^2}\
\mathrm{Re}\,\gamma_\mu \right| &\le& 1.37, \label{SM10conditions-d}
\\
\left| \frac{m_W}{m_\mu}\, \sqrt{1-x_2^2}\
\mathrm{Im}\,\gamma_\mu \right| &\le& 0.3. \label{SM10conditions-e}
\ea
\es
These conditions constitute quite strong constraints
on $x_2$ and on the Yukawa couplings $\gamma_\tau$ and $\gamma_\mu$.
Conditions~\eqref{SM10conditions-b} and~\eqref{SM10conditions-d}
are displayed in Fig.~\ref{fig-kappas}.
In the experimental papers,
for any given decay mode a coupling modifier is defined as
$\kappa_i^2 = \Gamma^i \! \left/ \Gamma_\mathrm{SM}^i \right.$,
therefore in our analysis
we allow for either positive or negative
$\mathrm{Re}\, \kappa_\mu$ and $\mathrm{Re}\, \kappa_\tau$,
as illustrated in Fig.~\ref{fig-kappas}.

We parameterize the vertex of $S_2^0$ with two charged scalars
through Eq.~\eqref{3vifogp}.
We already know from Eqs.~\eqref{vjfifdop0} that
\bs
\label{lambdaCoeff_2HDM}
\ba
\lambda_{212} = \lambda_{122}
&=& \frac{e \left( m_{H^+}^2 - m_h^2 \right)}{2 s_w m_W}\, \sqrt{1 - x_2^2},
\\
\lambda_{112} &=& - \frac{e m_h^2}{2 s_w m_W}\, x_2.
\ea
\es
The value of $\lambda_{222}$,
\textit{i.e.}\ of the coupling $H^- H^+ S_2^0$,
depends on the scalar potential.
If we write the quartic part of the scalar potential of the 2HDM
in the standard notation~\cite{Branco:2011iw}
\ba
V_4 &=&
\frac{\lambda_1}{2} \left( \Phi_1^\dagger \Phi_1 \right)^2
+ \frac{\lambda_2}{2} \left( \Phi_2^\dagger \Phi_2 \right)^2
+ \lambda_3\, \Phi_1^\dagger \Phi_1\, \Phi_2^\dagger \Phi_2
+ \lambda_4\, \Phi_1^\dagger \Phi_2\, \Phi_2^\dagger \Phi_1
\no & &
+ \left[ \frac{\lambda_5}{2} \left( \Phi_1^\dagger \Phi_2 \right)^2
  + \left( \lambda_6 \Phi_1^\dagger \Phi_1 + \lambda_7 \Phi_2^\dagger \Phi_2
  \right) \Phi_1^\dagger \Phi_2 + \mathrm{H.c.} \right],
\label{v4}
\ea
then~\cite{Lavoura:1994fv}
%
\be
\label{lambdaCoeff_2HDM_2}
\lambda_{222} = - \frac{2 s_w m_W}{e} \left(
x_2 \lambda_3 + \sqrt{1-x_2^2}\ \mathrm{Re}\, \lambda_7 \right).
\ee
The coupling $\lambda_{222}$ is important for $S_2^0 \to \ell_1^+ \ell_2^-$;
there is a diagram for that decay wherein $S_2^0$ attaches to $H^- H^+$.
However,
in practice that diagram gives amplitudes~\eqref{jdisofi}
that are always much smaller than
the dominant amplitudes~\eqref{vmfr0395} and~\eqref{7a}.
We have found that,
for $-1 < \lambda_3 < 7$
and $\left| \mathrm{Re}\, \lambda_7 \right| < 1.5$~\cite{Jurciukonis:2018skr},
the branching ratios $\mathrm{BR} \left( S_2^0 \to \ell_1^+ \ell_2^- \right)$
are almost completely independent of $\lambda_{222}$.\footnote{There is
an exception to this behaviour
when $1 - x_2 \lesssim 10^{-7}$,
\textit{i.e.}\ when one is extremely close
to the `alignment' situation $x_2 = 1$.
In this case the amplitudes~\eqref{vmfr0395} and~\eqref{7a}
are strongly suppressed and the exact value of $\lambda_{222}$
becomes quite relevant.
However,
in that very contrived case the branching ratios
of $S_2^0 \to \ell_1^+ \ell_2^-$
become very close to zero and,
therefore,
uninteresting to us,
since in this paper we are looking for the possibility
of largish LFV branching ratios.}
Thereafter we have kept $\lambda_3 = \mathrm{Re}\, \lambda_7 = 1$ fixed.

\subsection{Fit to the lepton-mixing data}
\label{Sect_NeutrSect}

The lepton mixing matrix $U$
is in the charged-current Lagrangian~\eqref{ncksod}.
It is a $3 \times 6$ matrix.
We must connect it to the standard PMNS $3 \times 3$ unitary matrix.
In order to make this connection we use
the seesaw approximation~\cite{Minkowski:1977sc,
  Yanagida:1979as,Glashow:1979nm,GellMann:1980vs,Mohapatra:1979ia},
which is valid when the eigenvalues of $M_R$
are very much larger than the (diagonal) matrix elements of $M_D$.
The $3 \times 3$ symmetric matrix
\be
\label{m}
\mathcal{M}_\nu = - M_D^T M_R^{-1} M_D
\ee
is diagonalized by an unitary matrix $V$ as
\be
\label{v5}
V^T \mathcal{M}_\nu V = \mathrm{diag} \left( n_1, n_2, n_3 \right)
:= \hat n,
\ee
where the $n_p$ ($p = 1, 2, 3$) are real and positive.
It follows from Eqs.~\eqref{m} and~\eqref{v5} that
\be
\label{mr}
M_R = - M_D V \hat n^{-1} V^T M_D^T.
\ee
In our fitting program we input
the PMNS matrix $V$,\footnote{Recall that in our model there is
conservation of the flavour lepton numbers in the Yukawa couplings
and therefore the charged-lepton mass matrix is diagonal from the start.}
the Yukawa couplings $d_{e,\mu,\tau}$,
and the
$n_p$.
We firstly write the matrix $M_D$ given by Eq.~\eqref{ivufo}.
We then determine $M_R$ through Eq.~\eqref{mr}.
We use $M_R$ and $M_D$
to construct the $6 \times 6$ matrix
\be
\label{vmfodd06}
\left( \begin{array}{cc} 0_{3 \times 3} & M_D^T \\ M_D & M_R \end{array} \right).
\ee
We diagonalize the matrix~\eqref{vmfodd06} through the unitary matrix $U_6$
as in Eq.~\eqref{dnksps}.
We thus find both $U$,
\textit{viz.}\ the $3 \times 6$ upper submatrix of $U_6$,
and the neutrino masses $m_i$
($i = 1, \ldots, 6$).
Because the seesaw approximation is very good,
one obtains $m_i \approx n_i$ for $i = 1, 2, 3$
and moreover the $3 \times 3$ left submatrix of $U$
turns out approximately equal to $V$.
Finally, we order the heavy-neutrino masses
as $m_4 \le m_5 \le m_6$.

Since the inputted $n_p$ are many orders of magnitude below the Fermi scale,
the matrix elements of $M_R$ are much above the Fermi scale
unless the Yukawa couplings $d_\ell$ are extremely small.
Therefore,
\emph{when we lower the inputted $d_\ell$, we lower the heavy-neutrino masses}.

For the $n_p$ we use the light-neutrino masses.
The cosmological bound~\cite{Planck:2018nkj} is
\be
\label{cosmological}
\sum_{p=1}^3 n_p \approx \sum_{\mathrm{light\, neutrinos}} m_\nu < 0.12\,\mathrm{eV},
\ee
together with the squared-mass differences
$\Delta_\mathrm{solar} = n_2^2 - n_1^2$
and $\Delta_\mathrm{atmospheric} = \left| n_3^2 - n_1^2 \right|$,
that are taken from phenomenology.
The lightest-neutrino mass is kept free;
we let it vary in between $10^{-5}$\,eV and $\sim 0.03$\,eV
for normal ordering ($n_1 < n_3$),
and in between $10^{-5}$\,eV and $\sim 0.015$\,eV
for inverted ordering ($n_3 < n_1$);
the upper bound on the lightest-neutrino mass is indirectly provided
by the cosmological bound~\eqref{cosmological}.
The smallest $n_p$ cannot be allowed to be zero
because $\hat n^{-1}$ appears in Eq.~\eqref{mr}.
For the matrix $V$ we use the parameterization~\cite{PDG2020}
\be
\label{vvv}
V = \left( \begin{array}{ccc}
  c_{12} c_{13} &
  s_{12} c_{13} &
  \epsilon^\ast \\
  - s_{12} c_{23} - c_{12} s_{23} \epsilon &
  c_{12} c_{23} - s_{12} s_{23} \epsilon &
  s_{23} c_{13} \\
  s_{12} s_{23} - c_{12} c_{23} \epsilon &
  - c_{12} s_{23} - s_{12} c_{23} \epsilon &
  c_{23} c_{13} 
\end{array} \right)
\times \mathrm{diag} \left( 1,\ e^{i \alpha_{21} / 2},\ e^{i \alpha_{31} / 2} \right),
\ee
where $\epsilon \equiv s_{13} \exp{\left( i \delta \right)}$,
$c_{pq} = \cos{\theta_{pq}}$,
and $s_{pq} = \sin{\theta_{pq}}$ for $(pq) = (12), (13), (23)$.

Three different groups~\cite{deSalas:2017kay,Capozzi:2018ubv,Esteban:2018azc}
have derived,
from the data provided by various neutrino-oscillation experiments,
values for the mixing angles $\theta_{pq}$,
for the phase $\delta$,
and for $\Delta_\mathrm{solar}$ and $\Delta_\mathrm{atmospheric}$.
The results of the three groups
(especially the $1\sigma$ bounds)
are different,
but in Ref.~\cite{deSalas:2017kay} the values of the observables
are in between the bounds
of Refs.~\cite{Capozzi:2018ubv} and~\cite{Esteban:2018azc}. 
In this paper we use the $3\sigma$ data from Ref.~\cite{deSalas:2017kay}
that are summarised in table~\ref{tab:expNeutParam}.
\begin{table}[h!]\centering
   \begin{tabular}{lccc}
    \hline\hline
    Quantity & Best fit &
    \hphantom{x} 1$\sigma$ range  \hphantom{x} &
    \hphantom{x} 3$\sigma$ range  \hphantom{x}
    \\  
    \hline 
    $\Delta_\mathrm{solar} \left/ \left( 10^{-5}\mathrm{eV}^2 \right) \right.$
    \phantom{$1^{1^{1^a}}$}
    & 7.55 & 7.39--7.75 & 7.05--8.14 \\[1.5mm]  
    $\Delta_\mathrm{atmospheric}
    \left/ \left( 10^{-3}\mathrm{eV}^2 \right) \right.$ (NO)
    &  2.50 &  2.47--2.53 &  2.41--2.60\\
    $\Delta_\mathrm{atmospheric}
    \left/ \left( 10^{-3}\mathrm{eV}^2 \right) \right.$ (IO)
    &  2.42 &  2.34--2.47 &  2.31-2.51 \\[1.5mm]  
    $\sin^2\theta_{12} \left/ 10^{-1} \right.$
    & 3.20 & 3.04--3.40 & 2.73--3.79\\ [1.5mm] 
     $\sin^2\theta_{23} \left/ 10^{-1} \right.$ (NO)
              &	5.47 & 5.17--5.67 & 4.45--5.99 \\
     $\sin^2\theta_{23} \left/ 10^{-1} \right.$ (IO)
              & 5.51 & 5.21--5.69 & 4.53--5.98\\[1.5mm] 
    $\sin^2\theta_{13} \left/ 10^{-2} \right.$ (NO)
    & 2.160 &  2.091--2.243 & 1.96--2.41 \\
    $\sin^2\theta_{13} \left/ 10^{-2} \right.$ (IO)
    & 2.220 & 2.144--2.146 & 1.99--2.44 \\ [1.5mm]  
   $\delta \left/ \mathrm{rad} \right.$ (NO)
   	& 3.80 & 3.33--4.46 & 2.73--6.09 \\
    $\delta \left/ \mathrm{rad} \right.$ (IO)	
   	& 4.90 & 4.43--5.31 & 3.52--6.09 \\
       \hline\hline
     \end{tabular}
     \caption{ \label{tab:expNeutParam} 
       The neutrino-oscillation parameters
       used in our fits~\cite{deSalas:2017kay}.        
     }
\end{table}
The Majorana phases $\alpha_{21}$ and $\alpha_{31}$
are kept free in our analysis.

\section{Numerical results}
\label{Sect_NumResults}

\subsection{Details of the computation}
\label{Sect_Details}

We have generated the complete set of diagrams for each process
in Feynman gauge by using the package {\tt FeynMaster}~\cite{Fontes:2019wqh}
(that package combines {\tt FeynRules}~\cite{Christensen:2008py,Alloul:2013bka},
{\tt QGRAF}~\cite{Nogueira:1991ex},
and {\tt FeynCalc}~\cite{Mertig:1990an,Shtabovenko:2016sxi})
with a modified version of the {\tt FeynRules} Standard-Model file
to account for the six neutrinos,
for lepton flavour mixing,
and for the additional Higgs doublet.
The amplitudes generated automatically by {\tt FeynMaster}
were expressed through Passarino--Veltman (PV) functions
by using the package {\tt FeynCalc} and specific functions of {\tt FeynMaster}. 
All the amplitudes were checked by performing the computations manually.
The results of these computations are presented
in Appendices~\ref{Sect_taumugamma},
\ref{Sect_taumuZ},
and~\ref{Sect_taumuS}.

For numerical calculations we made two separate programs,
one with {\tt Mathematica} and another one with {\tt Fortran}. 
Because of the very large differences among
\begin{description}
\item the mass scale of the light neutrinos,
  between $10^{-5}$\,eV and $0.1$\,eV,
\item the mass scale of the charged leptons,
  between $100$\,keV and $1$\,GeV,
\item and the mass scale of the heavy neutrinos,
  between $100$\,GeV and $10^{16}$\,GeV,
\end{description}
there are both numerical instabilities
and delicate cancellations in the calculations. 
These numerical problems could be solved
with the high-precision numbers that {\tt Mathematica} allows.
However,
this strongly slows down the calculations.
Fortunately,
numerical inaccuracies
occur only for very small values
(less than $10^{-30}$)
of the branching ratios (BRs),
therefore we were able to use a program written with {\tt Fortran}
to implement the minimization procedure
and to find BRs within ranges relevant to experiment. 
Some parts of the {\tt Fortran} code
(such as the module for matrix diagonalization)
have used quadruple precision to avoid inaccuracies,
but most of the code has used just double precision
so that the computational speed was sufficient for minimization. 
The final results were checked with the high precision afforded by
{\tt Mathematica}.

The numerical computation of the PV functions was performed
by using the {\tt Fortran} library {\tt Collier}~\cite{Denner:2016kdg},
which is designed for the numerical evaluation
of one-loop scalar and tensor integrals. 
A major advantage of {\tt Collier}
over the {\tt LoopTools} package~\cite{Hahn:1998yk}
is that it avoids numerical
instabilities when the neutrino masses are very large,
even
when one only uses double precision. 
The integrals were checked with {\tt Mathematica}'s high-precision numbers
and {\tt Package-X}~\cite{Patel:2015tea}
analytic expressions of one-loop integrals. 

In the fits of subsection~\ref{Sect_fitting},
in order to find adequate numerical values for the parameters
we have constructed a $\chi^2$ function to be minimized:
\be
\chi^2 = \sum_{i=1}^{n} \left[
\Theta \left( O_i^b - O_i^v \right) \left( \frac{O_i^b}{O_i^v}\right)^2 
+ \Theta \left( O_i^v - O_i^b \right)
\left( \frac{O_i^v - O_i^b}{k}\right)^2 \right].
\label{chi2}
\ee
%
In Eq.~\eqref{chi2},
\begin{itemize}
\item $n$ is the total number of observables to be fitted;
  this is usually nine,
  since we fit the BRs of the nine LFV decays
  in order to find them within the ranges accessible to experiment.
\item $\Theta$ is the Heaviside step function.
\item $O_i^v$ is the computed value of each observable.
\item $O_i^b$ is the experimental upper bound on the observable,
  which is given in table~\ref{LFV_limits}.
\item $k$ is an appropriately small number
  that short-circuits the minimization algorithm
  when $O_i^v$ turns out larger than $O_i^b$.\footnote{In practice,
  in each case we have tried
  various values of $k$ before settling on the one that worked best,
  \textit{i.e.}\ that maximized the efficiency of the minimization algorithm
  for each problem at hand.
  Since the observables \textit{viz.}\ the branching ratios are very small,
  $k \sim 10^{-20}$ was a typical order of magnitude.}
\end{itemize}
The $\chi^2$ function~\eqref{chi2} works well even when the calculated BRs
and the experimental upper bounds differ by many orders of magnitude. 
We have performed the fits in subsection~\ref{Sect_fitting}
by minimizing $\chi^2$
with respect to the model parameters---the Yukawa couplings $d_\ell$,
$\delta_\ell$,
and $\gamma_\ell$,
and the PMNS-matrix parameters.
The mass of the lightest neutrino and the parameter $x_2$
were
randomly generated before the minimization of the $\chi^2$ function,
in order to be able
to explore the full range of the neutrino masses
and the full range of $x_2$.
In the fits of subsection~\ref{Sect_fitting}
the mass of the charged scalar $H^+$
was
usually kept fixed,
just as the parameters $\lambda_3$ and $\mathrm{Re}\, \lambda_7$
of the scalar potential~\eqref{v4}.

The minimization of $\chi^2$ is not an easy task
because of the large number of model parameters
that, moreover,
may differ by several orders of magnitude,
and because there is always a large number of local minima. 
However,
we don't try to find absolute minima,
{\it i.e.}\ BRs as close as possible to the experimental upper bound;
our purpose is rather to search under which circumstances
the decay rates may be in experimentally accessible ranges. 

The inputted values of the masses of the leptons and bosons
were taken from Ref.~\cite{PDG2020}.
We have used $s_w^2 = 0.22337$ and $e \equiv \sqrt{4 \pi \alpha}$,
where $\alpha = 1/137.036$ is the fine-structure constant.
The
neutrino-oscillation data
are in table~\ref{tab:expNeutParam}.

We introduce the shorthands BR($\ell$),
BR($Z$),
and BR($h$) for the branching ratios of the decays
$\ell_1^\pm \to \ell_2^\pm \gamma$,
$Z \to \ell_1^\pm \ell_2^\mp$,
and $h \to \ell_1^\pm \ell_2^\mp$,
respectively.
We also define
the lower bound $\Ymin = 10^{-6}$
and the upper bound $\Ymax = \sqrt{4 \pi} \approx 3.5$
on the moduli of the Yukawa coupling constants.

\subsection{Benchmark points}
\label{Sect_benchmark}

We produce in table~\ref{table-bench} two benchmark points
(BPs).
For those two BPs the neutrino mass ordering is normal,
the neutrino squared-mass differences and the lepton mixing angles
take their best-fit values in table~\ref{tab:expNeutParam},
the mass of the charged scalar is 750\,GeV,
and the parameters $\lambda_3$ and $\lambda_7$
of the scalar potential are both equal to 1.
\begin{table}[h!]
  \begin{center}
    \begin{tabular}{|c|c|c|} \hline
      $\phantom{\frac{1^1}{1^1}}$ & Point~1 \hypertarget{BP1}{(BP-1)}
      & Point~2 \hypertarget{BP2}{(BP-2)} \\ \hline
      $d_e$ & $10^{-6}$ & $0.25$ \\
      $d_\mu$ & $4 \times 10^{-6}$ & $0.7$ \\
      $d_\tau$ & $10^{-6}$ & $0.4$ \\
      $\delta_e$ & $3.5$ & $0.7$ \\
      $\delta_\mu$ & $3.5$ & $0.2$ \\
      $\delta_\tau$ & $3.5$ & $0.7$ \\
      $\gamma_e$ & $0.3$ & $0.3$ \\
      $\gamma_\mu$ & $10^{-4}$ & --\,$0.043$ \\
      $\gamma_\tau$ & --\,$0.24$ & --\,$0.73$ \\ \hline
      $\kappa_\mu$ & $1.01$ & --\,$1.04$  \\
      $\kappa_\tau$ & --\,$1.04$ & --\,$1.06$  \\ \hline
      $m_1$\,(meV) & 16.5 & 5.2 \\
      $\alpha_{21}$\,(rad) & $3.515$ & 0 \\ 
      $\alpha_{31}$\,(rad) & $1.060$ & 0 \\
      $1 - x_2$ & $0.0036$ & $0.0004$ \\ \hline
      $m_4$\,(TeV) & $1.16066$ & $1.33740 \times 10^{11}$ \\
      $m_5$\,(TeV) & $3.54866$ & $3.07494 \times 10^{11}$ \\
      $m_6$\,(TeV) & $6.67826$ & $12.5369 \times 10^{11}$ \\ \hline
      $\mathrm{BR} \left( \tau^\pm \to \mu^\pm \gamma \right) \times 10^9$ &
      $8.1$ & --- \\
      $\mathrm{BR} \left( \tau^\pm \to e^\pm \gamma \right) \times 10^9$ &
      $12$ & --- \\
      $\mathrm{BR} \left( \mu^\pm \to e^\pm \gamma \right) \times 10^{13}$ &
      $2.2$ & --- \\
      $\mathrm{BR} \left( Z \to \tau^\pm \mu^\mp \right) \times 10^{13}$ &
      $9.6$ & --- \\ 
      $\mathrm{BR} \left( Z \to \tau^\pm e^\mp \right) \times 10^{13}$ &
      $12$ & --- \\ 
      $\mathrm{BR} \left( Z \to \mu^\pm e^\mp \right) \times 10^{15}$ &
      $38$ & --- \\ 
      $\mathrm{BR} \left( h \to \tau^\pm \mu^\mp \right) \times 10^4$ &
      $3.6$ & $5.5$ \\ 
      $\mathrm{BR} \left( h \to \tau^\pm e^\mp \right) \times 10^4$ &
      $5.5$ & $8.9$ \\ 
      $\mathrm{BR} \left( h \to \mu^\pm e^\mp \right) \times 10^5$ &
      $3.6$ & $3.9$ \\ \hline
    \end{tabular}
  \end{center}
  \caption{Two benchmark points.
    In the third column,
    the symbol `---' stands for a tiny number $\lesssim 10^{-20}$.
    The values of the input parameters absent from the first column
    are given at the beginning of subsection~\ref{Sect_benchmark}.
  }
\label{table-bench}  
\end{table}
The first nine rows of table~\ref{table-bench}
contain the inputted values of the Yukawa couplings;
the next two lines
have the computed
values of $\kappa_{\mu}$ and $\kappa_{\tau}$;
in the next four lines
one finds the inputted values of the lightest-neutrino mass $m_1$,
of the Majorana phases $\alpha_{21}$ and $\alpha_{31}$,
and of the non-alignment parameter $1 - x_2$;
the next three lines have
the computed masses of the heavy neutrinos,
ordered as $m_4 \le m_5 \le m_6$;
the last nine lines display the computed branching ratios.

In benchmark point~2 (\hyperlink{BP2}{BP-2})
only the BR($h$) are sufficiently large to be observed in the future,
while the BR($\ell$) and BR($Z$) are negligibly small.
Benchmark point~1 (\hyperlink{BP1}{BP-1}) indicates
that very small values of the Yukawa couplings $d_\ell$
and large values of the Yukawa couplings $\delta_\ell$ are required
in order to obtain BR($\ell$)
in experimentally reachable ranges.
\hyperlink{BP2}{BP-2} shows that,
if only the BR($h$) are accessible,
then
the
Yukawa couplings may all be in the range
$\left[ 0.1,\ 1 \right]$;
in that case,
since the $d_\ell$ are not very small,
the heavy-neutrino masses are quite large.

\subsection{Evolution of BRs}
\label{Sect_evolution}

In this subsection we discuss the behaviour of the BRs
when we vary some input parameters
of the benchmark point~\hyperlink{BP1}{1} of the previous section,
while the other input parameters of that point remain fixed.

In order to visualize
the impact of the Yukawa coupling constants on the BRs,
we have fixed their ratios in the same way as in~\hyperlink{BP1}{BP-1},
\textit{viz.}
\be
\label{ratos}
d_\tau = d_e, \qquad d_\mu = 4\, d_\tau, \qquad
\delta _e = \delta_\mu = \delta_\tau, \qquad
\frac{0.3\, \gamma_\tau}{\gamma_e} = 
\frac{\gamma_\tau}{10^4\, \gamma_\mu} = -0.24.
\ee
We change either only $d_\tau$,
or only $\delta_\tau$,
or only $\gamma_\tau$,
and we let the other Yukawa couplings vary together with them
through the fixed ratios~\eqref{ratos}.
All the other input parameters keep the values of~\hyperlink{BP1}{BP-1}.

\begin{figure}[ht]	
\begin{center}
\includegraphics[width=1.0\textwidth]{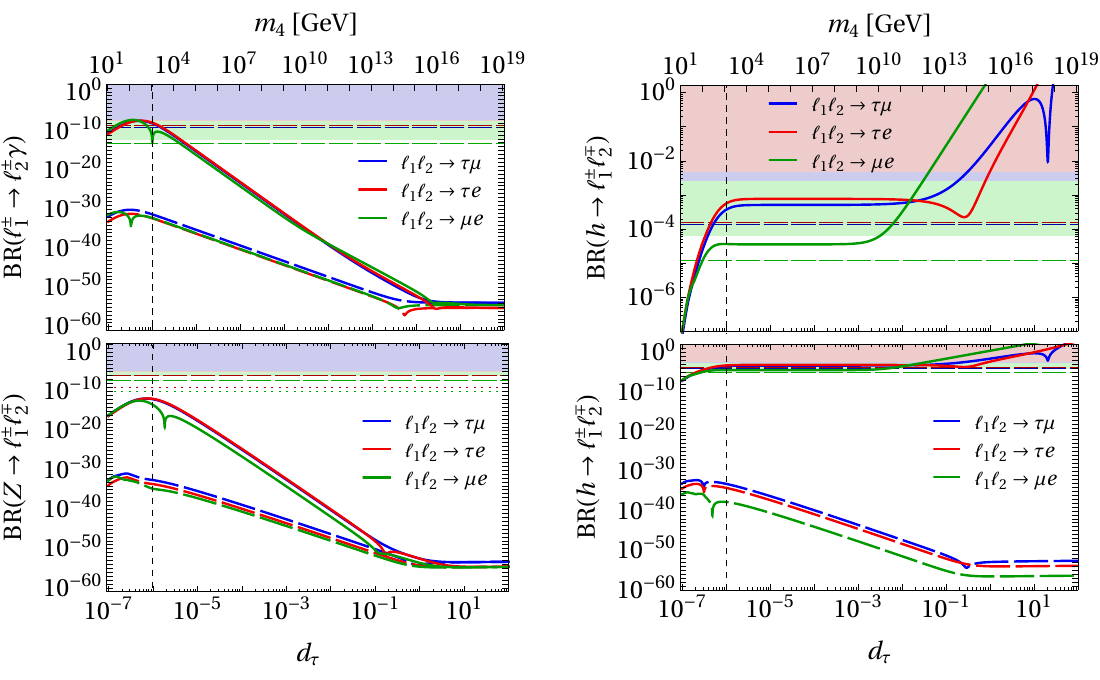}
\end{center}
\caption{The branching ratios as functions of the Yukawa coupling $d_\tau$.
  Full lines give the BRs computed with the contribution of all amplitudes;
  dashed lines represent the BRs computed only with the amplitudes
  that do not involve the charged scalars $H^\pm$.
  The shadowed bands are excluded by the present experimental data;
  the dashed/dotted horizontal lines show the future experimental sensitivities
  given in table~\ref{LFV_limits}.
  The colours of the shadowed bands coincide with the colours of the lines,
  {\it viz.}\
  blue for $\left( \ell_1,\, \ell_2 \right) = \left( \tau,\, \mu \right)$,
  red for $\left( \ell_1,\, \ell_2 \right) = \left( \tau,\, e \right)$,
  and green for $\left( \ell_1,\, \ell_2 \right) = \left( \mu,\, e \right)$.
  The vertical dashed lines mark the location
  of~\protect\hyperlink{BP1}{BP-1} of table~\ref{table-bench}.
  The upper-right panel is a zoom of the upper part of the bottom-right one.
  In the left panels,
  the blue lines almost always coincide with the red ones.}
\label{fig-dtau}
\end{figure}

In Fig.~\ref{fig-dtau} we display the BRs against $d_\tau$,
while the three $\delta_\ell$ and the three $\gamma_\ell$
are kept equal to their respective values of~\hyperlink{BP1}{BP-1}.
It should be noted,
in the upper and lower horizontal scales of Fig.~\ref{fig-dtau},
that the mass $m_4$ of the lightest heavy neutrino varies
as $m_4 \propto d_\tau^2$.
One observes,
in the top-left panel of Fig.~\ref{fig-dtau},
that the BR($\ell$) reach values close to their experimental upper bounds
for a narrow range of $d_\tau$,
{\it viz.} $2 \times 10^{-7} \lesssim d_\tau \lesssim 8 \times 10^{-6}$;
for these tiny values of $d_\tau$,
$m_4 \lesssim 10^4$\,GeV.
The behaviour of the BR($Z$)
is shown in the bottom-left panel of Fig.~\ref{fig-dtau};
it is similar to the behaviour of the BR($\ell$),
as one might foresee from the similarities in the amplitudes
for the two processes,
\textit{cf.}\ appendices~\ref{Sect_taumugamma} and~\ref{Sect_taumuZ}.
Unfortunately,
however,
because of a small factor in the decay width,
\textit{cf.}\ Eq.~\eqref{ZdecayWidth},
the predicted BR($Z$) are smaller by more than six orders of magnitude
than the present experimental upper bounds.

We observe a completely different behaviour of the BR($h$)
in the right panels of Fig.~\ref{fig-dtau}
(the top panel is a zoom of part of the bottom one):
the BR($h$) achieve values comparable to the experimental upper bounds
for a wide range of $d_\tau$,
\textit{viz.}\ even when the heavy-neutrino masses are quite large.

The main message of Fig.~\ref{fig-dtau} is that
\emph{all nine BRs would be very small
if there were no charged scalars $H^\pm$}.
The contributions to the amplitudes from the diagrams with $H^\pm$
increase some BRs in some circumstances by several orders of magnitude.

\begin{figure}[ht]
\begin{center}
\includegraphics[width=1.0\textwidth]{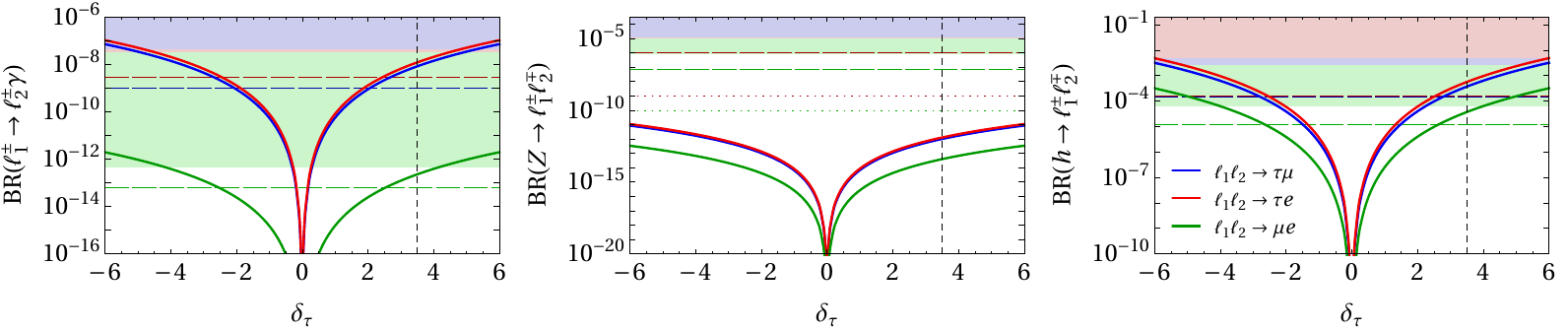}
\end{center}
\caption{The branching ratios as functions
  of the Yukawa couplings $\delta_\tau = \delta_\mu = \delta_e$.
  See the caption of Fig.~\ref{fig-dtau} for further explanations.
}
\label{fig-deltatau}
\end{figure}

The BRs behave differently when plotted
against the Yukawa couplings $\delta_\ell$,
as shown in Fig.~\ref{fig-deltatau}.
We see that all the BRs
increase with increasing absolute value of $\delta_\ell$;
the BR($\ell$) and BR($h$) become visible in planned experiments
when $\left| \delta_\ell \right| \gtrsim 2$
(for appropriate values of the other parameters,
especially very small $d_\ell$,
as they are in Fig.~\ref{fig-deltatau}).
With decreasing $\left| \delta_\ell \right|$
the BRs decrease monotonically for all decays
because of the decreasing values of all the amplitudes; 
when $\left| \delta_\tau \right| \sim 10^{-6}$ the BRs have minimum values
somewhere between $10^{-35}$ and $10^{-25}$.

\begin{figure}[ht]
\begin{center}
\includegraphics[width=1.0\textwidth]{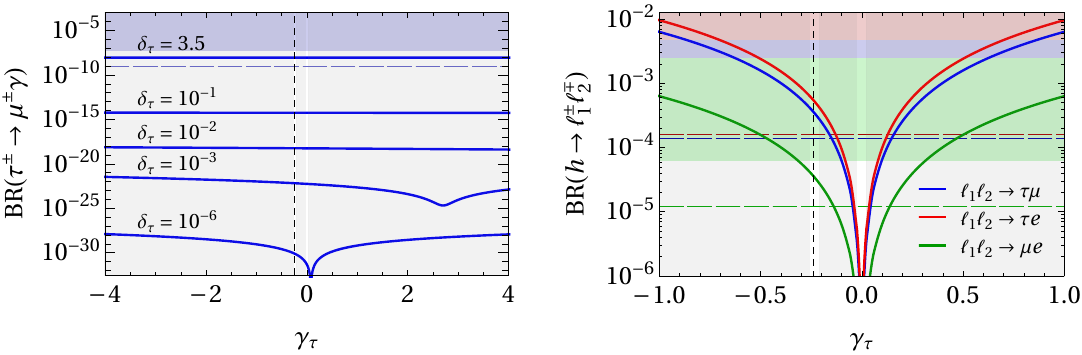}
\end{center}
\caption{Some branching ratios
  as functions of the Yukawa coupling $\gamma_\tau$.
  The vertical dashed lines mark the value $\gamma_\tau = -0.24$
  of benchmark point~\protect\hyperlink{BP1}{1}.
  Left panel: $\mathrm{BR} \left( \tau^\pm \to \mu^\pm \gamma \right)$
  {\it versus} $\gamma_\tau$ for various values of $\delta_\tau$.
  Right panel: the BR($h$) {\it versus} $\gamma_\tau$ for $\delta_\tau = 3.5$.
  The light-gray-shadowed bands are regions excluded
  by the inequalities~\eqref{SM10conditions}.
  }
\label{fig-gammatau}
\end{figure}

The dependence of the BR($\ell$) and BR($Z$)
from the Yukawa couplings $\gamma_\ell$ is weak,
as shown in the left panel of Fig.~\ref{fig-gammatau}. 
The reason for this is that in the dominant amplitudes,
\textit{viz.}\ the ones in Eq.~\eqref{coef1},
the $d_\ell$ and $\delta_\ell$ have much stronger impact
than the $\gamma_\ell$.
The relevance of the $\gamma_\ell$ is much stronger on BR($h$);
in the right panel of Fig.~\ref{fig-gammatau}
one sees that experimentally visible BR($h$)
may be reached when $\left| \gamma_\tau \right| \gtrsim 0.1$,
for appropriate values of the other parameters.  
The BR($h$) decrease with decreasing $\left| \gamma_\ell \right|$
because of the decreasing values of the dominant amplitudes,
\textit{viz.}\ $d_{l/rb,\mathrm{16}}$ in Eq.~\eqref{vmfr0395}
and $d_{l/rb,\mathrm{19}}$ in Eqs.~\eqref{7a}. 
However,
for $\left| \gamma_\tau \right| \lesssim 10^{-3}$
the amplitudes $d_{l/rb,\mathrm{17}}$ in Eq.~\eqref{jdisofi}  
become dominant and the BR($h$) do not decrease much any further.
 
\begin{figure}[ht]
\begin{center}
\includegraphics[width=0.5\textwidth]{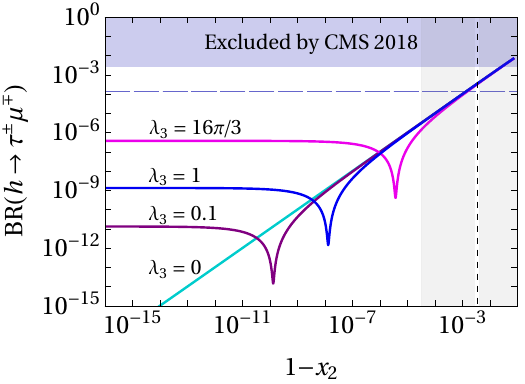}
\end{center}
\caption{$\mathrm{BR} \left( h \to \tau^\pm \mu^\mp \right)$ {\it versus}
  $1 - x_2$ for various values
  of the triple-scalar coupling $\lambda_3$;
  all the other input parameters are as in~\protect\hyperlink{BP1}{BP-1}.
  The bands shadowed in light gray are excluded by 
  inequalities~\eqref{SM10conditions}.
  The vertical dashed line marks the value $1 - x_2=0.0036$
  of~\protect\hyperlink{BP1}{BP-1}.
}
\label{fig-T11}
\end{figure}

In Figs.~\ref{fig-dtau} and~\ref{fig-gammatau}
we have seen that the behaviour of BR($h$)
is different from the one of BR($Z$) and BR($\ell$). 
This happens because of different amplitudes,
but also because of additional parameters,
{\it viz.}\ $x_2$ and the triple-scalar couplings $\lambda_3$ and $\lambda_7$,
that arise in the diagram of Fig.~\ref{fig5}
where $h$ attaches to two charged scalars
with couplings given by Eqs.~\eqref{lambdaCoeff_2HDM}
and~\eqref{lambdaCoeff_2HDM_2}. 
However,
due to the small factor $\sqrt{1-x_2^2}$
in the second term of Eq.~\eqref{lambdaCoeff_2HDM_2},
the impact of $\lambda_7$ on BR($h$) is almost imperceptible.
On the other hand,
$\lambda_3$ may have a strong influence on BR($h$).
This happens only for extremely small values of $1 - x_2$,
though;
and,
for such extremely small values of $1 - x_2$,
BR($h$) is anyway much too small to be measurable.
This is displayed in Fig.~\ref{fig-T11}.
In the cases that we are interested in,
\textit{viz.}\ when the BR($h$) are rather large,
the exact value of $\lambda_3$ is unimportant.
For the sake of simplicity,
from now one we assume $\lambda_3 = \lambda_7 = 1$ everywhere.

With decreasing $1 - x_2$,
the BRs in Fig.~\ref{fig-T11} decrease because
of the decreasing dominant amplitudes
$d_{l/rb,\mathrm{16}}$ in Eq.~\eqref{vmfr0395}
and $d_{l/rb,\mathrm{19}}$ in Eqs.~\eqref{7a}.
At some point,
though,
the amplitudes $d_{l/rb,\mathrm{17}}$ in Eqs.~\eqref{jdisofi}  
begin to dominate and then the BRs do not decrease much
any further.
The dips in the lines of Fig.~\ref{fig-T11}
arise from the partial cancellation of amplitudes
$d_{l/rb,\mathrm{16}}$ and $d_{l/rb,\mathrm{19}}$
with the amplitudes $d_{l/rb,\mathrm{17}}$. 

As shown in Fig.~\ref{fig-T11},
$1 - x_2$ has a strong impact on BR($h$).
It is also important for making BR($h$) and BR($\ell$)
simultaneously close to the experimental bounds.
Indeed,
the BR($h$) may be made sufficiently large,
for a wide range of the Yukawa couplings $d_{\ell}$
and for sufficiently large values of the $\delta_\ell$,
just by varying $1 - x_2$. 
The strong impact of $1 - x_2$ and of the $\gamma_\ell$
on BR($h$) allows one to adjust BR($h$),
together with BR($\ell$),
to be close to the experimental upper bounds---but for a quite restricted
range of $d_\ell$ and $\delta_\ell$,
because large BR($\ell$)
require extremely small $d_\ell$ and rather large $\delta_\ell$.
If,
on the other hand,
one attempts to fit only BR($h$),
then both $1 - x_2$ and the Yukawa couplings may be much more relaxed, 
as shown in~\hyperlink{BP2}{BP-2} of table~\ref{table-bench}.

\begin{figure}[h]
\begin{center}
\includegraphics[width=1.0\textwidth]{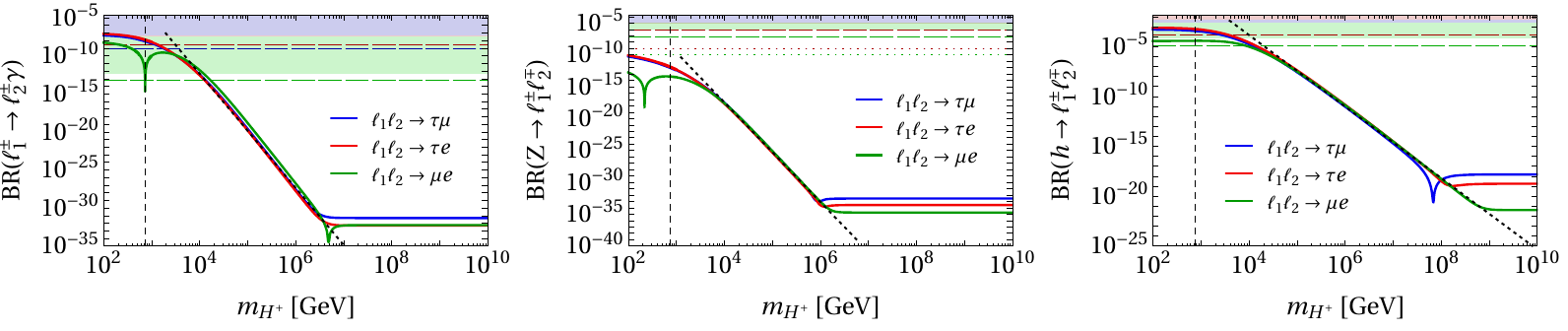}
\end{center}
\caption{The decay rates \textit{versus}\/
  the mass of the charged scalar $m_{H^+}$;
  all the other parameters are kept at their values
  of benchmark point~\protect\hyperlink{BP1}{1} in table~\ref{table-bench}.
  The vertical dashed lines mark $m_{H^+}=750$\,GeV.
  The oblique dashed lines are given by Eqs.~\eqref{l03393}.
}
\label{fig-mC}
\end{figure}

In Fig.~\ref{fig-mC} we illustrate the evolution of the BRs
when the mass of the charged scalar $m_{H^+}$ is changed,
while the other parameters are kept fixed at their values 
of~\hyperlink{BP1}{BP-1}. 
One observes that when $m_{H^+}$ increases
the BRs mostly decrease monotonically as
(for the other parameters fixed in their values
of~\hyperlink{BP1}{BP-1})
\bs
\label{l03393}
\ba
\mathrm{BR} \left( \ell \right) &\approx& 10^{17.72}
\left( \frac{m_{H^+}}{1\, \mathrm{GeV}} \right)^{-7.51},
\\
\mathrm{BR} \left( Z \right) &\approx& 10^{13.82}
\left( \frac{m_{H^+}}{1\, \mathrm{GeV}} \right)^{-7.82},
\\
\mathrm{BR} \left( h \right) &\approx& 10^{10.47}
\left( \frac{m_{H^+}}{1\, \mathrm{GeV}} \right)^{-3.58}.
\ea
\es
Eventually,
when $m_{H^+} \sim 10^7$\,GeV for BR($\ell$) and BR($Z$),
and when $m_{H^+} \sim 10^9$\,GeV for BR($h$),
the BRs settle at their SM values.
This illustrates the decoupling of $H^+$.
One also sees in Fig.~\ref{fig-mC} that,
for $\left( \ell_1, \ell_2 \right) = \left( \mu, e \right)$,
there is near $m_{H^+} = 750$\,GeV
a partial cancellation of amplitudes
that leads to a sudden drop of
$\mathrm{BR} \left( \mu^\pm \to e^\pm \gamma \right)$;
our benchmark point~\protect\hyperlink{BP1}{1} has profited from that effect
for attaining $\mathrm{BR} (\mu^\pm \to e^\pm \gamma)$
smaller than its experimental upper bound.

\begin{figure}[h]
\begin{center}
\includegraphics[width=1.0\textwidth]{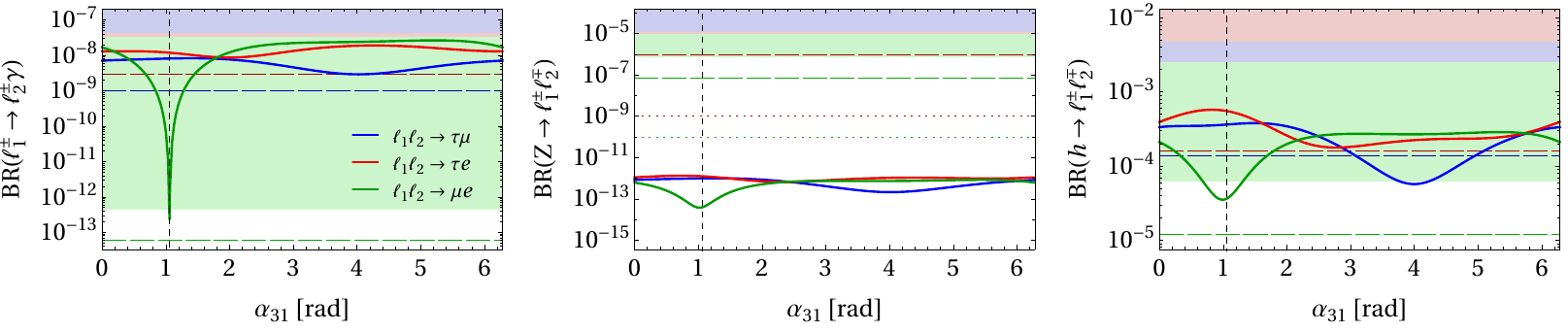}
\end{center}
\caption{The decay rates \textit{versus}\/
  the Majorana phase $\alpha_{31}$;
  all the other parameters are kept
  at their values of benchmark point~\protect\hyperlink{BP1}{1}
  of table~\ref{table-bench}.
  The vertical dashed lines mark the value $\alpha_{31} = 1.06$
  of~\protect\hyperlink{BP1}{BP-1}.}
\label{fig-alpha31}
\end{figure}

In Fig.~\ref{fig-alpha31} we display the BRs as functions
of the Majorana phase $\alpha_{31}$,
with the other input parameters kept fixed
at their values of~\hyperlink{BP1}{BP-1}. 
Here too,
for $\left( \ell_1, \ell_2 \right) = \left( \mu, e \right)$
there is a sudden drop of the branching ratios when $\alpha_{31}=1.06$,
which is precisely the value
that we have utilized in benchmark point~\hyperlink{BP1}{1}.
A similar behaviour of the green lines also occurs with other parameters,
besides $m_{H^+}$ (Fig.~\ref{fig-mC})
and $\alpha_{31}$ (Fig.~\ref{fig-alpha31}).
Hence,
the values of the parameters must be chosen very carefully
if we want to find all six BR($\ell$) and BR($h$)
simultaneously close to their experimental upper bounds. 
The main difficulty arises because
the upper bound on $\mathrm{BR} \left( \mu^\pm \to e^\pm \gamma \right)$
differs from the upper bound on
$\mathrm{BR} \left( \tau^\pm \to \mu^\pm \gamma \right)$
by five orders of magnitude.
Fortunately,
our minimization procedure allows this to be done quite efficiently.

\subsection{Fitting the BRs}
\label{Sect_fitting}

In this model there is a large number of input parameters.
We have performed a minimization procedure
in order to find adequate values for all of them.
For each set of input parameters,
we have computed the branching ratios of the nine LFV decays;
we have then selected sets of input parameters for which
all six BR($\ell$) and BR($h$) are simultaneously
between the current experimental upper bounds
and the future experimental sensitivities.\footnote{It is extremely difficult
to achieve values of the BR($Z$)
close to the future experimental sensitivities.
Still,
our minimization procedure also seeks to obtain
the highest possible values of the BR($Z$).}

Since in this subsection we use a fitting procedure,
we must enforce definite bounds on the input parameters,
lest they acquire either much too small or much too large values.
We adopt the following conditions:
\begin{itemize}
\item The neutrino-oscillation parameters,
  \textit{viz.}\ the mixing angles $\theta_{12}$,
  $\theta_{13}$,
  and $\theta_{23}$,
  the Dirac phase $\delta$,
  and the neutrino squared-mass differences,
  are varied within their respective $3\sigma$ ranges
  taken from Ref.~\cite{deSalas:2017kay}
  and reproduced in table~\ref{tab:expNeutParam}.
  The Majorana phases $\alpha_{21}$ and $\alpha_{31}$ are kept free,
  \textit{i.e.}\ we let them vary from 0 to $2\pi$.
\item The lightest-neutrino mass is varied in between $10^{-5}$\,eV
  and either $\sim 0.03$\,eV for normal ordering of the neutrino masses
  or $\sim 0.015$\,eV for inverted ordering.
  The precise upper bound on the lightest neutrino mass is fixed,
  for each pair of values of $\Delta_\mathrm{solar}$
  and $\Delta_\mathrm{atmospheric}$,
  by the {\it Planck}~2018 cosmological upper bound~\eqref{cosmological}.
  %
\item The Yukawa coupling constants $d_\ell$,
  $\delta_\ell$,
  and $\gamma_\ell$ are assumed to be real
  (either positive or negative).\footnote{We have also investigated
the case with complex Yukawa couplings.
We have found out that
its results do not differ much from the real case,
therefore we do not present fits with complex couplings.}
\item The moduli of the Yukawa coupling
  constants
  are varied between $\Ymin = 10^{-6}$
  (which is the order of magnitude
  of the Yukawa coupling of the electron)
  and a perturbativity bound $\Ymax = \sqrt{4 \pi} \approx 3.5$.
\item We enforce Eqs.~\eqref{SM10conditions}.
\end{itemize}

There are experimental and phenomenological constraints
on the mass of the charged scalar $m_{H^+}$,
as discussed in Appendix~\ref{Sect_mH+}.
The numerical study in the previous subsection
(see Fig.~\ref{fig-mC})
shows that,
when $m_{H^+}$ increases,
most BRs decrease.
Since we attempt to obtain largish BRs,
the fitting procedure always tends to produce
the lowest $m_{H^+}$ in the allowed range.
In our fits we have fixed $m_{H^+} = 750$\,GeV,
in agreement with the lower bounds of
recent global fits~\cite{Chowdhury:2017aav,Eberhardt:2020dat}.
We have also kept the triple-scalar couplings fixed,
\textit{viz.}\ $\lambda_3 = \lambda_7 = 1$,
because they do not change much the BRs. 
Finally,
we have checked that all the final points
meet the $3\sigma$ conditions on the $Z$ invisible decay width
in Eq.~\eqref{InvZwidthFull}.

In the figures of this subsection we display three different fits:
\begin{enumerate}
\item In the first fit
  (displayed through blue points and called `NO' from now on),
  we have assumed normal ordering of the light-neutrino masses.
\item In the second fit
  (displayed through red points and named `IO')
  there is inverted ordering of the light-neutrino masses.
\item In the numerical analysis\footnote{See
the histograms of Fig.~\ref{fig-hist_Yukawas}.}
  we have found that most points have $\left| d_\ell \right|$
  close to the lower bound $Y_{\mathrm{min}} = 10^{-6}$.
  Therefore,
  we have produced a third fit
  (displayed through green points and labelled `$\Ymin = 10^{-7}$')
  that has normal ordering like the first one,
  but where the lower bound on the moduli of the Yukawa couplings
  is $10^{-7}$ instead of $10^{-6}$.\footnote{The numerical analysis
  has also shown that most points have $\left| \delta_\ell \right|$
  close to the upper bound $Y_{\mathrm{max}} \approx 3.5$.
  Therefore we have made an extra fit
  where that upper bound was relaxed to 12.5.
  However,
  that extra fit,
  which we do not display,
  did \emph{not} produce much improvement on BR($\ell$) and BR($h$).
  It did produce larger BR($Z$),
  but they were still very much below the future experimental sensitivities.
  Thus,
  it appears to us that there are no advantages
  in allowing the moduli of the Yukawa couplings to be larger than 3.5.}
\end{enumerate}
\begin{figure}[h]
\begin{center}
\includegraphics[width=1.0\textwidth]{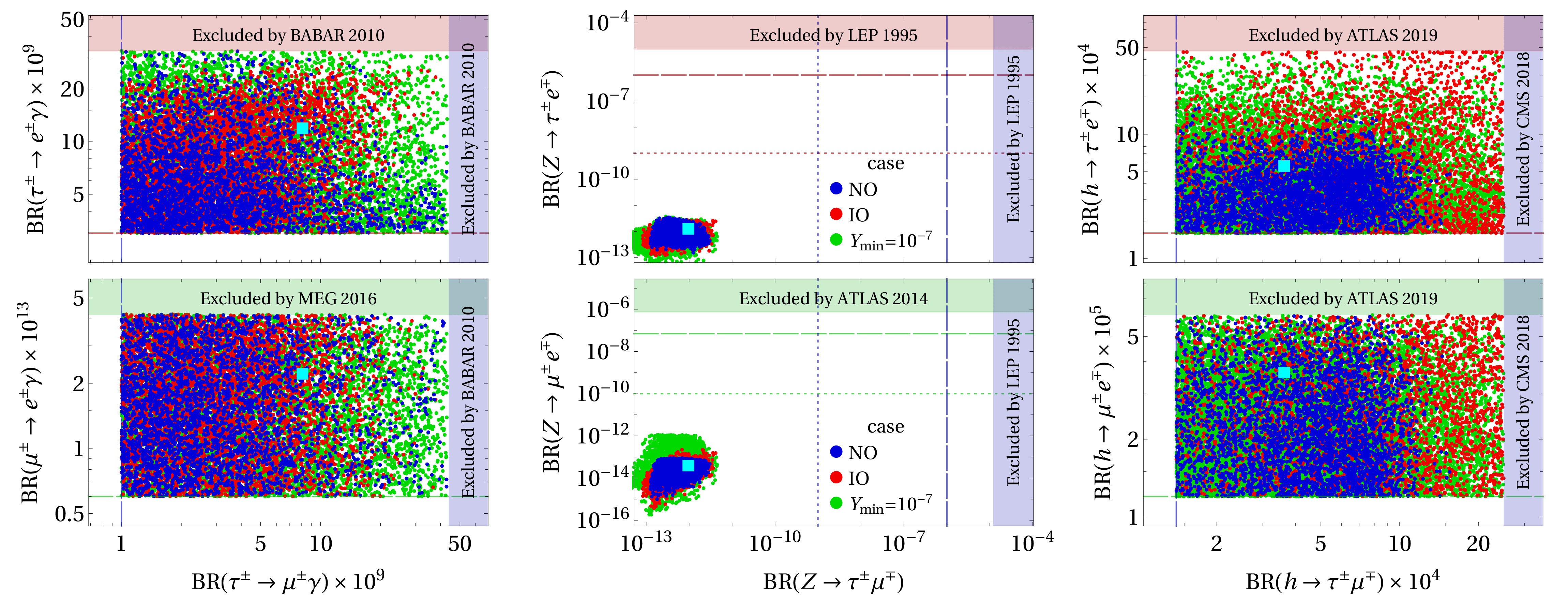}
\end{center}
\caption{Scatter plots of the branching ratios for three different fits.
  Blue points have normal ordering of the light-neutrino masses,
  red points have inverted ordering,
  and green points have a lower bound
  ($10^{-7}$ instead of $10^{-6}$)
  on the moduli of the Yukawa couplings.
  A cyan square marks~\protect\hyperlink{BP1}{BP-1} of table~\ref{table-bench}.
  The shadowed bands are the present experimental upper bounds on the BRs;
  dashed and/or dotted lines show the future experimental sensitivities,
  as described in table~\ref{LFV_limits}.}
\label{fig-scatter}
\end{figure} 
Figure~\ref{fig-scatter} shows that points for the NO and IO cases
are similarly distributed in what respects the BR($\ell$) and BR($Z$).
It is possible in both cases to find points
with the BR($\ell$) close to the experimental upper bounds,
while the BR($Z$) always remain much too suppressed.
For the BR($h$),
on the other hand,
NO usually leads to smaller values than IO.
The larger freedom of the third fit
(with $\Ymin = 10^{-7}$ instead of $\Ymin = 10^{-6}$)
facilitates larger BR($\ell$),
as shown by the green points in Fig.~\ref{fig-scatter}.

Most points in Fig.~\ref{fig-scatter} have negative $\kappa_\tau$.
This allows larger $\mathrm{BR} \left( h \to \tau^\pm \mu^\mp \right)$
and $\mathrm{BR} \left( h \to \tau^\pm e^\mp \right)$.
If one only allows positive $\kappa_\tau$,
then in NO it is not possible to reach the future sensitivity
for $\mathrm{BR} \left( h \to \tau^\pm \mu^\mp \right)$,
except if one allows complex Yukawa couplings.
On the other hand,
in both the IO and $\Ymin = 10^{-7}$ cases
it is still possible to get all three BR($h$) above their future sensitivities
with positive $\kappa_\mu$ and $\kappa_\tau$.

We have found that free Majorana phases
permit larger BRs for the decays
with $\tau^\pm$.
Thus,
it is advantageous to fit the Majorana phases
instead of fixing them at any pre-assigned values.

\begin{figure}[h]
\begin{center}
\includegraphics[width=1.0\textwidth]{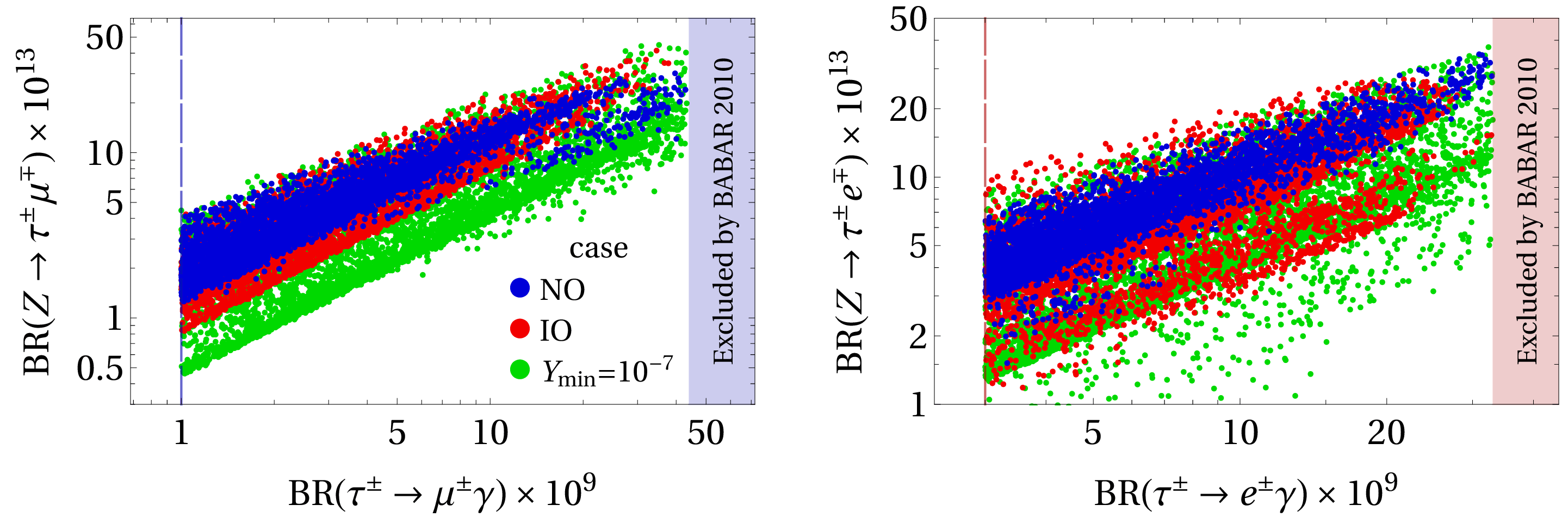}
\end{center}
\caption{Correlation plots between BR($\ell$) and BR($Z$)
  for the three fits of Fig.~\ref{fig-scatter};
  the points and the notation are the same as in that figure.
}
\label{fig-correlation}
\end{figure}

\begin{figure}[h]
\begin{center}
\includegraphics[width=1.0\textwidth]{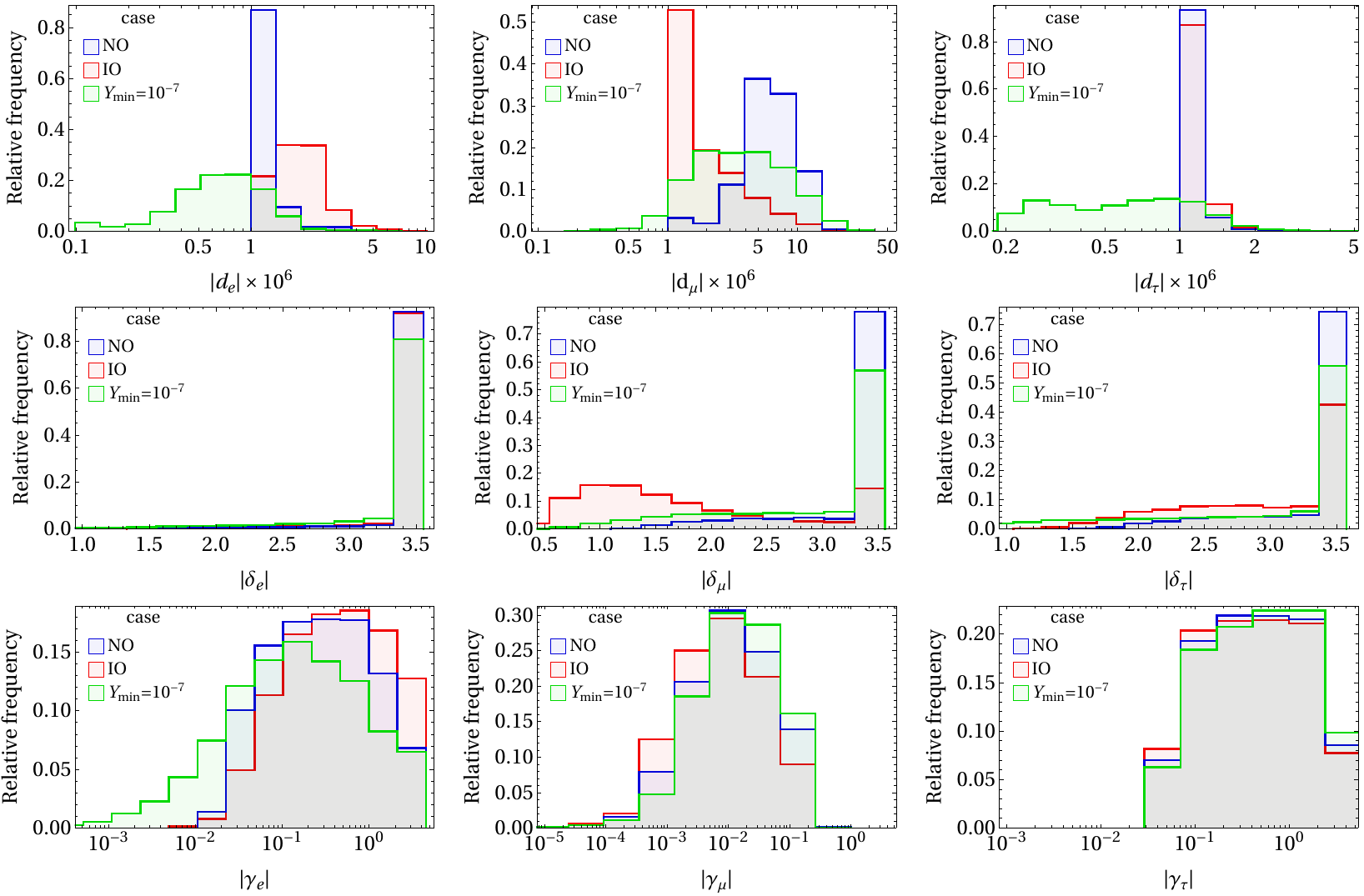}
\end{center}
\caption{The distributions of the moduli of the Yukawa couplings
  for the three fits.}
\label{fig-hist_Yukawas}
\end{figure}

In Fig.~\ref{fig-correlation}
we display correlation plots of BR($\ell$) and BR($Z$).
One sees that there is a correlation
between $\mathrm{BR} \left( \tau^\pm \to \mu^\pm \gamma \right)$
and $\mathrm{BR} \left( Z \to \tau^\pm \mu^\mp \right)$,
and a correlation between
$\mathrm{BR} \left( \tau^\pm \to e^\pm \gamma \right)$
and $\mathrm{BR} \left( Z \to \tau^\pm e^\mp \right)$.
These correlations are one of the main reasons for the small BR($Z$)
in our model;
if we want to keep the BR($\ell$) below their experimental upper bounds,
then we necessarily obtain much too low BR($Z$).
Indeed,
one sees in Fig.~\ref{fig-correlation} that
the BABAR 2010 upper bounds on the BR($\tau$) lead,
in our model,
to $\mathrm{BR} \left( Z \to \tau^\pm \mu^\mp \right) \lesssim 6 \times 10^{-12}$
and $\mathrm{BR} \left( Z \to \tau^\pm e^\mp \right) \lesssim 4 \times 10^{-12}$;
those values
are much lower than the future experimental sensitivity.
We point out that in other models
(see for instance Refs.~\cite{DeRomeri:2016gum},
\cite{Iltan:2001au},
and~\cite{CortesMaldonado:2011uh})
there are also correlations between the BR($Z$) and BR($\ell$),
and also with the branching ratios for three-body LFV decays
$\ell^\pm_1 \to \ell^\pm_2 \ell^+_3 \ell^-_3$.

In some models there are correlations between
$\mathrm{BR} \left( h \to \tau^\pm \mu^\mp \right)$
and either~\cite{Sierra:2014nqa, Crivellin:2015hha, Liu:2015oaa,
  Altmannshofer:2016oaq}
$\mathrm{BR} \left( \tau^\pm \to \mu^\pm \gamma \right)$
or~\cite{Botella:2015hoa}
$\mathrm{BR} \left( \mu^\pm \to e^\pm \gamma \right)$.
In our model we did \emph{not} find correlations
between the BR($\ell$) and BR($h$).

In Fig.~\ref{fig-hist_Yukawas} we display histograms
of the moduli of the Yukawa couplings for our three fits.
In the first row of panels one sees that,
in order to get BR($\ell$) in experimentally reachable ranges,
our fits always have very small
$\left| d_\ell \right| \lesssim 10^{-5}$.
If we had set $Y_{\mathrm{min}}$ much larger than $10^{-6}$,
then it might not have been possible to obtain BR($\ell$)
visible in the next generation of experiments.
In the fit with relaxed $Y_{\mathrm{min}} = 10^{-7}$
the distribution of the $\left| d_\ell \right|$ is more uniform.
The second row of Fig.~\ref{fig-hist_Yukawas} shows that,
in all three fits,
$\left| \delta_{e, \mu, \tau} \right|$ have values close
to the allowed upper bound $Y_{\mathrm{max}}$.
In the third row
one sees that the $\left| \gamma_{e, \mu, \tau} \right|$
vary in rather wide ranges,
from $\sim 10^{-4}$ to $Y_{\mathrm{max}}$.
This happens because the parameter $1 - x_2$ has a strong impact on BR($h$);
for smaller values of the $\gamma_\ell$,
larger values of $1 - x_2$ still allow BR($h$)
to reach experimentally reachable ranges.

\begin{figure}[h]
\begin{center}
\includegraphics[width=1.0\textwidth]{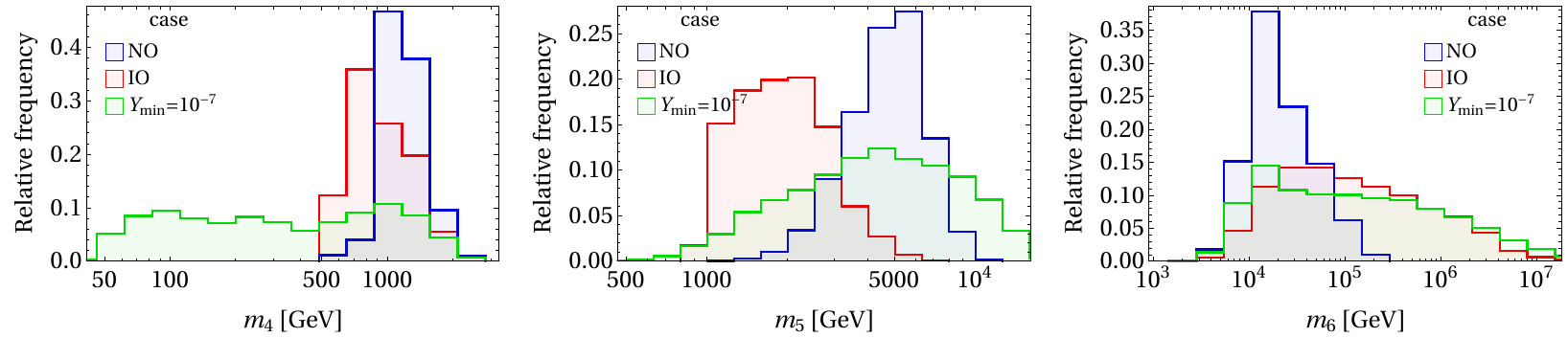}
\end{center}
\caption{Histograms of the heavy-neutrino masses for the three fits.}
\label{fig-hist_MR}
\end{figure}

In Fig.~\ref{fig-hist_MR} we display the heavy-neutrino masses
$m_4$,
$m_5$,
and $m_6$ for our three fits.
Because the $\left| d_\ell \right|$ are always so small in the fits,
the heavy-neutrino masses are very small too.
Thus,
in NO and IO $m_4$ lies in between $\sim 0.5$\,TeV and $\sim 2.5$\,TeV,
and in $Y_{\mathrm{min}} = 10^{-7}$
it may be as small as 45\,GeV.\footnote{We note
the recent paper~\cite{Bensalem:2021wfr}
that analyzes a model including a Majorana neutrino with mass of order 100~GeV.
That model apparently gives rise to lepton-number-violating signatures
that might be visible at the LHC.}
The mass $m_5$ is in between $\sim 1$\,TeV and $\sim 10$\,TeV for all cases,
and the mass $m_6$ is in between $\sim 2.5$\,TeV and $\sim 100$\,TeV for NO,
and $\sim 10^4$\,TeV for both IO and $\Ymin = 10^{-7}$.

In all three fits,
it is found that the mixing angles $\theta_{12,23,13}$,
the Dirac phase $\delta$,
and the Majorana phases $\alpha_{21}$ and $\alpha_{31}$
may have values anywhere in their ranges.

\subsection{Single decays}
\label{Sect_singlededays}

In the previous subsection we have discussed the results
that are obtained when all six BR($\ell$) and BR($h$)
are simultaneously between the current experimental upper bounds
and the future experimental sensitivities.
Here we consider the case where \emph{only one}
of those six BRs is above the future sensitivity.

We have found that requiring just one BR($\ell$)
to be above the future sensitivity still restricts
the Yukawa couplings $\left| d_\ell \right| \lesssim 10^{-3}$
and $\left| \delta_\ell \right| \gtrsim 0.5$.
Then,
because of the small $\left| d_\ell \right|$,
the heavy-neutrino mass $m_4$ is always of order 1\,TeV
(except in $\mu^\pm \to e^\pm \gamma$
for which $m_4$ may be of order 50\,TeV).

Requiring just one BR($h$) to be above the future sensitivity restricts
$\left| \delta_\ell \right| \gtrsim 0.1$.
The $\left| d_\ell \right|$
and the heavy-neutrino masses do not need to be very small,
as one sees for instance in~\hyperlink{BP1}{BP-2} of table~\ref{table-bench}.

In our model the decay $Z \to \mu^\pm e^\mp$ might be observed
at the FCC-ee collider in a very restricted range of circumstances,
\textit{viz.}\ with large $\left| \delta_{e, \mu} \right| \gtrsim 4$,
small $\left| d_e \right| \lesssim 10^{-6}$
and $\left| d_{\mu, \tau} \right| \lesssim 5 \times 10^{-6}$,
and small $m_{H^+} \lesssim 500$\,GeV.
Moreover,
a very precise finetuning is required,
wherein the Yukawa couplings are such that
on the hand the decay $\mu^\pm \to e^\pm \gamma$
has a cancellation of amplitudes leading to its BR being below
the experimental upper bound,
and on the other hand $\mathrm{BR} \left( Z \to \mu^\pm e^\mp \right)$
still remains a little above the FCC-ee sensitivity.
The other two
LFV $Z$ decays $Z \to \tau^\pm e^\mp$ and $Z \to \tau^\pm \mu^\mp$
are in our model always much too suppressed to be visible.

\subsection{Amplitudes}
\label{Sect_amplitudes}

Numerically,
we have found that only a few amplitudes
have a substantial impact on the BRs.

For $\ell_1^\pm \to \ell_2^\pm \gamma$ the amplitudes~\eqref{coef1} are dominant.
Specifically,
$a_{l,H}$ in Eq.~\eqref{al1} gives the main impact.
Therefore,
the approximate decay width is
\be
\label{DecWidthapp-lept}
\Gamma \left( \ell_1^\pm \to \ell_2^\pm \gamma \right)
\approx \frac{m_{\ell_1}^4 - m_{\ell_2}^4}{16 \pi m_{\ell_1}^3}\,
\left| \mathcal{S}\, e \right|^2 \left| a_{l,H} \right|^2.
\ee
This yields the following approximate formulas for the BRs:
\bs
\label{BRapp-lept}
\ba
\mathrm{BR} \left( \tau^\pm \to \mu^\pm \gamma \right)
&\approx& \left( 5.733 \times 10^4 \right) \left| a_{l,H} \right|^2,
\\
\mathrm{BR} \left( \tau^\pm \to e^\pm \gamma \right)
&\approx& \left( 5.733 \times 10^4 \right) \left| a_{l,H} \right|^2,
\\
\mathrm{BR} \left( \mu^\pm \to e^\pm \gamma \right) 
&\approx& \left( 2.580 \times 10^{10} \right)
\left| a_{l,H} \right|^2.
\ea
\es

The amplitude $\bar{a}_{l,H}$ in Eq.~\eqref{al1Z}
is the most important one for the BR($Z$).~\footnote{Due
  to the similarities between $a_{l,H}$ in Eq.~\eqref{al1}
  and $\bar{a}_{l,H}$ in Eq.~\eqref{al1Z},
  there are correlations between
  $\mathrm{BR} \left( \ell_1^\pm \to \ell_2^\pm \gamma \right)$
  and $\mathrm{BR} \left( Z \to \ell_1^\pm \ell_2^\mp \right)$,
  as already seen in Fig.~\ref{fig-correlation}.}
Therefore,
\ba
\label{DecWidthapp-Z}
\Gamma \left( Z \to \ell_1^\pm \ell_2^\mp \right)
&\approx& \frac{\sqrt{m_Z^4 - 2 m_Z^2 \left( m_{\ell_1}^2 + m_{\ell_2}^2 \right)
+ \left( m_{\ell_1}^2 - m_{\ell_2}^2 \right)^2}}{48 \pi m_Z^5}\, \notag
\\ 
&&\times \left| \mathcal{S}\, e \right|^2
\left[ 2 m_Z^4 - m_Z^2 \left( m_{\ell_1}^2 + m_{\ell_2}^2 \right)
- \left( m_{\ell_1}^2 - m_{\ell_2}^2 \right)^2 \right]
\left| \bar{a}_{l,H} \right|^2.
\ea
Hence,
\be
\label{BRapp-Z}
\mathrm{BR} \left( Z \to \ell_1^\pm \ell_2^\mp \right)
\approx \left( 1.781 \times 10^{-6} \right)
\left| \bar{a}_{l,H} \right|^2.
\ee

The amplitudes for the Higgs decays differ from those for the other decays.
The amplitudes from the self-energy-like diagrams of Fig.~\ref{fig4},
with the charged scalar $H^\pm$,
give the strongest impact on the branching ratios. 
Specifically,
the amplitude $d_{rb,\mathrm{16(a,b)}}$ in Eq.~\eqref{49b}
is significant for largish values of the Yukawa couplings $d_\ell$
and the amplitude $d_{lb,\mathrm{16(a,b)}}$ in Eq.~\eqref{49a}
is significant for all values of the $d_\ell$. 
Moreover,
for the $h$ decays the amplitudes from diagrams
with two internal neutrino lines,
depicted in Fig.~\ref{fig7},
are important too.
Specifically,
the amplitudes $d_{lb,\mathrm{19(a)}}$ and $d_{rb,\mathrm{19(a)}}$ are relevant.
Thus,
defining
\be
\label{BRapp-H2}
\bar{d}_{lb} = d_{lb,\mathrm{16(a,b)}} + d_{lb,\mathrm{19(a)}},
\quad \quad
\bar{d}_{rb} = d_{rb,\mathrm{16(a,b)}} + d_{rb,\mathrm{19(a)}},
\ee
we have
\bs
\label{BRapp-H}
\ba
\mathrm{BR} \left( h \to \tau^\pm \mu^\mp \right)
&\approx&
\left( 2.451 \times 10^{-2} \right)
\left(\left| \bar{d}_{lb} \right|^2 + \left| \bar{d}_{rb} \right|^2\right)
\no & &
- \left( 1.176 \times 10^{-6} \right)
\mathrm{Re} \left(\bar{d}_{lb} \bar{d}_{rb}^\ast \right),
\\
\mathrm{BR}\left( h \to \tau^\pm e^\mp \right)
&\approx&
\left( 2.451 \times 10^{-2} \right)
\left(\left| \bar{d}_{lb} \right|^2 + \left| \bar{d}_{rb} \right|^2\right)
\no & &
-\left( 5.690 \times 10^{-9} \right)
\mathrm{Re} \left(\bar{d}_{lb} \bar{d}_{rb}^\ast \right),
\\
\mathrm{BR} \left( h \to \mu^\pm e^\mp \gamma \right)
&\approx&
\left( 2.452 \times 10^{-2} \right)
\left(\left| \bar{d}_{lb} \right|^2 + \left| \bar{d}_{rb} \right|^2\right)
\no & &
-\left( 3.384 \times 10^{-10} \right)
\mathrm{Re} \left(\bar{d}_{lb} \bar{d}_{rb}^\ast \right).
\ea
\es

\begin{figure}[h]
\begin{center}
\includegraphics[width=1.0\textwidth]{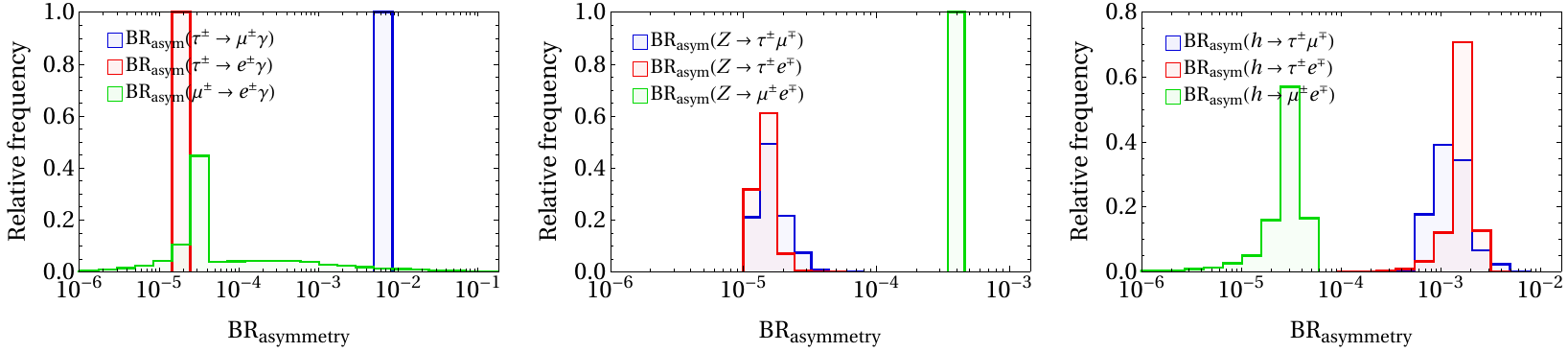}
\end{center}
\caption{Histograms of the asymmetries
  between the exact nine branching ratios
  and their approximate expressions in Eqs.~\eqref{BRapp-lept},
\eqref{BRapp-Z},
and~\eqref{BRapp-H2}--\eqref{BRapp-H}.}
\label{fig-hist_BRasym}
\end{figure}

In order to check the correctness of the approximate BRs
of Eqs.~\eqref{BRapp-lept},
\eqref{BRapp-Z},
and~\eqref{BRapp-H} we have calculated
the asymmetry between the exact BRs and the approximate ones,
\be
\mathrm{BR}_\mathrm{asymmetry} \equiv
\frac{\left| \mathrm{BR}_\mathrm{exact}
  - \mathrm{BR}_\mathrm{approximate} \right|}{\mathrm{BR}_\mathrm{exact}
  + \mathrm{BR}_\mathrm{approximate}}.
\label{BRasym}
\ee
Using the points of case `NO',
these asymmetries are displayed in Fig.~\ref{fig-hist_BRasym}. 
One sees that $\mathrm{BR}_\mathrm{asymmetry} \lesssim 0.1$,
which means that the approximate formulas are quite accurate. 
These approximate expressions for the BRs
may be very useful for intermediate calculations of the fitting procedure,
where the calculations need to be repeated many times,
before the final result is calculated by using the exact expressions. 
This computational trick has saved us a lot of
time.

\section{Summary and conclusions}
\label{Sect_Conclusions}

Here we summarize our main findings:

\begin{itemize}

\item The amplitudes with the charged scalar are crucial in order to obtain
  $\mathrm{BR} \left( \ell_1^\pm \to \ell_2^\pm \gamma \right)$
  and $\mathrm{BR} \left( h \to \ell_1^\pm \ell_2^\mp \right)$
  in experimentally accessible ranges.

\item Because the experimental upper bound on
  $\mathrm{BR} \left( \mu^\pm \to e^\pm \gamma \right)$ is very small,
  it is often necessary to finetune the values of the input parameters
  of the model so that the largest amplitudes for that specific decay
  partially cancel among themselves,
  while the other five decays $\ell_1^\pm \to \ell_2^\pm \gamma$
  and $h \to \ell_1^\pm \ell_2^\mp$
  remain experimentally visible in the future.

\item The decays $\ell_1^\pm \to  \ell_2^\pm \gamma$
  necessitate large values of the Yukawa couplings
  $\left| \delta_{\ell} \right| \gtrsim 1$
  and extremely small values of the Yukawa couplings
  $\left| d_{\ell} \right| \lesssim 10^{-5}$
  in order to be visible.
  The latter imply a very low seesaw scale,
  \textit{i.e.}\ rather light right-handed neutrinos.

\item In our model the decays $Z \to \ell_1^\pm \ell_2^\mp$
  correlate with the decays $\ell_1^\pm \to  \ell_2^\pm \gamma$,
  \textit{i.e.}\ they behave similarly as functions of the parameters.
  Because of this correlation,
  the experimental upper bounds on
  $\mathrm{BR} \left( \ell_1^\pm \to  \ell_2^\pm \gamma \right)$
  imply that the decays $Z \to \ell_1^\pm \ell_2^\mp$
  will remain invisible in all the planned experiments.

\item The decays $h \to \ell_1^\pm \ell_2^\mp$
  behave quite differently from $\ell_1^\pm \to \ell_2^\pm \gamma$
  and $Z \to \ell_1^\pm \ell_2^\mp$.
  They might be visible in future experiments
  without the need to choose
  either
  very large or very small Yukawa couplings.

\item The Majorana phases have a significant impact
  on the branching ratios of all the decays.
  One should refrain from fixing them at
  any
  pre-assigned values.

\item Both
  normal and inverted ordering of the light-neutrino masses
  may yield decay rates of adequate orders of magnitude.

\item When the mass of the charged scalar increases,
  most BRs decrease.
  Still,
  for $m_{H^+} \lesssim 1.5$\,TeV
  the decays $\ell_1^\pm \to \ell_2^\pm \gamma$ and $h \to \ell_1^\pm \ell_2^\mp$
  might be visible in future experiments.

\end{itemize}

\vspace*{5mm}

\paragraph{Acknowledgements:}
D.J.\ thanks both Jorge C.\ Rom\~ao and Duarte Fontes for useful discussions.
He also thanks the Lithuanian Academy of Sciences
for financial support through projects DaFi2019 and DaFi2021;
he was also supported by a COST STSM grant through action CA16201.
L.L.\ warmly thanks the Institute of Theoretical Physics and Astronomy
of the University of Vilnius for the hospitality extended during a visit
where part of this work has been done.
L.L.\ also thanks the Portuguese Foundation for Science and Technology
for support through projects CERN/FIS-PAR/0004/2019,
CERN/FIS-PAR/0008/2019,
PTDC/FIS-PAR/29436/2017,
UIDB/00777/2020,
and UIDP/00777/2020.


\newpage
\appendix

\setcounter{equation}{0}
\renewcommand{\theequation}{A\arabic{equation}}

\section{Passarino--Veltman functions}
\label{Sect_PVfunctions}

The relevant Passarino--Veltman (PV) functions
are defined in the following way.
Let the dimension of space--time be $d = 4 - \epsilon$
with $\epsilon \to 0$.
We define
\be
\label{S}
\mathcal{D} k := \mu^\epsilon\,
\frac{\mathrm{d}^d k}{\left( 2 \pi \right)^d},
\quad \quad
\mathcal{S} := \frac{i}{16 \pi^2}.
\ee
Then,
\bs
\label{b}
\ba
\int \mathcal{D} k\ \frac{1}{k^2 - A}\
\frac{1}{\left( k + p \right)^2 - B}
&=& \mathcal{S}\, B_0 \left( p^2, A, B \right),
\\
\int \mathcal{D} k\ k^\theta\,
\frac{1}{k^2 - A}\ \frac{1}{\left( k + p \right)^2 - B}
&=& \mathcal{S}\, p^\theta\, B_1 \left( p^2, A, B \right),
\label{ncjidso}
\ea
\es
and
\bs
\label{c}
\ba
\int \mathcal{D} k\ \frac{1}{k^2 - A}\
\frac{1}{\left( k + p \right)^2 - B}\
\frac{1}{\left( k + q \right)^2 - C}
&=& \mathcal{S}\, C_0
\left[ p^2, \left( p - q \right)^2, q^2, A, B, C \right],
\\
\int \mathcal{D} k\ k^\theta\, \frac{1}{k^2 - A}\
\frac{1}{\left( k + p \right)^2 - B}\
\frac{1}{\left( k + q \right)^2 - C}
&=&
\mathcal{S} \left\{
p^\theta\, C_1
\left[ p^2, \left( p - q \right)^2, q^2, A, B, C \right]
\right. \no & & \left.
+ q^\theta\, C_2
\left[ p^2, \left( p - q \right)^2, q^2, A, B, C \right]
\right\},
\no & &
\\
\int \mathcal{D} k\ k^\theta k^\psi\, \frac{1}{k^2 - A}\
\frac{1}{\left( k + p \right)^2 - B}\
\frac{1}{\left( k + q \right)^2 - C}
&=&
\mathcal{S} \left\{
p^\theta p^\psi\,
C_{11} \left[ p^2, \left( p - q \right)^2, q^2, A, B, C \right]
\right. \no & &
+ q^\theta q^\psi\,
C_{22} \left[ p^2, \left( p - q \right)^2, q^2, A, B, C \right]
\no & &
+ \left( p^\theta q^\psi + q^\theta p^\psi \right) C_{12}
\left[ p^2, \left( p - q \right)^2, q^2,
  \right. \no & & \left.
  A, B, C \right]
\no & & \left.
+ g^{\theta \psi}\, C_{00}
\left[ p^2, \left( p - q \right)^2, q^2, A, B, C \right]
\right\}.
\no & &
\label{cbufio}
\ea
\es

Some PV functions in Eqs.~\eqref{b} and~\eqref{c}
have $1 \! \left/ \epsilon \right.$ divergences
that are independent of the arguments of the PV functions.
Thus,
\bs
\ba
B_0 \left( p^2, A, B \right) &=&
\frac{2}{\epsilon} + \mathrm{finite\ terms},
\\
B_1 \left( p^2, A, B \right) &=&
- \frac{1}{\epsilon} + \mathrm{finite\ terms},
\\
C_{00} \left[ p^2, \left( p - q \right)^2, q^2, A, B, C \right]
&=& \frac{1}{2 \epsilon} + \mathrm{finite\ terms}.
\label{c00}
\ea
\es
All other PV functions in Eqs.~\eqref{b} and~\eqref{c}
converge when $\epsilon \to 0$.

We next introduce specific notations for some PV functions
that are used in appendices~\ref{Sect_taumugamma},
\ref{Sect_taumuZ},
and~\ref{Sect_taumuS}.
Thus,
\bs
\label{efunctions}
\ba
e_0 &:=& C_0 \left( m_{\ell_1}^2, q^2, m_{\ell_2}^2,
m_i^2, m_a^2, m_a^2 \right), \\
e_1 &:=& C_1 \left( m_{\ell_1}^2, q^2, m_{\ell_2}^2,
m_i^2, m_a^2, m_a^2 \right), \\
e_2 &:=& C_2 \left( m_{\ell_1}^2, q^2, m_{\ell_2}^2,
m_i^2, m_a^2, m_a^2 \right), \\
e_{11} &:=& C_{11} \left( m_{\ell_1}^2, q^2, m_{\ell_2}^2,
m_i^2, m_a^2, m_a^2 \right), \\
e_{22} &:=& C_{22} \left( m_{\ell_1}^2, q^2, m_{\ell_2}^2,
m_i^2, m_a^2, m_a^2 \right), \\
e_{12} &:=& C_{12} \left( m_{\ell_1}^2, q^2, m_{\ell_2}^2,
m_i^2, m_a^2, m_a^2 \right), \\
e_{00} &:=& C_{00} \left( m_{\ell_1}^2, q^2, m_{\ell_2}^2,
m_i^2, m_a^2, m_a^2 \right),
\ea
\es
\vspace*{-4mm}
\bs
\label{ffunctions}
\ba
f_0 &:=& C_0 \left( m_{\ell_1}^2, q^2, m_{\ell_2}^2,
m_i^2, m_W^2, m_W^2 \right), \\
f_1 &:=& C_1 \left( m_{\ell_1}^2, q^2, m_{\ell_2}^2,
m_i^2, m_W^2, m_W^2 \right), \\
f_2 &:=& C_2 \left( m_{\ell_1}^2, q^2, m_{\ell_2}^2,
m_i^2, m_W^2, m_W^2 \right), \\
f_{11} &:=& C_{11} \left( m_{\ell_1}^2, q^2, m_{\ell_2}^2,
m_i^2, m_W^2, m_W^2 \right), \\
f_{22} &:=& C_{22} \left( m_{\ell_1}^2, q^2, m_{\ell_2}^2,
m_i^2, m_W^2, m_W^2 \right), \\
f_{12} &:=& C_{12} \left( m_{\ell_1}^2, q^2, m_{\ell_2}^2,
m_i^2, m_W^2, m_W^2 \right), \\
f_{00} &:=& C_{00} \left( m_{\ell_1}^2, q^2, m_{\ell_2}^2,
m_i^2, m_W^2, m_W^2 \right),
\ea
\es
\vspace*{-4mm}
\bs
\label{gfunctions}
\ba
g_0 &:=& C_0 \left( m_{\ell_1}^2, q^2, m_{\ell_2}^2,
m_a^2, m_i^2, m_j^2 \right), \\
g_1 &:=& C_1 \left( m_{\ell_1}^2, q^2, m_{\ell_2}^2,
m_a^2, m_i^2, m_j^2 \right), \\
g_2 &:=& C_2 \left( m_{\ell_1}^2, q^2, m_{\ell_2}^2,
m_a^2, m_i^2, m_j^2 \right), \\
g_{11} &:=& C_{11} \left( m_{\ell_1}^2, q^2, m_{\ell_2}^2,
m_a^2, m_i^2, m_j^2 \right), \\
g_{22} &:=& C_{22} \left( m_{\ell_1}^2, q^2, m_{\ell_2}^2,
m_a^2, m_i^2, m_j^2 \right), \\
g_{12} &:=& C_{12} \left( m_{\ell_1}^2, q^2, m_{\ell_2}^2,
m_a^2, m_i^2, m_j^2 \right), \\
g_{00} &:=& C_{00} \left( m_{\ell_1}^2, q^2, m_{\ell_2}^2,
m_a^2, m_i^2, m_j^2 \right),
\ea
\es
\vspace*{-4mm}
\bs
\label{hfunctions}
\ba
h_0 &:=& C_0 \left( m_{\ell_1}^2, q^2, m_{\ell_2}^2,
m_W^2, m_i^2, m_j^2 \right), \\
h_1 &:=& C_1 \left( m_{\ell_1}^2, q^2, m_{\ell_2}^2,
m_W^2, m_i^2, m_j^2 \right), \\
h_2 &:=& C_2 \left( m_{\ell_1}^2, q^2, m_{\ell_2}^2,
m_W^2, m_i^2, m_j^2 \right), \\
h_{11} &:=& C_{11} \left( m_{\ell_1}^2, q^2, m_{\ell_2}^2,
m_W^2, m_i^2, m_j^2 \right), \\
h_{22} &:=& C_{22} \left( m_{\ell_1}^2, q^2, m_{\ell_2}^2,
m_W^2, m_i^2, m_j^2 \right), \\
h_{12} &:=& C_{12} \left( m_{\ell_1}^2, q^2, m_{\ell_2}^2,
m_W^2, m_i^2, m_j^2 \right), \\
h_{00} &:=& C_{00} \left( m_{\ell_1}^2, q^2, m_{\ell_2}^2,
m_W^2, m_i^2, m_j^2 \right),
\ea
\es
\vspace*{-4mm}
\bs
\label{jfunctions}
\ba
j_0 &:=& C_0 \left( m_{\ell_1}^2, q^2, m_{\ell_2}^2,
m_i^2, m_a^2, m_{a^\prime}^2 \right),
\\
j_1 &:=& C_1 \left( m_{\ell_1}^2, q^2, m_{\ell_2}^2,
m_i^2, m_a^2, m_{a^\prime}^2 \right),
\\
j_2 &:=& C_2 \left( m_{\ell_1}^2, q^2, m_{\ell_2}^2,
m_i^2, m_a^2, m_{a^\prime}^2 \right),
\ea
\es
\vspace*{-4mm}
\bs
\label{kfunctions}
\ba
k_0 &:=& C_0 \left( m_{\ell_1}^2, q^2, m_{\ell_2}^2,
m_i^2, m_W^2, m_a^2 \right), \\
k_1 &:=& C_1 \left( m_{\ell_1}^2, q^2, m_{\ell_2}^2,
m_i^2, m_W^2, m_a^2 \right), \\
k_2 &:=& C_2 \left( m_{\ell_1}^2, q^2, m_{\ell_2}^2,
m_i^2, m_W^2, m_a^2 \right), \\
k_{00} &:=& C_{00} \left( m_{\ell_1}^2, q^2, m_{\ell_2}^2,
m_i^2, m_W^2, m_a^2 \right), \\
k_{11} &:=& C_{11} \left( m_{\ell_1}^2, q^2, m_{\ell_2}^2,
m_i^2, m_W^2, m_a^2 \right), \\
k_{22} &:=& C_{22} \left( m_{\ell_1}^2, q^2, m_{\ell_2}^2,
m_i^2, m_W^2, m_a^2 \right), \\
k_{12} &:=& C_{12} \left( m_{\ell_1}^2, q^2, m_{\ell_2}^2,
m_i^2, m_W^2, m_a^2 \right),
\ea
\es
\vspace*{-4mm}
\bs
\label{lfunctions}
\ba
l_0 &:=& C_0 \left( m_{\ell_1}^2, q^2, m_{\ell_2}^2,
m_i^2, m_a^2, m_W^2 \right), \\
l_1 &:=& C_1 \left( m_{\ell_1}^2, q^2, m_{\ell_2}^2,
m_i^2, m_a^2, m_W^2 \right), \\
l_2 &:=& C_2 \left( m_{\ell_1}^2, q^2, m_{\ell_2}^2,
m_i^2, m_a^2, m_W^2 \right), \\
l_{00} &:=& C_{00} \left( m_{\ell_1}^2, q^2, m_{\ell_2}^2,
m_i^2, m_a^2, m_W^2 \right), \\
l_{11} &:=& C_{11} \left( m_{\ell_1}^2, q^2, m_{\ell_2}^2,
m_i^2, m_a^2, m_W^2 \right), \\
l_{22} &:=& C_{22} \left( m_{\ell_1}^2, q^2, m_{\ell_2}^2,
m_i^2, m_a^2, m_W^2 \right), \\
l_{12} &:=& C_{12} \left( m_{\ell_1}^2, q^2, m_{\ell_2}^2,
m_i^2, m_a^2, m_W^2 \right),
\ea
\es
where $m_{\ell_1}$ and $m_{\ell_2}$
are the masses of the charged leptons
$\ell_1^\pm$ and $\ell_2^\pm$,
respectively,
$m_i$ and $m_j$ are the masses of the neutrinos
$\nu_i$ and $\nu_j$,
respectively,
$m_a$ and $m_{a^\prime}$ are the masses of the charged scalars
$H_a^\pm$ and $H_{a^\prime}^\pm$,
respectively,
and $m_W$ is the mass of the gauge bosons $W^\pm$.

\newpage

\section{$\ell_1^\pm \to \ell_2^\pm \gamma$}
\label{Sect_taumugamma}
\setcounter{equation}{0}
\renewcommand{\theequation}{B\arabic{equation}}

We compute the process $\ell_1^- \left( p_1 \right)
\to \ell_2^- \left( p_2 \right) \gamma \left( q \right)$,
where $q = p_1 - p_2$.
Obviously,
\be
\label{q2}
p_1^2 = m_{\ell_1}^2, \quad \quad
p_2^2 = m_{\ell_2}^2, \quad \quad
2\, p_1 \cdot p_2 = m_{\ell_1}^2 + m_{\ell_2}^2 - q^2.
\ee
If the outgoing photon is physical,
then $q^2 = 0$;
but we keep $q^2 \neq 0$ for generality.
The amplitude for a photon with polarization $\sigma$ is
\be
T^\sigma = \mathcal{S}\, e\,
\bar u_{\ell_2} \left( p_2 \right) \left[
  \gamma^\sigma \left( a_l P_L + a_r P_R \right)
  + p_1^\sigma \left( b_l P_L + b_r P_R \right)
  + p_2^\sigma \left( c_l P_L + c_r P_R \right)
  \right] u_{\ell_1} \left( p_1 \right),
\label{ugiot}
\ee
where $\mathcal{S}$ has been defined in Eq.~\eqref{S}
and $e$ is the electric charge of the proton.
Clearly,
\be
\label{judio2}
\not \! p_1\, u_{\ell_1} \left( p_1 \right)
= m_{\ell_1}\, u_{\ell_1} \left( p_1 \right),
\quad \quad
\overline u_{\ell_2} \left( p_2 \right) \! \not \! p_2
= m_{\ell_2}\, \overline u_{\ell_2} \left( p_2 \right).
\ee
If $T^\sigma$ in Eq.~\eqref{ugiot}
is multiplied by $q_\sigma$
and then Eqs.~\eqref{q2} and~\eqref{judio2} are utilized,
one must obtain zero because of gauge invariance.
Thus,
\bs
\label{gi2}
\ba
2 m_{\ell_1} a_r - 2 m_{\ell_2} a_l
+ \left( m_{\ell_1}^2 - m_{\ell_2}^2 \right) \left( b_l + c_l \right)
+ q^2 \left( b_l - c_l \right) &=& 0,
\\*[1mm]
2 m_{\ell_1} a_l - 2 m_{\ell_2} a_r
+ \left( m_{\ell_1}^2 - m_{\ell_2}^2 \right) \left( b_r + c_r \right)
+ q^2 \left( b_r - c_r \right) &=& 0.
\ea
\es
We have used Eqs.~\eqref{gi2}---that hold
even when $q^2 \neq 0$---as a check on our calculations.

The decay width is,
in the rest frame of the decaying $\ell_1^-$,\footnote{Instead
  of Eq.~\eqref{ibvuop} there is another way to express the decay width,
  \textit{viz.}
  \be
  \Gamma=\frac{\left(m_{\ell_1}^2-m_{\ell_2}^2\right)^3}{64 \pi m_{\ell_1}^3}
  \left| \mathcal{S}\, e \right|^2 \left( \left| b_l + c_l \right|^2
  + \left| b_r + c_r \right|^2 \right).
  \ee
  This agrees with Eq.~(7) of Ref.~\cite{Lavoura:2003xp},
  that has a factor $e^2$ missing, though.}
\be
\Gamma = \frac{m_{\ell_1}^2 - m_{\ell_2}^2}{16 \pi m_{\ell_1}^3}\,
\left| \mathcal{S}\, e \right|^2
\left[ \left( m_{\ell_1}^2 + m_{\ell_2}^2 \right)
  \left( \left| a_l \right|^2 + \left| a_r \right|^2 \right)
  - 4\, m_{\ell_1} m_{\ell_2}\, \mathrm{Re} \left( a_l^\ast a_r \right) \right].
\label{ibvuop}
\ee

In our model each of the coefficients $a_l, \ldots, c_r$
is the sum of two contributions,
\textit{viz.}\
\be
a_l = a_{l,H} + a_{l,W},\quad \hdots,\quad c_r = c_{r,H} + c_{r,W}.
\ee
The contributions with sub-index $H$
arise from the diagrams in Fig.~\ref{fig1}
and are given in Eqs.~\eqref{coef1} below,
and the contributions with sub-index $W$
come from the diagrams in Fig.~\ref{fig2}
and are given in Eqs.~\eqref{jguif} below.
Notice that Fig.~\ref{fig1}
includes diagrams with the charged Goldstone bosons
$G^\pm \equiv H_1^\pm$.

In all our calculations we utilize Feynman's gauge.
Let $m_a$ denote the mass of $H_a^\pm$;
for $a=1$ one must use $m_{a=1} = m_W$ because we are in Feynman's gauge.

\subsection{$H_a^\pm$}

The charged scalars $H_a^\pm$
couple to the charged leptons and the neutrinos
according to Eq.~\eqref{mvjckp}.
The charged scalars include as a particular case
the charged Goldstone bosons.
For $G^\pm = H_1^\pm$,
one has~\cite{Grimus:2002ux}
\be
\label{r1l1}
R_{1 \ell i} = \frac{e}{\sqrt{2} s_w m_W}\, U_{\ell i} m_i,
\quad \quad
L_{1 \ell i} = \frac{e}{\sqrt{2} s_w m_W}\, U_{\ell i} m_\ell,
\ee
where $U$ is the lepton mixing matrix
and $s_w$ is the sine of the weak mixing angle.

\begin{figure}[h]
\begin{center}
\includegraphics[width=0.75\textwidth]{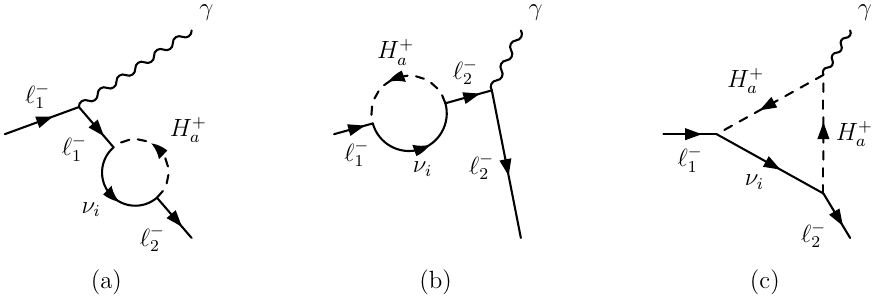}
\end{center}
\caption{The three diagrams for ${\ell_1}^- \to {\ell_2}^- \gamma$
  with a loop containing $H_a^\pm$.
  \label{fig1}}
\end{figure}

The diagrams in Figs.~\ref{fig1}(a) and~\ref{fig1}(b) produce
\ba
T_\mathrm{13(a)}^\sigma + T_{13(b)}^\sigma &=&
\mathcal{S}\, e\,
\sum_{i=1}^6 \sum_{a=1}^n
\left[ \bar u_{\ell_2}
  \left( p_2 \right) \gamma^\sigma
  \left( x_{ia} P_R + y_{ia} P_L \right) u_{\ell_1} \left( p_1 \right)
  m_{\ell_2} \mathbf{b}_0
  \right. \no & &
  + \bar u_{\ell_2} \left( p_2 \right) \gamma^\sigma
  \left( x_{ia} P_L + y_{ia} P_R \right) u_{\ell_1} \left( p_1 \right)
  m_{\ell_1} \mathbf{b}_0
  \no & &
  + \bar u_{\ell_2} \left( p_2 \right) \gamma^\sigma
  \left( z_{ia} P_L + w_{ia} P_R \right) u_{\ell_1} \left( p_1 \right) \mathbf{b}_1
  \no & & \left.
  + \bar u_{\ell_2} \left( p_2 \right) \gamma^\sigma
  \left( z_{ia} P_R + w_{ia} P_L \right) u_{\ell_1} \left( p_1 \right)
  m_{\ell_2} m_{\ell_1} \mathbf{b}_2
  \right],
\label{ncidosp}
\ea
where
\bs
\label{xyzw}
\ba
x_{ia} &:=& R_{a {\ell_2} i} L^\ast_{a {\ell_1} i} m_i, \\
y_{ia} &:=& L_{a {\ell_2} i} R^\ast_{a {\ell_1} i} m_i, \\
z_{ia} &:=& R_{a {\ell_2} i} R^\ast_{a {\ell_1} i}, \\
w_{ia} &:=& L_{a {\ell_2} i} L^\ast_{a {\ell_1} i},
\ea
\es
and
\bs
\label{mathbfb}
\ba
\mathbf{b}_0 &:=& \frac{B_0 \left( m_{\ell_1}^2, m_i^2, m_a^2 \right)
- B_0 \left( m_{\ell_2}^2, m_i^2, m_a^2 \right)}{m_{\ell_1}^2 - m_{\ell_2}^2},
\\
\mathbf{b}_1 &:=&
\frac{m_{\ell_1}^2\, B_1 \left( m_{\ell_1}^2, m_i^2, m_a^2 \right)
  - m_{\ell_2}^2 B_1 \left( m_{\ell_2}^2, m_i^2, m_a^2 \right)}
     {m_{\ell_1}^2 - m_{\ell_2}^2},
\\
\mathbf{b}_2 &:=& \frac{B_1 \left( m_{\ell_1}^2, m_i^2, m_a^2 \right)
  - B_1 \left( m_{\ell_2}^2, m_i^2, m_a^2 \right)}{m_{\ell_1}^2 - m_{\ell_2}^2}.
\ea
\es
Notice that in our model
\bs
\label{jskapc}
\ba
\sum_{i=1}^6 z_{ia} = \sum_{i=1}^6 R_{a \ell_2 i} R_{a \ell_1 i}^\ast &=&
\sum_{k, k^\prime = 1}^{n_d} \mathcal{U}_{ka}^\ast \mathcal{U}_{k^\prime a}
\left( \Delta_k^\dagger \Delta_{k^\prime} \right)_{\ell_2 \ell_1},
\\
\sum_{i=1}^6 w_{ia} = \sum_{i=1}^6 L_{a \ell_2 i} L_{a \ell_1 i}^\ast &=&
\sum_{k, k^\prime = 1}^{n_d} \mathcal{U}_{ka}^\ast \mathcal{U}_{k^\prime a}
\left( \Gamma_k \Gamma^\dagger_{k^\prime} \right)_{\ell_2 \ell_1},
\\
\sum_{i=1}^6 x_{ia} = \sum_{i=1}^6 R_{a \ell_2 i} L_{a \ell_1 i}^\ast m_i &=&
\sum_{k, k^\prime = 1}^{n_d} \mathcal{U}_{ka}^\ast \mathcal{U}_{k^\prime a}
\left( \Delta_k^\dagger M_D \Gamma_{k^\prime}^\dagger
\right)_{\ell_2 \ell_1},
\\
\sum_{i=1}^6 y_{ia} = \sum_{i=1}^6 L_{a \ell_2 i} R_{a \ell_1 i}^\ast m_i &=&
\sum_{k, k^\prime = 1}^{n_d} \mathcal{U}_{ka}^\ast \mathcal{U}_{k^\prime a}
\left( \Gamma_k M_D^\dagger \Delta_{k^\prime} \right)_{\ell_2 \ell_1}
\ea
\es
vanish when $\ell_2 \neq \ell_1$ by virtue of the matrices $\Gamma_k$,
$\Delta_k$,
and $M_D$ being diagonal.

The diagram~\ref{fig1}(c) produces
\ba
T_\mathrm{13(c)}^\sigma &=&
\mathcal{S}\, e\, \sum_{i=1}^6 \sum_{a=1}^n
\bar u_{\ell_2} \left( p_2 \right)
\left\{ \left[
  \left( e_0 + 2 e_1 \right) p_1^\sigma
  + \left( e_0 + 2 e_2 \right) p_2^\sigma
  \right] \left( x_{ia} P_R + y_{ia} P_L \right)
\right. \no & &
+ \left[ \left( e_2 + 2 e_{22} \right) p_2^\sigma
  + \left( e_2 + 2 e_{12} \right) p_1^\sigma
  \right] m_{\ell_2} \left( z_{ia} P_L + w_{ia} P_R \right)
\no & &
+ \left[ \left( e_1 + 2 e_{11} \right) p_1^\sigma
  + \left( e_1 + 2 e_{12} \right) p_2^\sigma
  \right] m_{\ell_1} \left( z_{ia} P_R + w_{ia} P_L \right)
\no & & \left.
+ 2 e_{00}\, \gamma^\sigma \left( z_{ia} P_L + w_{ia} P_R \right)
\right\} u_{\ell_1} \left( p_1 \right),
\label{bvjicofp}
\ea
where the $e$ functions have been defined in Eqs.~\eqref{efunctions}.

Thus,
adding Eqs.~\eqref{ncidosp} and~\eqref{bvjicofp},
one obtains
\bs
\label{coef1}
\ba
a_{l,H} &=&
\sum_{i=1}^6 \sum_{a=1}^n \left[
  \mathbf{b}_0 m_{\ell_1} x_{ia}
  + \mathbf{b}_0 m_{\ell_2} y_{ia}
  + \left( 2 e_{00}+ \mathbf{b}_1 \right) z_{ia}
  + \mathbf{b}_2 m_{\ell_2} m_{\ell_1} w_{ia}
  \right], \hspace*{7mm}
\label{al1} \\
a_{r,H} &=&
\sum_{i=1}^6 \sum_{a=1}^n \left[
  \mathbf{b}_0 m_{\ell_2} x_{ia}
  + \mathbf{b}_0 m_{\ell_1} y_{ia}
  + \mathbf{b}_2 m_{\ell_2} m_{\ell_1} z_{ia}
  + \left( 2 e_{00}+ \mathbf{b}_1 \right) w_{ia}
  \right],
\label{ar1} \\
b_{l,H} &=& \sum_{i=1}^6 \sum_{a=1}^n \left[
  \left( e_0 + 2 e_1 \right) y_{ia}
  + \left( e_2 + 2 e_{12} \right) m_{\ell_2} z_{ia}
  + \left( e_1 + 2 e_{11} \right) m_{\ell_1} w_{ia}
  \right],
\\
b_{r,H} &=& \sum_{i=1}^6 \sum_{a=1}^n \left[
  \left( e_0 + 2 e_1 \right) x_{ia}
  + \left( e_1 + 2 e_{11} \right) m_{\ell_1} z_{ia}
  + \left( e_2 + 2 e_{12}\right) m_{\ell_2} w_{ia}
  \right],
\\
c_{l,H} &=& \sum_{i=1}^6 \sum_{a=1}^n \left[
  \left( e_0 + 2 e_2 \right) y_{ia}
  + \left( e_2 + 2 e_{22} \right) m_{\ell_2} z_{ia}
  + \left( e_1 + 2 e_{12} \right) m_{\ell_1} w_{ia}
  \right],
\\
c_{r,H} &=& \sum_{i=1}^6 \sum_{a=1}^n \left[
  \left( e_0 + 2 e_2 \right) x_{ia}
  + \left( e_1 + 2 e_{12} \right) m_{\ell_1} z_{ia}
  + \left( e_2 + 2\, e_{22} \right) m_{\ell_2} w_{ia}
  \right].
\ea
\es

One may use
\bs
\label{b0b1}
\ba
B_0 \left( m_{\ell_1}^2, m_i^2, m_a^2 \right) &=&
d\, e_{00} - m_a^2\, e_0
+ m_{\ell_1}^2 \left( e_1 + e_{11} + e_{12} \right)
\no & &
+ m_{\ell_2}^2 \left( e_0 + e_1 + 2\, e_2 + e_{22} + e_{12} \right)
- q^2 \left( e_1 + e_{12} \right),
\\
B_0 \left( m_{\ell_2}^2, m_i^2, m_a^2 \right) &=&
d\, e_{00} - m_a^2\, e_0
+ m_{\ell_1}^2 \left( e_0 + 2\, e_1 + e_2 + e_{11} + e_{12} \right)
\no & &
+ m_{\ell_2}^2 \left( e_2 + e_{22} + e_{12} \right)
- q^2 \left( e_2 + e_{12} \right),
\\
B_1 \left( m_{\ell_1}^2, m_i^2, m_a^2 \right) &=&
\left( m_{\ell_2}^2 - m_{\ell_1}^2 \right) \left( e_1 + e_{11} + e_{12} \right)
- 2 e_{00} + q^2 \left( e_{12} - e_{11} \right),
\\
B_1 \left( m_{\ell_2}^2, m_i^2, m_a^2 \right) &=&
\left( m_{\ell_1}^2 - m_{\ell_2}^2 \right) \left( e_2 + e_{22} + e_{12} \right)
- 2 e_{00} + q^2 \left( e_{12} - e_{22} \right),
\ea
\es
where $d$ is the dimension of space--time,
to derive
\bs
\label{jispska}
\ba
\mathbf{b}_0 &=&
- e_0 - e_1 - e_2
+ \frac{q^2}{m_{\ell_1}^2 - m_{\ell_2}^2} \left( e_2 - e_1 \right),
\label{sjdisps}
\\
2 e_{00} + \mathbf{b}_1 &=&
- m_{\ell_1}^2 \left( e_1 + e_{11} \right)
- m_{\ell_2}^2 \left( e_2 + e_{22} \right)
+ \left( q^2 - m_{\ell_1}^2 - m_{\ell_2}^2 \right) e_{12}
\no & &
+ \frac{q^2}{m_{\ell_1}^2 - m_{\ell_2}^2}
\left( m_{\ell_2}^2\, e_{22} - m_{\ell_1}^2\, e_{11} \right),
\label{int_b1}
\\
\mathbf{b}_2 &=& - e_1 - e_2 - e_{11} - e_{22} - 2 e_{12}
+ \frac{q^2}{m_{\ell_1}^2 - m_{\ell_2}^2} \left( e_{22} - e_{11} \right).
\hspace*{7mm}
\ea
\es
As a consequence of Eqs.~\eqref{jispska},
the constraints~\eqref{gi2} hold for the contributions with sub-index $H$.

\subsection{$W^\pm$}

\begin{figure}[h]
\begin{center}
\includegraphics[width=0.75\textwidth]{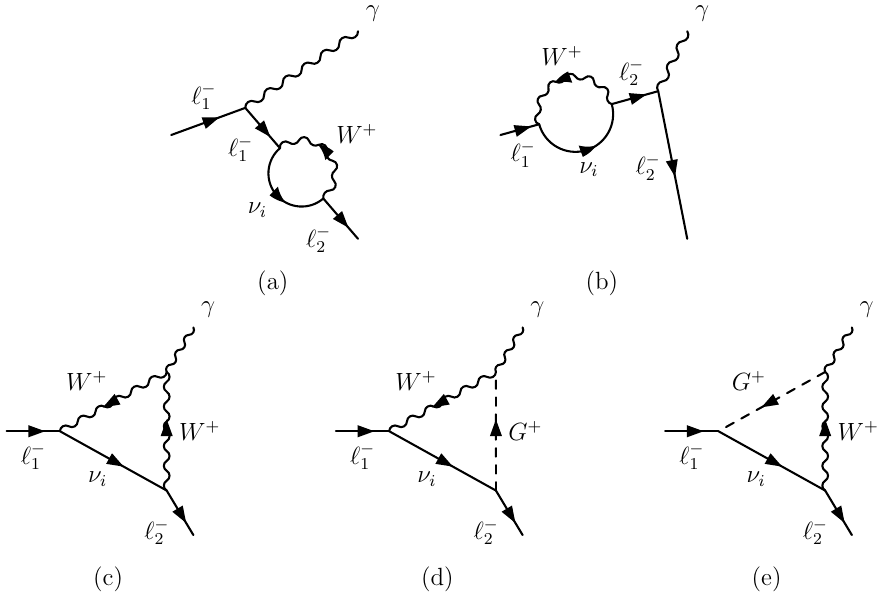}
\end{center}
\caption{The five diagrams for ${\ell_1}^- \to {\ell_2}^- \gamma$
  with a loop containing $W^\pm$.
  \label{fig2}}
\end{figure}

Besides the diagrams exclusively with $G^\pm$,
there are five diagrams with $W^\pm$,
\textit{cf.}~Fig.~\ref{fig2}.
Figures~\ref{fig2}(d) and~\ref{fig2}(e)
have the outgoing photon attaching to $W^\pm G^\mp$.
Those diagrams produce
\bs
\label{GW}
\ba
a_{l,de} &=& - \frac{e^2}{2 s_w^2}\,
\sum_{i=1}^6 U_{{\ell_2} i} U_{{\ell_1} i}^\ast
\left( 2 m_i^2 f_0 + m_{\ell_1}^2 f_1 + m_{\ell_2}^2 f_2 \right),
\\
a_{r,de} &=& \frac{e^2 m_{\ell_2} m_{\ell_1}}{2 s_w^2}\,
\sum_{i=1}^6 U_{{\ell_2} i} U_{{\ell_1} i}^\ast \left( f_1 + f_2 \right),
\\
b_{l,de} &=& - \frac{e^2 m_{\ell_2}}{s_w^2}\,
\sum_{i=1}^6 U_{{\ell_2} i} U_{{\ell_1} i}^\ast f_1,
\\
b_{r,de} &=& 0,
\\
c_{l,de} &=& 0,
\\
c_{r,de} &=& - \frac{e^2 m_{\ell_1}}{s_w^2}\,
\sum_{i=1}^6 U_{{\ell_2} i} U_{{\ell_1} i}^\ast f_2,
\ea
\es
where the $f$ functions have been defined in Eqs.~\eqref{ffunctions}.
The diagram of Fig.~\ref{fig2}(c) produces
\bs
\label{WW}
\ba
a_{l,c} &=& \frac{e^2}{2 s_w^2}\,
\sum_{i=1}^6 U_{{\ell_2} i} U_{{\ell_1} i}^\ast \left[
  \left( 3 m_{\ell_1}^2 + 2 m_{\ell_2}^2 - 2 q^2 \right) f_1
  + \left( 2 m_{\ell_1}^2 + 3 m_{\ell_2}^2 - 2 q^2 \right) f_2
  \right. \no & & \left.
  + 2 m_{\ell_1}^2 f_{11} + 2 m_{\ell_2}^2 f_{22}
  + 2 \left( m_{\ell_1}^2 + m_{\ell_2}^2 - q^2 \right) f_{12}
  + 12 f_{00} \right],
\hspace*{8mm} \label{sdhjsoa} \\
a_{r,c} &=& \frac{3 e^2 m_{\ell_2} m_{\ell_1}}{2 s_w^2}\,
\sum_{i=1}^6 U_{{\ell_2} i} U_{{\ell_1} i}^\ast \left( f_1 + f_2 \right),
\\
b_{l,c} &=& - \frac{e^2 m_{\ell_2}}{s_w^2}\,
\sum_{i=1}^6 U_{{\ell_2} i} U_{{\ell_1} i}^\ast \left( f_1 + f_2 - 2 f_{12} \right),
\\
b_{r,c} &=& \frac{e^2 m_{\ell_1}}{s_w^2}\,
\sum_{i=1}^6 U_{{\ell_2} i} U_{{\ell_1} i}^\ast \left( f_1 + 2 f_{11} \right),
\\
c_{l,c} &=& \frac{e^2 m_{\ell_2}}{s_w^2}\,
\sum_{i=1}^6 U_{{\ell_2} i} U_{{\ell_1} i}^\ast \left( f_2 + 2 f_{22} \right),
\\
c_{r,c} &=& - \frac{e^2 m_{\ell_1}}{s_w^2}\,
\sum_{i=1}^6 U_{{\ell_2} i} U_{{\ell_1} i}^\ast \left( f_1 + f_2 - 2 f_{12} \right).
\ea
\es
A crucial property of the lepton mixing matrix $U$ in our model is
\be
\left( U U^\dagger \right)_{\ell_2 \ell_1}
= \sum_{i=1}^6 U_{{\ell_2} i}\, U_{{\ell_1} i}^\ast = 0,
\label{judio}
\ee
\textit{cf.}\ Eq.~\eqref{vklspac1}.
In spite of $f_{00}$ containing a divergence
$1 \! \left/ \left( 2 \epsilon \right) \right.$,
$a_{l,c}$ in Eq.~\eqref{sdhjsoa}
is finite because of Eq.~\eqref{judio}.

Finally,
there are the diagrams of Figs.~\ref{fig2}(a) and~\ref{fig2}(b),
producing
\bs
\label{alrab}
\ba
a_{l,ab} &=& \frac{e^2}{s_w^2}\,
\sum_{i=1}^6 U_{{\ell_2} i} U_{{\ell_1} i}^\ast \mathbf{b}_4,
\label{al2ab}
\\
a_{r,ab} &=& \frac{e^2 m_{\ell_2} m_{\ell_1}}{s_w^2}\,
\sum_{i=1}^6 U_{{\ell_2} i} U_{{\ell_1} i}^\ast \mathbf{b}_5,
\label{ar2ab}
\\
b_{l,ab} = b_{r,ab} = c_{l,ab} = c_{r,ab} &=& 0,
\ea
\es
where
\bs
\label{cnvhdusi}
\ba
\mathbf{b}_4 &:=& \frac{m_{\ell_1}^2\,
  B_1 \left( m_{\ell_1}^2, m_i^2, m_W^2 \right)
  - m_{\ell_2}^2
  B_1 \left( m_{\ell_2}^2, m_i^2, m_W^2 \right)}{m_{\ell_1}^2 - m_{\ell_2}^2}
\\ &=&
- 2 f_{00}
- m_{\ell_1}^2 \left( f_1 + f_{11} + f_{12} \right)
- m_{\ell_2}^2 \left( f_2 + f_{22} + f_{12} \right)
\no & &
+ q^2\, f_{12}
+ \frac{q^2}{m_{\ell_1}^2 - m_{\ell_2}^2} \left( m_{\ell_2}^2 f_{22}
- m_{\ell_1}^2 f_{11} \right),
\label{vmjfodsp}
\\
\mathbf{b}_5 &:=& \frac{B_1 \left( m_{\ell_1}^2, m_i^2, m_W^2 \right)
  - B_1 \left( m_{\ell_2}^2, m_i^2, m_W^2 \right)}{m_{\ell_1}^2 - m_{\ell_2}^2}
\\ &=&
- f_1 - f_2 - f_{11} - f_{22} - 2 f_{12}
+ \frac{q^2}{m_{\ell_1}^2 - m_{\ell_2}^2} \left( f_{22} - f_{11} \right).
\label{mclapeo}
\ea
\es

Thus,
the sum total of the diagrams of Fig.~\ref{fig2} is
\bs
\label{jguif}
\ba
a_{l,W} &=& \frac{e^2}{s_w^2}\,
  \sum_{i=1}^6 U_{{\ell_2} i} U_{{\ell_1} i}^\ast \left[ - m_i^2 f_0
  + m_{\ell_2}^2 f_1 + m_{\ell_1}^2 f_2 + 4 f_{00}
  \right. \no & & \left.
  - q^2 \left( f_1 + f_2 \right)
  + \frac{q^2}{m_{\ell_1}^2 - m_{\ell_2}^2} \left( m_{\ell_2}^2 f_{22}
  - m_{\ell_1}^2 f_{11} \right)
  \right],
\\
a_{r,W} &=& \frac{e^2 m_{\ell_2} m_{\ell_1}}{s_w^2}\,
\sum_{i=1}^6 U_{{\ell_2} i} U_{{\ell_1} i}^\ast
\left[ f_1 + f_2 - f_{11} - f_{22} - 2 f_{12}
  \right. \no & & \left.
  + \frac{q^2}{m_{\ell_1}^2 - m_{\ell_2}^2} \left( f_{22} - f_{11} \right)
  \right],
\\
b_{l,W} &=& \frac{e^2 m_{\ell_2}}{s_w^2}\,
\sum_{i=1}^6 U_{{\ell_2} i} U_{{\ell_1} i}^\ast
\left( 2 f_{12} - 2 f_1 - f_2 \right),
\\
b_{r,W} &=& \frac{e^2 m_{\ell_1}}{s_w^2}\,
\sum_{i=1}^6 U_{{\ell_2} i} U_{{\ell_1} i}^\ast
\left( f_1 + 2 f_{11}\right),
\\
c_{l,W} &=& \frac{e^2 m_{\ell_2}}{s_w^2}\,
\sum_{i=1}^6 U_{{\ell_2} i} U_{{\ell_1} i}^\ast
\left( f_2 + 2 f_{22}\right),
\\
c_{r,W} &=& \frac{e^2 m_{\ell_1}}{s_w^2}\,
\sum_{i=1}^6 U_{{\ell_2} i} U_{{\ell_1} i}^\ast
\left( 2 f_{12} - f_1 - 2 f_2 \right).
\ea
\es

\newpage
\setcounter{equation}{0}
\renewcommand{\theequation}{C\arabic{equation}}
\section{$Z \to \ell_1^+ \ell_2^-$}
\label{Sect_taumuZ}

We compute the process $Z \left( q \right)
\to \ell_1^+ \left( p_1 \right) \ell_2^- \left( - p_2 \right)$,
where $q^2 = m_Z^2$ and Eqs.~\eqref{q2} hold.
The amplitude for a $Z$ with polarization $\sigma$ is written
\be
T^\sigma = \mathcal{S}\, e\,
\bar u_{\ell_2} \left( - p_2 \right) \left[
  \gamma^\sigma \left( \bar{a}_l P_L + \bar{a}_r P_R \right)
  + p_1^\sigma \left( \bar{b}_l P_L + \bar{b}_r P_R \right)
  + p_2^\sigma \left( \bar{c}_l P_L + \bar{c}_r P_R \right)
  \right] v_{\ell_1} \left( p_1 \right).
\ee
The decay width in the rest frame of the decaying $Z$ is
\be
\label{ZdecayWidth}
\Gamma = \frac{\sqrt{\lambda}}{16 \pi m_Z^3}\,
\left| \mathcal{S}\, e \right|^2
\left( \frac{\lambda \aleph_0}{12 m_Z^2}
+ \aleph_1 + \frac{\aleph_2}{3 m_Z^2} \right),
\ee
where
\be
\lambda := m_Z^4 + m_{\ell_1}^4 + m_{\ell_2}^4
- 2 \left( m_Z^2 m_{\ell_1}^2 + m_Z^2 m_{\ell_2}^2
+ m_{\ell_1}^2 m_{\ell_2}^2 \right)
\ee
and
\bs
\label{jfuisi}
\ba
\aleph_0
&=&
\left( m_Z^2 - m_{\ell_1}^2 - m_{\ell_2}^2 \right)
\left( \left| \bar{b}_l + \bar{c}_l \right|^2
+ \left| \bar{b}_r + \bar{c}_r \right|^2 \right)
- 4\, m_{\ell_1} m_{\ell_2}\,
\mathrm{Re} \left[ \left( \bar{b}_l + \bar{c}_l \right)
  \left( \bar{b}_r^\ast + \bar{c}_r^\ast \right) \right]
\no & &
- 4\, m_{\ell_1}\, \mathrm{Re} \left[
  \bar{a}_r^\ast \left( \bar{b}_l + \bar{c}_l \right)
  + \bar{a}_l^\ast \left( \bar{b}_r + \bar{c}_r \right)
  \right]
\no & &
- 4\, m_{\ell_2}\, \mathrm{Re} \left[
  \bar{a}_l^\ast \left( \bar{b}_l + \bar{c}_l \right)
  + \bar{a}_r^\ast \left( \bar{b}_r + \bar{c}_r \right)
  \right],
\\
\aleph_1 &=& 4\, m_{\ell_1} m_{\ell_2}\,
\mathrm{Re} \left( \bar{a}_l \bar{a}_r^\ast \right),
\\
\aleph_2 &=&
\left[ 2 m_Z^4
- m_Z^2 \left( m_{\ell_1}^2 + m_{\ell_2}^2 \right)
- \left( m_{\ell_1}^2 - m_{\ell_2}^2 \right)^2 \right]
\left( \left| \bar{a}_l \right|^2
+ \left| \bar{a}_r \right|^2 \right).
\ea
\es

We define
\be
t_l := \frac{s_w^2 - c_w^2}{2 c_w s_w}, \quad \quad t_r := \frac{s_w}{c_w},
\label{n3}
\ee
so that the coupling of the $Z$ to the charged leptons is given by
\be
\mathcal{L}_\mathrm{nc} = \cdots +
e Z_\sigma \sum_{\ell = e, \mu, \tau} \bar{\ell}\, \gamma^\sigma
\left( t_l P_L + t_r P_R \right) \ell,
\ee
\textit{cf.}\ Eq.~\eqref{n2}.
Notice that
\be
- \frac{1}{t_r} = 2 t_l - t_r.
\label{jf9t0ot}
\ee

We shall write the coefficients $\bar{a}_l, \ldots, \bar{c}_r$
as the sum of three pieces,
\textit{viz.}\
\ba
\bar{a}_l &=& \bar{a}_{l,H} + \bar{a}_{l,W} + \bar{a}_{l,2\nu}, \no
 &\vdots& \\
\bar{c}_r &=& \bar{c}_{r,H} + \bar{c}_{r,W} + \bar{c}_{r,2\nu}. \nonumber
\ea

\subsection{$H_a^\pm$}

We recover the diagrams of Fig.~\ref{fig1},
with the photon substituted by a $Z$.
Diagrams~\ref{fig1}(a) and~\ref{fig1}(b)
produce the result in Eq.~\eqref{ncidosp}
with the transformations $P_L \to t_l P_L$ and $P_R \to t_r P_R$.
Diagram~\ref{fig1}(c) produces the result
in Eq.~\eqref{bvjicofp} multiplied by $t_l$.
Thus,
the full result of Fig.~\ref{fig1} with $Z$ instead of $\gamma$ is
\bs
\label{coefZ1}
\ba
\bar{a}_{l,H} &=& t_l\, \sum_{i=1}^6 \sum_{a=1}^n \left[
\left( m_{\ell_1} x_{ia} + m_{\ell_2} y_{ia} \right) \mathbf{b}_0
+ z_{ia} \left( 2 e_{00} + \mathbf{b}_1 \right)
+ m_{\ell_2} m_{\ell_1} w_{ia}\, \mathbf{b}_2
\right],
\label{al1Z} \\
\bar{a}_{r,H} &=&  \sum_{i=1}^6 \sum_{a=1}^n \left\{ t_r \left[
  \left( m_{\ell_2} x_{ia} + m_{\ell_1} y_{ia} \right) \mathbf{b}_0
  + m_{\ell_2} m_{\ell_1} z_{ia}\, \mathbf{b}_2
  + w_{ia}\, \mathbf{b}_1 \right]
+ 2 t_l w_{ia} e_{00} \right\},
\label{ar1Z} \\
\bar{b}_{l,H} &=& t_l\, \sum_{i=1}^6 \sum_{a=1}^n
\left[ y_{ia} \left( e_0 + 2 e_1 \right)
+ m_{\ell_2} z_{ia} \left( e_2 + 2 e_{12} \right)
+ m_{\ell_1} w_{ia} \left( e_1 + 2 e_{11}\right) \right],
\\
\bar{b}_{r,H} &=& t_l\, \sum_{i=1}^6 \sum_{a=1}^n
\left[ x_{ia} \left( e_0 + 2 e_1 \right)
+ m_{\ell_1} z_{ia} \left( e_1 + 2 e_{11} \right)
+ m_{\ell_2} w_{ia} \left( e_2 + 2 e_{12} \right) \right],
\\
\bar{c}_{l,H} &=& t_l\, \sum_{i=1}^6 \sum_{a=1}^n
\left[ y_{ia} \left( e_0 + 2 e_2 \right)
+ m_{\ell_2} z_{ia} \left( e_2 + 2 e_{22} \right)
+ m_{\ell_1} w_{ia} \left( e_1 + 2 e_{12} \right) \right],
\\
\bar{c}_{r,H} &=& t_l\, \sum_{i=1}^6 \sum_{a=1}^n
\left[ x_{ia} \left( e_0 + 2 e_2 \right)
+ m_{\ell_1} z_{ia} \left( e_1 + 2 e_{12} \right)
+ m_{\ell_2} w_{ia} \left( e_2 + 2 e_{22} \right) \right].
\ea
\es

\subsection{$W^\pm$}

We consider the diagrams of Fig.~\ref{fig2}
with a $Z$ instead of the $\gamma$.
They produce
\ba
\bar{a}_{l,W} &=& \bar{a}_{l,de} + \bar{a}_{l,c} + \bar{a}_{l,ab}, \no
 &\vdots& \label{jdospsao} \\
\bar{c}_{r,W} &=& \bar{c}_{r,de} + \bar{c}_{r,c} + \bar{c}_{r,ab}. \nonumber
\ea
Figures~\ref{fig2}(d) and~\ref{fig2}(e) yield
\bs
\label{nvjdxpdo}
\ba
\bar{a}_{l,de} &=& t_r\, a_{l,de},
\\
\bar{a}_{r,de} &=& t_r\, a_{r,de},
\\
\bar{b}_{l,de} &=& t_r\, b_{l,de},
\\
\bar{b}_{r,de} &=& 0,
\\
\bar{c}_{l,de} &=& 0,
\\
\bar{c}_{r,de} &=& t_r\, c_{r,de}.
\ea
\es
Figure~\ref{fig2}(c) produces
\bs
\label{njvckxp}
\ba
\bar{a}_{l,c} &=& - \frac{1}{t_r}\, a_{l,c},
\label{djvcifo} \\
\bar{a}_{r,c} &=& - \frac{1}{t_r}\, a_{r,c},
\\
\bar{b}_{l,c} &=& - \frac{1}{t_r}\, c_{l,c},
\\
\bar{b}_{r,c} &=& - \frac{1}{t_r}\, b_{r,c},
\\
\bar{c}_{l,c} &=& - \frac{1}{t_r}\, c_{l,c},
\\
\bar{c}_{r,c} &=& - \frac{1}{t_r}\, c_{r,c}.
\ea
\es
Notice that in Eqs.~\eqref{njvckxp} one may use Eq.~\eqref{jf9t0ot}.
Figures~\ref{fig2}(a) and~\ref{fig2}(b) give
\bs
\label{mbkogp}
\ba
\bar{a}_{l,ab} &=& t_l\, a_{l,ab},
\\
\bar{a}_{r,ab} &=& t_r\, a_{r,ab},
\\
\bar{b}_{l,ab} = \bar{b}_{r,ab} = \bar{c}_{l,ab} = \bar{c}_{r,ab} &=& 0.
\ea
\es

\subsection{Diagrams with two neutrino internal lines}

\begin{figure}[h]
\begin{center}
\includegraphics[width=0.5\textwidth]{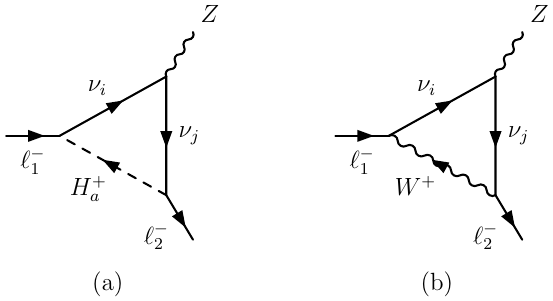}
\end{center}
\caption{Two diagrams for $Z \to {\ell_1}^+ {\ell_2}^-$
  where the $Z$ attaches to two neutrino lines.
  \label{fig3}}
\end{figure}

There are also diagrams where the $Z$ boson attaches to the neutrino line
as depicted in Fig.~\ref{fig3}.
The relevant vertex is given in Eq.~\eqref{Z}.
We have
\ba
\bar{a}_{l,2\nu} &=&
\bar{a}_{l,\mathrm{15(a)}} + \bar{a}_{l,\mathrm{15(b)}}, \no
 &\vdots& \\
\bar{c}_{r,2\nu} &=&
\bar{c}_{r,\mathrm{15(a)}} + \bar{c}_{r,\mathrm{15(b)}}. \nonumber
\ea

From the diagram~\ref{fig3}(a) one obtains
\bs
\label{coefZ2}
\ba
\bar{a}_{l,\mathrm{15(a)}} &=& \frac{1}{2 c_w s_w}
\sum_{a=1}^n \sum_{i,j=1}^6
\left\{
m_{\ell_1} x_{ija} \left[ m_i q_{ij} g_1
  + m_j q_{ji} \left( g_0 + g_1 \right) \right]
\right. \no & &
+ m_{\ell_2} y_{ija} \left[ m_j q_{ij} g_2
  + m_i q_{ji} \left( g_0 + g_2 \right) \right]
\no & &
+ z_{ija} q_{ij} \left[ - 2 g_{00}
  + q^2 g_{12} 
  - m^2_{\ell_2} \left( g_2 + g_{12} + g_{22} \right)
  - m^2_{\ell_1} \left( g_1 + g_{12} + g_{11} \right) \right]
\no & & \left.
- z_{ija} q_{ji} m_i m_j g_0
-  w_{ija} q_{ji} m_{\ell_2} m_{\ell_1} \left( g_0 + g_1 + g_2 \right)
\right\},
\label{T4Zal2} \\
\bar{a}_{r,\mathrm{15(a)}} &=& \frac{1}{2 c_w s_w}
\sum_{a=1}^n \sum_{i,j=1}^6 \left\{
- m_{\ell_2} x_{ija} \left[ m_i q_{ij} \left( g_0 + g_2 \right)
  + m_j q_{ji} g_2 \right]
+ m_i m_j w_{ija} q_{ij} g_0
\right. \no & &
- m_{\ell_1} y_{ija} \left[ m_j q_{ij} \left( g_0 + g_1 \right)
  + m_i q_{ji} g_1 \right]
+ m_{\ell_2} m_{\ell_1} z_{ija} q_{ij} \left( g_0 + g_1 + g_2 \right)
\no & & \left.
+ w_{ija} q_{ji} \left[ 2 g_{00} - q^2 g_{12}
  + m_{\ell_2}^2 \left( g_2 + g_{12} + g_{22} \right)
  + m_{\ell_1}^2 \left( g_1 + g_{12} + g_{11} \right) \right]
\right\}, \hspace*{7mm}
\label{T4Zar2} \\
\bar{b}_{l,\mathrm{15(a)}} &=& \frac{1}{c_w s_w}
\sum_{a=1}^n \sum_{i,j=1}^6
\left[
  m_i y_{ija} q_{ji} g_1
  + m_{\ell_2} z_{ija} q_{ij} g_{12}
  - m_{\ell_1} w_{ija} q_{ji} \left( g_1 + g_{11} \right)
  \right],
\label{T4Zbl2}\\
\bar{b}_{r,\mathrm{15(a)}} &=& \frac{1}{c_w s_w}
\sum_{a=1}^n \sum_{i,j=1}^6
\left[
  - m_i x_{ija} q_{ij} g_1
  + m_{\ell_1} z_{ija} q_{ij} \left( g_1 + g_{11} \right)
  - m_{\ell_2} w_{ija} q_{ji} g_{12}
  \right],
\hspace*{7mm}
\label{T4Zbr2}\\
\bar{c}_{l,\mathrm{15(a)}} &=& \frac{1}{c_w s_w}
\sum_{a=1}^n \sum_{i,j=1}^6
\left[
  - m_j y_{ija} q_{ij} g_2
  + m_{\ell_2} z_{ija} q_{ij} \left( g_2 + g_{22} \right)
  - m_{\ell_1} w_{ija} q_{ji} g_{12} \right],
\label{T4Zcl2}\\
\bar{c}_{r,\mathrm{15(a)}} &=& \frac{1}{c_w s_w}
\sum_{a=1}^n \sum_{i,j=1}^6
\left[ m_j x_{ija} q_{ji} g_2
  + m_{\ell_1} z_{ija} q_{ij} g_{12}
  - m_{\ell_2} w_{ija} q_{ji} \left( g_2 + g_{22} \right) \right].
\label{T4Zcr2}
\ea
\es
In Eqs.~\eqref{coefZ2},
\bs
\label{Zxyzw}
\ba
x_{ija} &:=& R_{a {\ell_2} j} L^\ast_{a {\ell_1} i}, \\
y_{ija} &:=& L_{a {\ell_2} j} R^\ast_{a {\ell_1} i}, \\
z_{ija} &:=& R_{a {\ell_2} j} R^\ast_{a {\ell_1} i}, \\
w_{ija} &:=& L_{a {\ell_2} j} L^\ast_{a {\ell_1} i},
\ea
\es
and the $g$ functions are defined in Eqs.~\eqref{gfunctions}.
Note that the divergences cancel out in
$\bar{a}_{l,\mathrm{15(a)}}$ and $\bar{a}_{r,\mathrm{15(a)}}$.
Indeed,
\ba
\sum_{i,j=1}^6 z_{ija} q_{ij} =
\sum_{i,j=1}^6 R_{a \ell_2 j} R^\ast_{a \ell_1 i} q_{ij} &=&
\sum_{k, k^\prime = 1}^{n_d} \sum_{i,j=1}^6
\mathcal{U}_{ka}^\ast \left( \Delta_k^\dagger X \right)_{\ell_2 j}
\mathcal{U}_{k^\prime a} \left( X^\dagger \Delta_{k^\prime} \right)_{i \ell_1}
\left( U^\dagger U \right)_{ij}
\no &=&
\sum_{k, k^\prime = 1}^{n_d} \mathcal{U}_{ka}^\ast \mathcal{U}_{k^\prime a}
\left( \Delta_k^\dagger X U^T U^\ast X^\dagger
\Delta_{k^\prime} \right)_{\ell_2 \ell_1}
\no &=& 0,
\label{prop1}
\ea 
because the unitarity of the matrix $U_6$ of Eq.~\eqref{U6}
implies $U X^T = 0_{3 \times 3}$;
and
\ba
\sum_{i,j=1}^6 w_{ija} q_{ji} =
\sum_{i,j=1}^6 L_{a \ell_2 j} L^\ast_{a \ell_1 i} q_{ji} &=&
\sum_{k, k^\prime = 1}^{n_d} \sum_{i,j=1}^6
\mathcal{U}_{ka}^\ast \left( \Gamma_k U \right)_{\ell_2 j}
\mathcal{U}_{k^\prime a} \left( U^\dagger \Gamma_{k^\prime}^\dagger \right)_{i \ell_1}
\left( U^\dagger U \right)_{ji}
\no &=&
\sum_{k, k^\prime = 1}^{n_d} \mathcal{U}_{ka}^\ast \mathcal{U}_{k^\prime a}
\left( \Gamma_k U U^\dagger U U^\dagger
\Gamma_{k^\prime}^\dagger \right)_{\ell_2 \ell_1}
\no &=&
\sum_{k, k^\prime = 1}^{n_d} \mathcal{U}_{ka}^\ast \mathcal{U}_{k^\prime a}
\left( \Gamma_k \Gamma_{k^\prime}^\dagger \right)_{\ell_2 \ell_1}
\label{prop2}
\ea 
vanishes if $\ell_2 \neq \ell_1$ because the matrices $\Gamma_k$
are diagonal.

The diagram~\ref{fig3}(b) yields
\bs
\label{jvidosp}
\ba
\bar{a}_{l,\mathrm{15(b)}} &=& \frac{e^2}{2 c_w s_w^3} \sum_{i, j = 1}^6
U_{{\ell_2} j} U_{{\ell_1} i}^\ast \left[ q_{ij} m_i m_j h_0
  + 2 q_{ji} h_{00}
  \right. \no & &
  + q_{ji} \left( m_{\ell_1}^2 + m_{\ell_2}^2 - q^2 \right)
  \left( h_0 + h_1 + h_2 + h_{12} \right)
  \no & & \left.
  + q_{ji} m_{\ell_1}^2 \left( h_1 + h_{11} \right)
  + q_{ji} m_{\ell_2}^2 \left( h_2 + h_{22} \right)
  \right],
\label{vjfkops} \\
\bar{a}_{r,\mathrm{15(b)}} &=& \frac{e^2 m_{\ell_2} m_{\ell_1}}{2 c_w s_w^3}
\sum_{i, j = 1}^6
U_{{\ell_2} j} U_{{\ell_1} i}^\ast q_{ji} \left( h_0 + h_1 + h_2 \right),
\\
\bar{b}_{l,\mathrm{15(b)}} &=& - \frac{e^2 m_{\ell_2}}{c_w s_w^3} \sum_{i, j = 1}^6
U_{{\ell_2} j} U_{{\ell_1} i}^\ast q_{ji} \left( h_0 + h_1 + h_2 + h_{12} \right),
\\
\bar{b}_{r,\mathrm{15(b)}} &=& - \frac{e^2 m_{\ell_1}}{c_w s_w^3} \sum_{i, j = 1}^6
U_{{\ell_2} j} U_{{\ell_1} i}^\ast q_{ji} \left( h_1 + h_{11} \right),
\\
\bar{c}_{l,\mathrm{15(b)}} &=& - \frac{e^2 m_{\ell_2}}{c_w s_w^3} \sum_{i, j = 1}^6
U_{{\ell_2} j} U_{{\ell_1} i}^\ast q_{ji} \left( h_2 + h_{22} \right),
\\
\bar{c}_{r,\mathrm{15(b)}} &=& - \frac{e^2 m_{\ell_1}}{c_w s_w^3} \sum_{i, j = 1}^6
U_{{\ell_2} j} U_{{\ell_1} i}^\ast q_{ji} \left( h_0 + h_1 + h_2 + h_{12} \right),
\ea
\es
where the $h$ functions are defined in Eqs.~\eqref{hfunctions}.
When writing Eq.~\eqref{vjfkops} we have used the fact that
\be
\sum_{i,j=1}^6 U_{\ell_2 j} U^\ast_{\ell_1 i} q_{ji}
= \sum_{i,j=1}^6 U_{\ell_2 j} U^\dagger_{i \ell_1} \left( U^\dagger U \right)_{ji}
= \delta_{\ell_2 \ell_1},
\ee
because $U U^\dagger$ is the $3 \times 3$ unit matrix,
\textit{cf.}\ Eq.~\eqref{vklspac1}.

\newpage
\setcounter{equation}{0}
\renewcommand{\theequation}{D\arabic{equation}}
\section{$S^0_b \to \ell_1^+ \ell_2^-$}
\label{Sect_taumuS}

We compute the process $S^0_b \left( q \right)
\to \ell_1^+ \left( p_1 \right) \ell_2^- \left( - p_2 \right)$,
where $S^0_b$ is a physical neutral scalar,
\textit{i.e.} $b \neq 1$.
Equations~\eqref{q2} hold and $q^2 = m_b^2$.
The amplitude is written
\ba
T_b &=& \mathcal{S}\, \bar u_{\ell_2} \left( - p_2 \right)
\left( d_{lb} P_L + d_{rb} P_R \right) v_{\ell_1} \left( p_1 \right),
\label{pvcifus}
\ea
where $\mathcal{S}$ was defined in Eq.~\eqref{S}.
The decay width in the rest frame of $S^0_b$ is
\be
\label{Gamma_Higgs}
\Gamma = \frac{\sqrt{\lambda}}{16 \pi m_b^3}\,
\left| \mathcal{S} \right|^2
\left[ \left( m_b^2 - m_{\ell_1}^2 - m_{\ell_2}^2 \right)
  \left( \left| d_{lb} \right|^2
  + \left| d_{rb} \right|^2 \right)
  - 4\, m_{\ell_1} m_{\ell_2}\,
  \mathrm{Re} \left( d_{lb} d_{rb}^\ast \right) \right],
\ee
where
\be
\lambda := m_b^4 + m_{\ell_1}^4 + m_{\ell_2}^4
- 2 \left( m_b^2 m_{\ell_1}^2 + m_b^2 m_{\ell_2}^2 + m_{\ell_1}^2 m_{\ell_2}^2 \right).
\ee

\subsection{Diagrams in which $S^0_b$ attaches to charged leptons}

\begin{figure}[h]
  \begin{center}
    \includegraphics[width=0.5\textwidth]{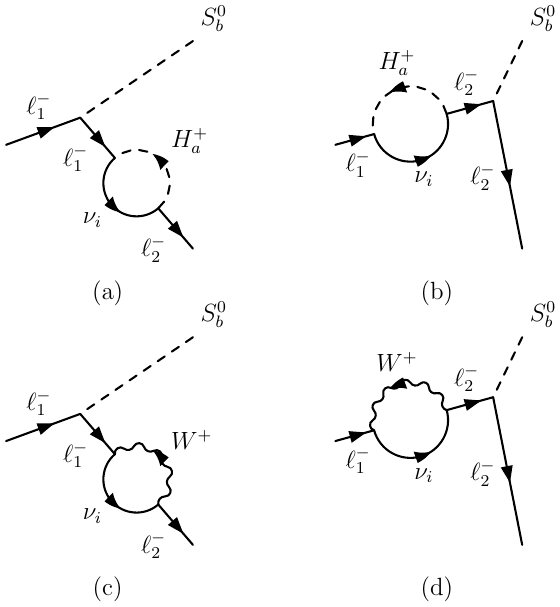}
  \end{center}
  \caption{The four self-energy-like diagrams
    for $S^0_b \to \ell_1^+ \ell_2^-$.
    \label{fig4}}
\end{figure}

There are self-energy-like diagrams with a loop of either
$H_a^\pm$---diagrams~(a) and~(b) in Fig.~\ref{fig4}---or
$W^\pm$---diagrams~(c) and~(d) in Fig.~\ref{fig4}.
The vertex of $S_b^0$ with the charged leptons
is given by Eq.~\eqref{nvjfui}.
One obtains
\bs
\label{vmfr0395}
\ba
d_{lb,\mathrm{16(a,b)}} &=&
\frac{1}{\sqrt{2} \left( m_{\ell_1}^2 - m_{\ell_2}^2 \right)} 
\sum_{a=1}^n \sum_{i=1}^6
\left\{ g_{b {\ell_1}}
\left( m_{\ell_2} x_{ia} + m_{\ell_1} y_{ia} \right)
B_0 \left( m_{\ell_2}^2, m_i^2, m_a^2 \right)
\right. \no & &
+ g_{b {\ell_1}} m_{\ell_2}
\left( m_{\ell_1} z_{ia} + m_{\ell_2} w_{ia} \right)
B_1 \left( m_{\ell_2}^2, m_i^2, m_a^2 \right)
\no & &
- g_{b {\ell_2}}
\left( m_{\ell_1} x_{ia} + m_{\ell_2} y_{ia} \right)
B_0 \left( m_{\ell_1}^2, m_i^2, m_a^2 \right)
\no & & \left.
- g_{b {\ell_2}} m_{\ell_1}
\left( m_{\ell_1} z_{ia} + m_{\ell_2} w_{ia} \right)
B_1 \left( m_{\ell_1}^2, m_i^2, m_a^2 \right)
\right\},
\label{49a} \\
d_{rb,\mathrm{16(a,b)}} &=&
\frac{1}{\sqrt{2} \left( m_{\ell_1}^2 - m_{\ell_2}^2 \right)} 
\sum_{a=1}^n \sum_{i=1}^6 \left\{
g_{b {\ell_1}}^\ast
\left( m_{\ell_1} x_{ia} + m_{\ell_2} y_{ia} \right)
B_0 \left( m_{\ell_2}^2, m_i^2, m_a^2 \right)
\right. \no & &
+ g_{b {\ell_1}}^\ast m_{\ell_2}
\left( m_{\ell_2} z_{ia} + m_{\ell_1} w_{ia} \right)
B_1 \left( m_{\ell_2}^2, m_i^2, m_a^2 \right)
\no & &
- g_{b {\ell_2}}^\ast
\left( m_{\ell_2} x_{ia} + m_{\ell_1} y_{ia} \right)
B_0 \left( m_{\ell_1}^2, m_i^2, m_a^2 \right)
\no & & \left.
- g_{b {\ell_2}}^\ast m_{\ell_1}
\left( m_{\ell_2} z_{ia} + m_{\ell_1} w_{ia} \right)
B_1 \left( m_{\ell_1}^2, m_i^2, m_a^2 \right)
\right\},
\label{49b} \\
d_{lb,\mathrm{16(c,d)}} &=&
\frac{e^2 m_{\ell_1}}{\sqrt{2} s_w^2 \left( m_{\ell_1}^2 - m_{\ell_2}^2 \right)}
\sum_{i=1}^6 U_{{\ell_2} i} U_{{\ell_1} i}^\ast
\left[ g_{b {\ell_1}} m_{\ell_2}\,
  B_1 \left( m_{\ell_2}^2, m_i^2, m_W^2 \right)
  \right. \no & & \left.
  - g_{b {\ell_2}} m_{\ell_1}\,
  B_1 \left( m_{\ell_1}^2, m_i^2, m_W^2 \right)
\right],
\label{49c} \\
d_{rb,\mathrm{16(c,d)}} &=&
\frac{e^2 m_{\ell_2}}{\sqrt{2} s_w^2 \left( m_{\ell_1}^2 - m_{\ell_2}^2 \right)}
\sum_{i=1}^6 U_{{\ell_2} i} U_{{\ell_1} i}^\ast
\left[ g_{b {\ell_1}}^\ast m_{\ell_2}\,
  B_1 \left( m_{\ell_2}^2, m_i^2, m_W^2 \right)
  \right. \no & & \left.
  - g_{b {\ell_2}}^\ast m_{\ell_1}\,
  B_1 \left( m_{\ell_1}^2, m_i^2, m_W^2 \right)
\right]. \hspace*{7mm}
\label{49d}
\ea
\es

\subsection{Diagrams in which $S^0_b$ attaches to charged scalars}

\begin{figure}[h]
  \begin{center}
    \includegraphics[width=0.22\textwidth]{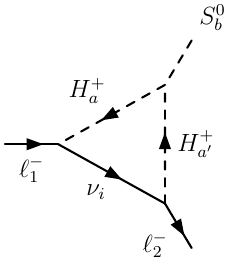}
  \end{center}
  \caption{Diagram for $S^0_b \to \ell_1^+ \ell_2^-$
    with the neutral scalar attaching to two charged scalars.
    \label{fig5}}
\end{figure}

There is a diagram,
depicted in Fig.~\ref{fig5},
wherein the $S^0_b$ attaches to two charged scalars
that may in principle be different.
We parameterize the vertex of the three scalars
through Eq.~\eqref{3vifogp},
where the coefficients $\lambda_{a a^\prime b}$ are in general complex
but obey $\lambda_{a a^\prime b} = \lambda_{a^\prime a b}^\ast$
because of the Hermiticity of the Lagrangian.
The values of the $\lambda_{a a^\prime b}$ depend on the scalar potential
and are unconstrained by gauge invariance,
unless either $a = 1$ or $a^\prime = 1$.
The diagram of Fig.~\ref{fig5} yields
\bs
\label{jdisofi}
\ba
d_{lb,\mathrm{17}} &=& \sum_{a, a^\prime = 1}^n \sum_{i=1}^6
\lambda_{a a^\prime b} \left(
R_{a^\prime {\ell_2} i} R_{a {\ell_1} i}^\ast m_{\ell_2}\, j_2
+ L_{a^\prime {\ell_2} i} L_{a {\ell_1} i}^\ast m_{\ell_1}\, j_1
+ L_{a^\prime {\ell_2} i} R_{a {\ell_1} i}^\ast m_i\, j_0
\right),
\\
d_{rb,\mathrm{17}} &=& \sum_{a, a^\prime = 1}^n \sum_{i=1}^6
\lambda_{a a^\prime b} \left(
R_{a^\prime {\ell_2} i} R_{a {\ell_1} i}^\ast m_{\ell_1}\, j_1
+ L_{a^\prime {\ell_2} i} L_{a {\ell_1} i}^\ast m_{\ell_2}\, j_2
+ R_{a^\prime {\ell_2} i} L_{a {\ell_1} i}^\ast m_i\, j_0
\right), \hspace*{7mm}
\ea
\es
where $j_{0,1,2}$ are defined by Eqs.~\eqref{jfunctions}.

Note that the diagram of Fig.~\ref{fig5} implicitly contains
the cases where either $H_a^\pm$ or $H_{a^\prime}^\pm$ (or both)
coincide with the charged Goldstone bosons $G^\pm := H_1^\pm$.
In those cases one must use $m_{a=1} = m_W$ together with Eqs.~\eqref{vjfifdop0}.

\subsection{Diagrams in which $S^0_b$ attaches to $W$ bosons}

\begin{figure}[h]
  \begin{center}
    \includegraphics[width=0.75\textwidth]{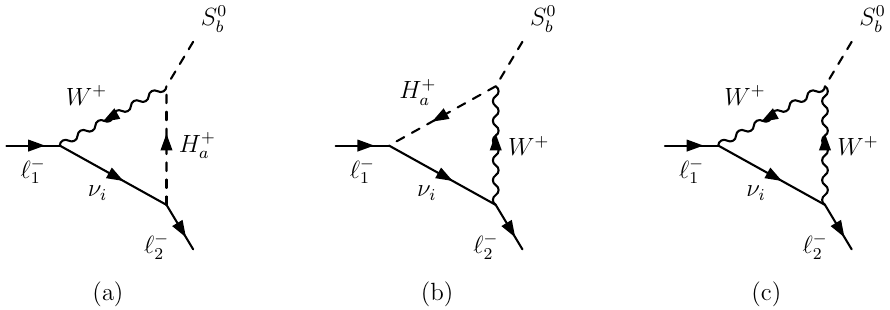}
  \end{center}
  \caption{Three diagrams for $S^0_b \to \ell_1^+ \ell_2^-$
    with the neutral scalar attaching to a gauge boson.
    \label{fig6}}
\end{figure}

We next compute the diagrams in Fig.~\ref{fig6}.
The relevant terms of the Lagrangian are
the ones of Eq.~\eqref{vfi493586}~\cite{Grimus:2002ux,Grimus:2007if}.
Diagram~\ref{fig6}(a) produces
\bs
\label{6a}
\ba
d_{lb,\mathrm{18(a)}} &=&
\frac{e^2}{2 \sqrt{2}\, s_w^2} \sum_{a=1}^n \sum_{i=1}^6
U_{{\ell_1} i}^\ast \left( \mathcal{V}^\dagger \mathcal{U} \right)_{ba} \left\{
- R_{a {\ell_2} i} m_i m_{\ell_2} \left( 2\, k_0 + k_2 \right)
\right. \no & &
- L_{a {\ell_2} i} \left[ 4\, k_{00}
  + m_{\ell_1}^2 \left( k_{11} + k_{12} + k_1 \right)
  + m_{\ell_2}^2 \left( k_{22} + k_{12} + 2\, k_1 + 2\, k_2 \right)
  \right. \no & & \left. \left.
  - q^2 \left( k_{12} + 2\, k_1 \right)
  \right]
\right\},
\\
d_{rb,\mathrm{18(a)}} &=&
\frac{e^2}{2 \sqrt{2}\, s_w^2} \sum_{a=1}^n \sum_{i=1}^6
U_{{\ell_1} i}^\ast \left( \mathcal{V}^\dagger \mathcal{U} \right)_{ba} m_{\ell_1}
\left[ R_{a {\ell_2} i} m_i \left( k_0 - k_1 \right)
  + L_{a {\ell_2} i} m_{\ell_2} \left( 2\, k_1 + k_2 \right) \right],
\no & &
\ea
\es
with the $k$ functions defined in Eqs.~\eqref{kfunctions}.
Diagram~\ref{fig6}(b) produces
\bs
\label{6b}
\ba
d_{lb,\mathrm{18(b)}} &=&
\frac{e^2}{2 \sqrt{2}\, s_w^2} \sum_{a=1}^n \sum_{i=1}^6
U_{{\ell_2} i} \left( \mathcal{U}^\dagger \mathcal{V} \right)_{ab} m_{\ell_2}
\left[
R_{a {\ell_1} i}^\ast m_i \left( l_0 - l_2 \right)
+ L_{a {\ell_1} i}^\ast m_{\ell_1} \left( l_1 + 2\, l_2 \right) \right],
\no & &
\\
d_{rb,\mathrm{18(b)}} &=&
\frac{e^2}{2 \sqrt{2}\, s_w^2} \sum_{a=1}^n \sum_{i=1}^6
U_{{\ell_2} i} \left( \mathcal{U}^\dagger \mathcal{V} \right)_{ab} \left\{
- R_{a {\ell_1} i}^\ast m_i m_{\ell_1} \left( 2\, l_0 + l_1 \right)
\right. \no & &
- L_{a {\ell_1} i}^\ast \left[ 4\, l_{00}
  + m_{\ell_1}^2 \left( l_{11} + l_{12} + 2\, l_1 + 2\, l_2 \right)
  + m_{\ell_2}^2 \left( l_{22} + l_{12} + l_2 \right)
  \right. \no & & \left. \left.
  - q^2 \left( l_{12} + 2\, l_2 \right)
  \right]
\right\},
\ea
\es
with the $l$ functions defined in Eqs.~\eqref{lfunctions}.
Equations~\eqref{6a} and~\eqref{6b} contain no divergences
because
\be
\label{djapdo}
\sum_{i=1}^6 L_{a \ell i} U_{\ell^\prime i}^\ast
= \sum_{k=1}^{n_d} \mathcal{U}_{ka}^\ast
\left( \Gamma_k U U^\dagger \right)_{\ell \ell^\prime}
= \sum_{k=1}^{n_d} \mathcal{U}_{ka}^\ast
\left( \Gamma_k \right)_{\ell \ell^\prime}
\ee
%
vanishes if $\ell \neq \ell^\prime$,
since the matrices $\Gamma_k$ are diagonal.

Equations~\eqref{6a} and~\eqref{6b}
include the particular case where $a=1$;
then,
$H_a^\pm$ coincides with the Goldstone bosons $G^\pm$.
In that particular case one must use $m_{a=1} = m_W$
together with Eqs.~\eqref{r1l1} and
\be
\left( \mathcal{U}^\dagger \mathcal{V} \right)_{1b}
= \left( \mathcal{V}^\dagger \mathcal{U} \right)_{b1}
= x_b,
\ee
where $x_b$ is the real number defined in Eq.~\eqref{xb}.

In order to compute diagram~\ref{fig6}(c)
one must know the vertex of a neutral scalar with two $W^\pm$ gauge bosons,
which is given by Eq.~\eqref{vuffsid9}~\cite{Grimus:2007if}.
One then obtains
\bs
\label{jcidops}
\ba
d_{lb, \mathrm{18(c)}} &=&
- \frac{e^3 m_W m_{\ell_2} x_b}{s_w^3}\,
\sum_{i=1}^6 U_{{\ell_2} i} U_{{\ell_1} i}^\ast\, f_2,
\\
d_{rb, \mathrm{18(c)}} &=&
- \frac{e^3 m_W m_{\ell_1} x_b}{s_w^3}\,
\sum_{i=1}^6 U_{{\ell_2} i} U_{{\ell_1} i}^\ast\, f_1,
\ea
\es
where $f_1$ and $f_2$ are defined in Eqs.~\eqref{ffunctions}.
Notice that Eqs.~\eqref{jcidops} only hold for $b \neq 1$;
indeed,
$d_{l1, \mathrm{18(c)}} = d_{r1, \mathrm{18(c)}} = 0$
because the vertex $W^+ W^- G^0$ does not exist.

\subsection{Diagrams where $S^0_b$ attaches to neutrino lines}

\begin{figure}[h]
  \begin{center}
    \includegraphics[width=0.5\textwidth]{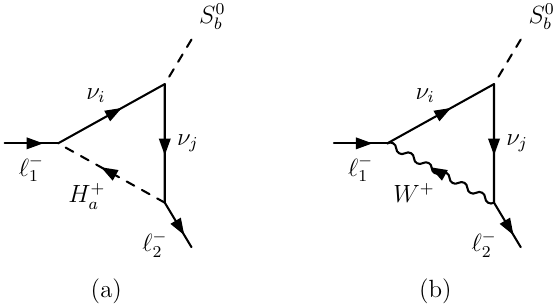}
  \end{center}
  \caption{Two diagrams for $S^0_b \to \ell_1^+ \ell_2^-$
    where $S^0_b$ attaches to neutrinos.
    \label{fig7}}
\end{figure}

The neutral scalar $S^0_b$ may also attach to two internal neutrino lines.
The relevant diagrams are displayed in Fig.~\ref{fig7}.
The vertex of the neutral scalars with the neutrinos
is given by Eq.~\eqref{jdsiao}.
From Fig.~\ref{fig7}(a) one obtains
\bs
\label{7a}
\ba
d_{lb,\mathrm{19(a)}} &=&
\frac{1}{\sqrt{2}}\, \sum_{i, j = 1}^6 \sum_{a=1}^n \left\{
- x_{ija} f_{bji} m_{\ell_2} m_{\ell_1} \left( g_0 + g_1 + g_2 \right)
- y_{ija} f_{bji} m_i m_j g_0
\right. \no & &
- y_{ija} f_{bji}^\ast \left[ 4 g_{00}
  + m_{\ell_1}^2 \left( g_1 + g_{11} + g_{12} \right)
  + m_{\ell_2}^2 \left( g_2 + g_{22} + g_{12} \right)
  - q^2 g_{12} \right]
\no & &
+ z_{ija} f_{bji} m_i m_{\ell_2} \left( g_0 + g_2 \right)
+ z_{ija} f_{bji}^\ast m_j m_{\ell_2} g_2 
\no & & \left.
+ w_{ija} f_{bji}^\ast m_i m_{\ell_1} g_1
+ w_{ija} f_{bji} m_j m_{\ell_1} \left( g_0 + g_1 \right)
\right\},
\\
d_{rb,\mathrm{19(a)}} &=&
\frac{1}{\sqrt{2}}\, \sum_{i, j = 1}^6 \sum_{a=1}^n \left\{
- y_{ija} f_{bji}^\ast m_{\ell_2} m_{\ell_1} \left( g_0 + g_1 + g_2 \right)
- x_{ija} f_{bji}^\ast m_i m_j g_0
\right. \no & &
- x_{ija} f_{bji} \left[ 4 g_{00}
  + m_{\ell_1}^2 \left( g_1 + g_{11} + g_{12} \right)
  + m_{\ell_2}^2 \left( g_2 + g_{22} + g_{12} \right)
  - q^2 g_{12} \right]
\no & &
+ z_{ija} f_{bji}^\ast m_j m_{\ell_1} \left( g_0 + g_1 \right)
+ z_{ija} f_{bji} m_i m_{\ell_1} g_1 
\no & & \left.
+ w_{ija} f_{bji} m_j m_{\ell_2} g_2
+ w_{ija} f_{bji}^\ast m_i m_{\ell_2} \left( g_0 + g_2 \right)
\right\},
\ea
\es
where the relevant symbols are defined in Eqs.~\eqref{Zxyzw}
and~\eqref{gfunctions}.
The divergences originating in the function $g_{00}$
vanish in Eqs.~\eqref{7a} because
\bs
\ba
\sum_{i, j = 1}^6 \sum_{a=1}^n y_{ija} f_{bji}^\ast
&=&
\sum_{i, j = 1}^6 \sum_{a=1}^n L_{a {\ell_2} j} R_{a {\ell_1} i}^\ast f_{bji}^\ast
\no &=&
\sum_{i, j = 1}^6 \sum_{a=1}^n \sum_{k, k^\prime, k^{\prime\prime} = 1}^{n_d}
\mathcal{U}^\ast_{ka} \left( \Gamma_k U \right)_{{\ell_2} j}
\mathcal{U}_{k^\prime a} \left( X^\dagger \Delta_{k^\prime} \right)_{i {\ell_1}}
\no & & \times
\mathcal{V}^\ast_{k^{\prime\prime} b} \left( U^\dagger \Delta_{k^{\prime\prime}}^\ast X
+ X^T \Delta_{k^{\prime\prime}}^\ast U^\ast \right)_{ij}
\no &=&
\sum_{i, j = 1}^6 \sum_{a=1}^n \sum_{k, k^\prime, k^{\prime\prime} = 1}^{n_d}
\left( \mathcal{U} \mathcal{U}^\dagger \right)_{k^\prime k}
\mathcal{V}^\ast_{k^{\prime\prime} b}
\no & & \times
\left( \Gamma_k U U^\dagger \Delta_{k^{\prime\prime}}^\ast X X^\dagger \Delta_{k^\prime}
+ \Gamma_k U X^T \Delta_{k^{\prime\prime}}^\ast U^\ast X^\dagger
\Delta_{k^\prime} \right)_{{\ell_2} {\ell_1}}
\no &=&
\sum_{i, j = 1}^6 \sum_{a=1}^n \sum_{k, k^\prime, k^{\prime\prime} = 1}^{n_d}
\delta_{k^\prime k} \mathcal{V}^\ast_{k^{\prime\prime} b}
\left( \Gamma_k \Delta_{k^{\prime\prime}}^\ast \Delta_{k^\prime} \right)_{{\ell_2} {\ell_1}}
\no &=& 0,
\ea
\es
since the Yukawa-coupling matrices $\Gamma_k$ and $\Delta_k$
are all diagonal.
In a similar fashion one easily demonstrates that
\be
\sum_{i, j = 1}^6 \sum_{a=1}^n x_{ija} f_{bji} = 0.
\ee

From Fig.~\ref{fig7}(b) one obtains
\bs
\label{7b}
\ba
d_{lb,\mathrm{19(b)}} &=&
\frac{e^2 m_{\ell_2}}{\sqrt{2} s_w^2}
\sum_{i, j = 1}^6 U_{{\ell_2} j} U_{{\ell_1} i}^\ast
\left[ f_{bji} m_j\, h_2 + f_{bji}^\ast m_i \left( h_0 + h_2 \right) \right],
\\
d_{rb,\mathrm{19(b)}} &=& \frac{e^2 m_{\ell_1}}{\sqrt{2} s_w^2}
\sum_{i, j = 1}^6 U_{{\ell_2} j} U_{{\ell_1} i}^\ast
\left[ f_{bji}^\ast m_i\, h_1 + f_{bji} m_j \left( h_0 + h_1 \right) \right],
\ea
\es
\textit{cf.}~Eqs~\eqref{hfunctions}.

\newpage
\setcounter{equation}{0}
\renewcommand{\theequation}{E\arabic{equation}}
\section{The $Z$ invisible decay width}
\label{Sect_Z}

The determination by LEP of the number of light active neutrinos
provides a constraint to heavy-neutrino mixing.
The $Z$ invisible decay width was measured by LEP~\cite{ALEPH:2005ab, PDG2020}
to be
\be
\Gamma \left( Z \rightarrow \mathrm{invisible} \right)_{\mathrm{experimental}}
= \left( 0.499 \pm 0.0015 \right) \mathrm{GeV}. \label{InvZwidthExp}
\ee
This is almost $2\sigma$ below the SM theoretical expectation
\be
\Gamma \left( Z \rightarrow \mathrm{invisible} \right)_{\mathrm{SM}} = 
\sum_\nu \Gamma
\left( Z \rightarrow \nu \bar{\nu} \right)_{\mathrm{SM}}
= \left( 0.50169 \pm 0.00006 \right) \mathrm{GeV}.
\ee

The tree-level $Z$ invisible decay width
in the presence of six Majorana neutrinos with masses $m_i$
reads~\cite{Abada:2013aba}
\ba
\Gamma \left( Z \rightarrow \mathrm{invisible} \right)_{\mathrm{tree}} &=&
\sum_{i=1}^6 \sum_{j=i}^6 \Delta_{ij}\,
\Theta \left( m_Z-m_i-m_j \right)
\frac{\sqrt{\lambda \left( m_Z, m_i, m_j \right)}}{24 \pi m_Z v^2}
\left\{ \left[
  \vphantom{\frac{\left( m_i^2 - m_j^2\right)^2}{m_Z^2}}
  2 m_Z^2
  \right. \right. \no & & \left. \left.
  - m_i^2 - m_j^2 - \frac{\left( m_i^2 - m_j^2\right)^2}{m_Z^2} \right]
\left| x^{ij} \right|^2
- 6\, m_i m_j\, \mathrm{Re} \left(x^{ij}\right)^2 \right\},
\label{InvZwidthTree}
\ea 
where $\Theta$ is the Heaviside step function,
\textit{i.e.}\ the sum in Eq.~\eqref{InvZwidthTree}
extends over pairs of neutrinos $\nu_i$ and $\nu_j$
that have masses $m_i$ and $m_j$,
respectively,
such that $m_i + m_j$ is smaller than the mass $m_Z$ of the $Z$;
the kinematical function $\lambda$ is defined as
\be
\lambda \left( a,\, b,\, c \right) = a^4 + b^4 + c^4
- 2 \left( a^2 b^2 + a^2 c^2 + b^2 c^2 \right).
\label{kinemFunct}
\ee
The factor $\Delta_{ij} = 1 - \left. \delta_{ij} \right/ 2$
in Eq.~\eqref{InvZwidthTree}
accounts for the Majorana character of the neutrinos.
The coupling $x^{ij}$ is
\be
x^{ij} = \sum_{\ell = e, \mu, \tau} U_{\ell i}^\ast\, U_{\ell j}
= \left( U^\dagger U \right)_{ij},
\label{x^ij}
\ee
and the vacuum expectation value is defined through
$v \equiv \left( \sqrt{2}\, G_F \right)^{-1/2} \approx 246.22$\,GeV,
where $G_F$ is the Fermi coupling constant.

In a correct computation of the full invisible width of the $Z$
one must include a parameter $\rho$ that accounts
for that part of the radiative corrections coming from the SM loops.
Thus,
\be
\Gamma \left( Z \rightarrow \mathrm{invisible} \right) =
\rho \times
\Gamma \left( Z \rightarrow \mathrm{invisible} \right)_{\mathrm{tree}},
\label{InvZwidthFull}
\ee
where $\rho$ is evaluated as~\cite{DeRomeri:2016gum,Brdar:2020nbj}
\be
\rho =
\frac{\Gamma \left( Z \rightarrow \mathrm{invisible}
  \right)_{\mathrm{SM}}}{\sum_{\nu} \Gamma \left( Z \rightarrow \nu \bar{\nu}
  \right)_\mathrm{tree,\, SM}}
= \frac{8 \pi v^2\, \Gamma \left( Z \rightarrow \mathrm{invisible}
  \right)_{\mathrm{SM}}}{m_Z^3}. \label{rho}
\ee
After accounting for the uncertainties of 
$\Gamma \left( Z \rightarrow \mathrm{invisible} \right)_{\mathrm{SM}}$,
one obtains $\rho = 1.00812 \pm 0.00012$.

In our numerical results,
the tree-level $Z$ invisible decay width in Eq.~\eqref{InvZwidthTree}
is always within the $1\sigma$ experimental bands of Eq.~\eqref{InvZwidthExp},
while the decay width of Eq.~\eqref{InvZwidthFull},
including the corrections,
is within the $2\sigma$ experimental bands. 
Therefore,
the $Z$ invisible decay width does not effectively constrain
the branching ratios of LFV processes in our model.
This is distinct from LFV studies
in the inverse seesaw model~\cite{DeRomeri:2016gum}
or in the effective field theory of the seesaw~\cite{Coy:2018bxr}.
That happens because in our case the masses
$m_{4,5,6}$ of the heavy neutrinos are sufficiently high that
the $Z$ can never
decay into a heavy neutrino plus a light neutrino, except for very small values
$\left| d_\ell \right| \lesssim 10^{-7}$ of the
Yukawa couplings;
and because the non-unitarity of the matrix $U$ has a very weak impact
on the couplings of the active neutrinos to the $Z$ boson in our model.

\newpage
\setcounter{equation}{0}
\renewcommand{\theequation}{F\arabic{equation}}
\section{Constraints on the mass of the charged scalar}
\label{Sect_mH+}

Direct constraints on $m_{H^+}$
may be obtained from collider experiments
on the production and decay of on-shell charged Higgs bosons. 
The search sensitivity is limited by the kinematic reach of experiments,
but collider constraints have the advantage
of being robust and model-independent.
The bound obtained from direct searches at LEP
for any value of $\tan{\beta}$
is $m_{H^+} > 78.6$\,GeV at 95\%\,CL~\cite{Searches:2001ac}. 
Combining data of the four LEP experiments,
a limit of $m_{H^+} \gtrsim 80$\,GeV is obtained~\cite{PDG2020,Abbiendi:2013hk},
while $m_{H^+} \gtrsim 150$\,GeV may be derived from
the searches at LHC~\cite{Khachatryan:2015qxa,Arbey:2017gmh}. 
Stronger mass limits on $m_{H^+}$ may be obtained
for specific regions of $\tan{\beta}$.

Some constraints on $m_{H^+}$ from flavour physics
depend strongly on the 2HDM Yukawa type,
while others are type-independent. 
Among the flavour processes,
the constraints from $b \to s \gamma$ are most stringent
due to the
constructive interference
of the $H^\pm$ contribution with the SM contribution.
For a type-II 2HDM,
the lower limit $m_{H^+} > 480$\,GeV at 95\%\,CL~\cite{Misiak:2015xwa}
includes NNLO QCD corrections and is rather independent of $\tan{\beta}$.
In a recent study~\cite{Misiak:2017bgg},
the branching ratio of $b \to s \gamma$
enforces $m_{H^+} \gtrsim 580$\,GeV at 95\%\,CL
both for the type-II and for the flipped 2HDM.

The recent global fits
in Refs.~\cite{Arbey:2017gmh,Haller:2018nnx,Chowdhury:2017aav,Eberhardt:2020dat}
give bounds on the charged-Higgs mass for various 2HDM Yukawa types. 
In those studies only 2HDMs with a $\mathbb{Z}_2$-symmetric potential
are considered,
but one may suppose that the bounds would be similar for the general 2HDM.
In Ref.~\cite{Arbey:2017gmh} it is found that, for the type-II 2HDM,
flavour-physics observables impose a lower bound
$m_{H^+} \gtrsim 600$\,GeV that is independent of $\tan{\beta}$
when $\tan{\beta}> 1$ but increases to $m_{H^+} \gtrsim 650$\,GeV
when $\tan{\beta} < 1$. 
In Ref.~\cite{Chowdhury:2017aav},
$m_{H^+} > 740$\,GeV in both the type-II and flipped 2HDMs,
but $m_{H^+} \gtrsim 460$\,GeV for the lepton-specific 2HDM. 
In Ref.~\cite{Eberhardt:2020dat} on finds $m_{H^+} \gtrsim 500$\,GeV
or $m_{H^+} \gtrsim 750$\,GeV in the aligned 2HDM, 
depending on the fitted mass range. 
However,
for the type-I and lepton-specific 2HDMs
the restrictions on $m_{H^+}$ from flavour constraints
are weaker~\cite{Arbey:2017gmh,Haller:2018nnx,Sanyal:2019xcp}.

\newpage


\end{document}